\documentclass[12pt,preprint]{aastex}

\usepackage{graphicx}

\shorttitle{An observational study of SNe~2005ip and 2006jd.}
\shortauthors{Stritzinger et al.}
\begin{document}

\title{Multi-wavelength Observations of the Enduring Type~IIn Supernovae 2005ip and 2006jd\altaffilmark{1}}

\author{Maximilian Stritzinger\altaffilmark{2,3},
Francesco Taddia\altaffilmark{3},
Claes Fransson\altaffilmark{3},
Ori D. Fox\altaffilmark{4},
Nidia Morrell\altaffilmark{5},
M. M. Phillips\altaffilmark{5},
Jesper Sollerman\altaffilmark{3},
J. P. Anderson\altaffilmark{6},
Luis Boldt\altaffilmark{7}, 
Peter J. Brown\altaffilmark{8},
Abdo Campillay\altaffilmark{5},
Sergio Castellon\altaffilmark{5},
Carlos~Contreras\altaffilmark{5},
Gast\'on~Folatelli\altaffilmark{9},
S. M.  Habergham\altaffilmark{10},
Mario~Hamuy\altaffilmark{6},
Jens~Hjorth\altaffilmark{11},
Phil A. James\altaffilmark{10},
Wojtek~Krzeminski\altaffilmark{5},
Seppo Mattila\altaffilmark{12},
Sven~E.~Persson\altaffilmark{13},
Miguel Roth\altaffilmark{5} 
}

\altaffiltext{1}{This paper includes data gathered with the 6.5-m Magellan Telescopes, located at Las Campanas Observatory, Chile; the Gemini-North Telescope, Mauna Kea, USA (Gemini Program GN-2010B$-$Q$-$67, PI: Stritzinger); 
the ESO  NTT,
La Silla, Chile (Program 076.A-0156 and 078.D-0048, PI: Hamuy);  and the 
INT and the NOT (Proposal number 45$-$004, PI: Taddia), La Palma, Spain.}
\altaffiltext{2}{
 Department of Physics and Astronomy, Aarhus University, Ny Munkegade 120, DK-8000 Aarhus C, Denmark.}
\altaffiltext{3}{
The Oskar Klein Centre, Department of Astronomy, Stockholm University, AlbaNova, 10691 Stockholm, Sweden.} 
  \altaffiltext{4}{Astrophysics Science Division, Observational Cosmology Laboratory, NASA Goddard Space Flight Center, Greenbelt, MD 20771, USA.}

\altaffiltext{5}{Carnegie Observatories, Las Campanas Observatory, 
  Casilla 601, La Serena, Chile.}  \altaffiltext{6}{Departamento de Astronomia, Universidad de Chile, Casilla 36D, Santiago, Chile.}
\altaffiltext{7}{Argelander Institut f\"ur Astronomie, Universit\"at Bonn, Auf dem H\"ugel 71, D-53111 Bonn, Germany.}
  \altaffiltext{8}{Mitchell Institute for Fundamental Physics  \& Astronomy, Department of Physics \& Astronomy, Texas A\&M University, College Station, TX 77843, USA.}
\altaffiltext{9}{Institute for the Physics and Mathematics of the Universe (IPMU), University of Tokyo, 5-1-5 Kashiwanoha, Kashiwa, Chiba 277-8583, Japan.}  
  \altaffiltext{10}{Astrophysics Research Institute, Liverpool John Moores University,
Twelve Quays House, Egerton Wharf, Birkenhead CH41 1LD, UK.}
  \altaffiltext{11}{Dark Cosmology Centre, Niels Bohr Institute, University
of Copenhagen, Juliane Maries Vej 30, 2100 Copenhagen \O, Denmark.}  
\altaffiltext{12}{Tuorla Observatory, University of Turku, V\"ais\"al\"antie 20, FI-21500 Piikki\"o, Finland.}
\altaffiltext{13}{Observatories of the Carnegie Institution for
 Science, 813 Santa Barbara St., Pasadena, CA 91101, USA.}

\begin{abstract}
  \noindent We present an observational study of the Type IIn supernovae (SNe~IIn)  2005ip and 2006jd. 
 Broad-band UV, optical and near-IR photometry, and visual-wavelength 
spectroscopy of SN~2005ip complement and extend upon published observations 
to 6.5 years past discovery. 
 Our observations of SN~2006jd extend from UV to mid-infrared wavelengths, and like SN~2005ip, are compared to reported X-ray measurements to understand the nature of the progenitor. 
 Both objects display a number of similarities with the 1988Z-like subclass of 
 SN~IIn including: (i) remarkably similar early- and late-phase optical spectra, 
 (ii) a variety of high ionization coronal lines, (iii) long-duration optical and near-IR emission and, 
(iv) evidence of cold and warm dust components. 
 However, diversity is apparent including an unprecedented late-time $r$-band excess in SN~2006jd. 
 The observed differences are attributed to 
 differences between the mass-loss history of the 
 progenitor stars. 
 We conclude that the progenitor of SN~2006jd likely experienced a significant mass-loss event during its pre-SN evolution akin to the great 19th century eruption of $\eta$ Carinae.
 Contrarily, as advocated by \citet{smith09}, we find the circumstellar environment of SN~2005ip to be more consistent with a clumpy wind progenitor.

\end{abstract}

\section{INTRODUCTION}
\label{sec:intro}

Type IIn supernovae (SNe~IIn) are thought to originate from the 
core-collapse of massive stars enshrouded in hydrogen-rich circumstellar material (CSM). 
Members of this SN class display a variety of photometric and spectral properties, however each member shares the commonality of conspicuous narrow Balmer emission lines \citep{schlegel90,kiewe12}.\footnote{Narrow circumstellar lines have also been identified in some Type IIL SNe, e.g. SN~1979C \citep{branch81}, as well as in the SN~1984E in 
NGC~3169 \citep{dopita84,henry87}.} 
 H$\alpha$ emission typically dominates the spectrum and often exhibits 
multiple components characterized by a narrow core ($v_{\rm FWHM}$ $\sim$ 200 to 500 km~s$^{-1}$),
 an intermediate component ($v_{\rm FWHM}$ $\sim$ 2000 to 5000 km~s$^{-1}$),
 and a broad component ($v_{\rm FWHM}$ $\gtrsim$ 10$^{4}$ km~s$^{-1}$). 
Depending on the nature of the CSM and the dynamics of the explosion, these SNe can be observed from X-ray to radio wavelengths, and often display an infrared (IR) excess attributed to the reprocessing of energy by pre-existing dust, newly-formed dust, or a combination 
of the two \citep[e.g.][]{gerardy02,fox09}. 
The various observational characteristics of this SN class are therefore modulated by the SN emission, the mass-loss history of the progenitor star, and 
the physics of the SN--CSM interaction (e.g. \citealp{chevalier11,chevalier12}).

 Observations have now led to an emergent 
picture of a number of broadly different types of SNe~IIn.
At the top of the luminosity scale are the exceedingly bright ($M_V \sim -22$ mag) 
 objects, including amongst others; SN~2005ap \citep{quimby07}, SN~2006gy \citep{ofek07,smith07,woosley07,smithmccray07,agnoletto09}, SN~2006tf \citep{smith08},
SN~2008am \citep{chatzopoulos11} and SN~2008iy \citep{miller10}.
The exact nature of these objects is a matter of debate, but the general consensus 
seems to require one or more dense shells of CSM ejected from the progenitor 
during sequential luminous blue variable (LBV)-like eruptions in the decades prior to explosion.
Lower on the luminosity scale ($M_V \sim -$17 to $-19$ mag) are the 1988Z-like SNe~IIn \citep{turatto93,pastorello02}, which are thought to be powered for years by X-ray emission generated by the SN blast wave interacting with either 
 a highly asymmetric distribution of CSM or a clumpy-wind circumstellar medium
  \citep[e.g.][]{chugai94,fransson02}. 
Thirdly, are the 1994W-like SNe~IIn \citep{sollerman98,kankare12} that reach moderate peak brightnesses ($M_V \sim -18$ mag), and whose light curves exhibit a $\sim$ 100 day long plateau phase followed by a sharp drop in luminosity. 
 The origin of the 1994W-like objects is an open question with possibilities ranging from low-luminosity SNe interacting with CSM 
 \citep{chugai04}, to colliding shells of material \citep{dessart09}. 
 In addition to these groups there appears to be a variety of other SNe that interact with their CSM, ranging from amongst others, 
 SN~1997cy which was suggested to be  linked to a gamma-ray burst \citep{germany00,turatto00}, to 2002ic-like SNe that may or may not be associated with white dwarf progenitors
 \citep[see e.g.][and references therein] {hamuy03,deng04,benetti06}.
 Clearly, a diverse set of progenitor scenarios are required to plausibly explain the
 various objects classified under the SN~IIn designation. 
 
 SN~2005ip appeared as a 1988Z-like object whose
 SN--CSM interaction powered its spectral energy distribution (SED) for the duration of 
 years, and led to the formation of a rich forest of coronal lines observed over all epochs. 
 These coronal lines have been seen in only a handful of SNe~IIn, 
 e.g., SN~1988Z \citep{turatto93}, SN~1995N \citep{fransson02}, SN~1997ef \citep{hoffman08}, as well as in the Type~II SN~1987A \citep{groningsson06}, 
 however in the case of SN~2005ip their early appearance, high-ionization and long duration was unprecedented \citep{smith09}.
 Such coronal lines may be formed from the photoionization of dense gas by X rays produced from shocks, or as in the case of SN~1987A, hot gas with 
 plasma temperatures in excess of $\sim$~10$^{6}$~K \citep{groningsson06}.

Previous studies of SN~2005ip have documented its observational 
properties; however, differences of opinion are found  in the literature 
regarding the exact nature of the progenitor, and details of 
the dust emission \citep{smith09,fox09,fox10}. 
Based on their circumstellar (CS) wind speed and
dust mass estimations, \citet{fox10} 
have argued the progenitor was a LBV star.
Alternatively, \citet{smith09} suggest the progenitor 
was akin to a red supergiant (RSG) star --like VY~Canis Majoris (VY CMa) \citep[e.g.][]{chugai94,fransson02,smith09b}-- that exploded in the midst of (or soon after) a period of enhanced mass-loss characterized by eruptive 
mass ejections. In this instance the mass-loss mechanism might be linked to pulsation-driven super-winds \citep{heger97,yooncantiello10}. 

In this paper we present UltraViolet (UV), optical and near-IR photometry, and visual-wavelength spectroscopy of SN~2005ip obtained over the course of the {\em Carnegie Supernova Project} \citep[CSP;][]{hamuy06}. 
These observations complement the already existing dataset, and are used 
to interpret a related object,  SN~2006jd. 
Our observations of SN~2006jd cover UV, optical and IR wavelengths, and are combined with reported 
X-ray  \citep{chandra12} 
 and mid-IR 
measurements \citep{fox11} 
 to gain insight on the progenitor star, its mass-loss history, and details
 concerning the interaction of the (subsequent) SN with the circumstellar environment. 
The organization of this paper is as follows. Section~\ref{sec:obs} contains
a description of the observations obtained for both objects, 
Section~3 presents the final photometry and spectroscopic sequences, and an analysis concerning the dust emission, Section~4 contains the discussion, and this is followed by our conclusions in Section~5.

\section{OBSERVATIONS}
\label{sec:obs}

\subsection{Supernova 2005ip}

\begin{figure}[t]
\plotone{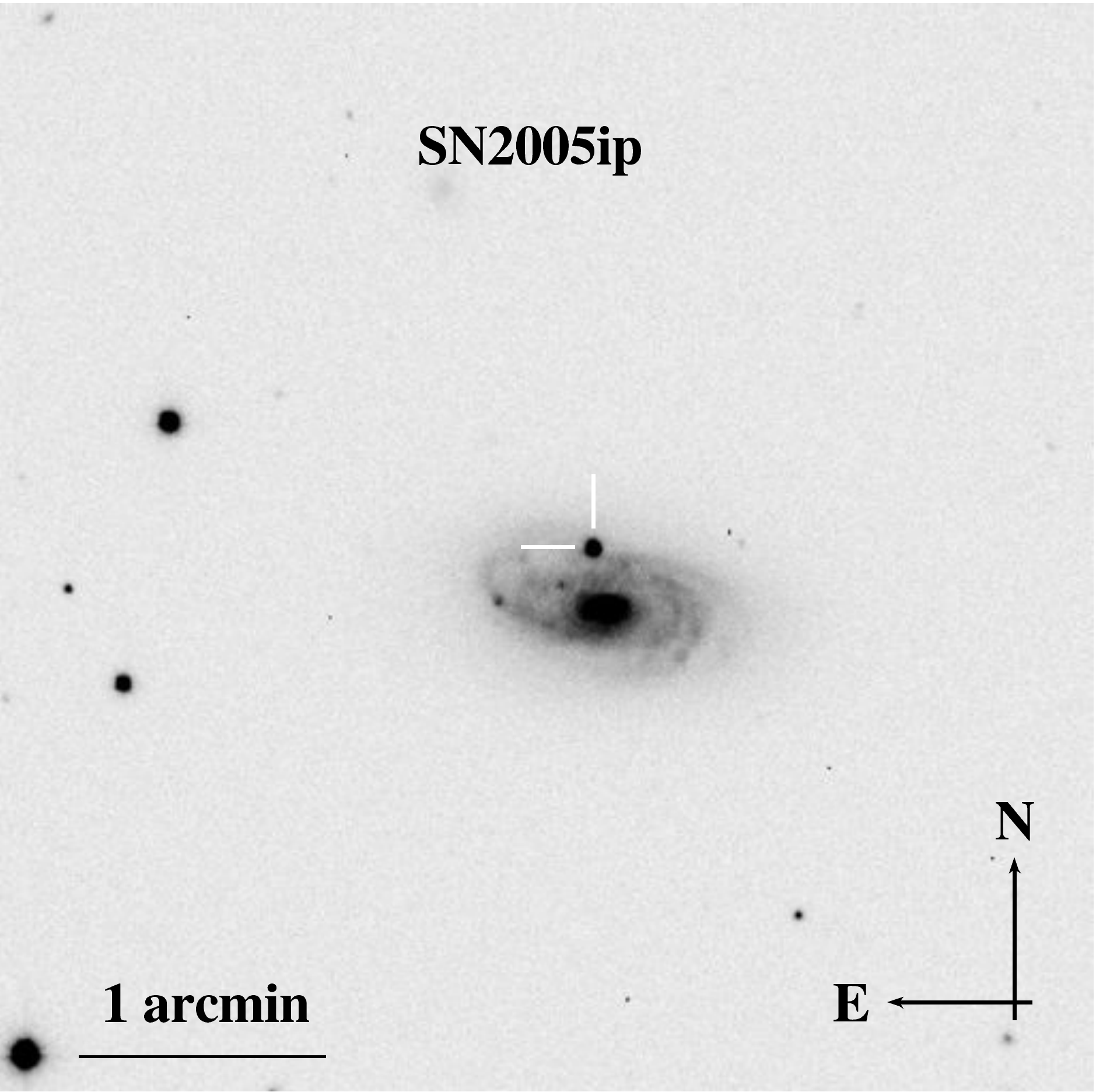}
\figcaption[]{Swope $V$-band image of NGC~2906 with SN~2005ip indicated.
 \label{fig1}} 
\end{figure}

SN~2005ip was discovered on 5.2 November 2005 UT 
 \citep{boles05} in the Scd galaxy NGC 2906 (see Figure~\ref{fig1}).
 With J2000.0 coordinates of 
 $\alpha =$09$^{\rm h}$32$^{\rm m}$06${\fs}$42 and
$\delta$ $= +$08$\arcdeg$26$\arcmin$44$\farcs$4, this SN was located 
2$\farcs$8 E and 14$\farcs$2 N of the center of the host. 
From previous non-detection images obtained on 21 January 2005 no precise constraints can be placed on the explosion epoch, however, from a black-body ($BB$) fit to their earliest spectrum,
\citet{smith09} claimed the explosion epoch to have occurred 8--10 days prior to discovery. In what follows we present our data with respect to the date of discovery, i.e. 
JD $=$ 2453679.66.
 According to the IR dust maps of \citet{schlegel98}, the Galactic color excess in the direction of SN~2005ip is $E(B-V)_{\rm MW} = 0.047$ mag. 
 The spectroscopic observations presented below show no signs of 
 \ion{Na}{1} D absorption at the redshift of the host galaxy, indicating 
 negligible host galaxy extinction.
 We therefore assume the total color excess to be solely attributed to Galactic reddening,
 which for a standard $R_V = 3.1$, yields a visual extinction of $A_{V} = 0.15$ mag.
 NED lists the host recession velocity as 2140 $\pm$ 6 km~s$^{-1}$. 
 Correcting for Virgo, Great Attractor and Shapley infall models, and assuming the WMAP five-year
 cosmological parameters of H$_{0} =$ 70.5~km~s$^{-1}$~Mpc$^{-1}$, $\Omega_m = $0.27 and $\Omega_\Lambda = 0.73$,
 we obtain a luminosity distance to the host galaxy of 34.9 $\pm$ 0.6 Mpc.

\citet{smith09} published  optical spectroscopy  and an unfiltered light curve of SN~2005ip extending $\sim$ 4 years. These observations were later complemented 
 with near-IR and {\em Spitzer} mid-IR photometry and spectrophotometry by
 \citet{fox09,fox10}. 
 Here we present CSP optical ($uBgVri$) and near-IR ($YJH$)
light curves of SN~2005ip.
Our photometric monitoring began four days after discovery,
and extend 5.3 years, with the last
near-IR observations obtained on 21.1 February 2011 UT. 
 
 The majority of our imaging of SN~2005ip (and SN~2006jd) was obtained at the Las Campanas Observatory (LCO) with the Henrietta Swope 1.0 m telescope
 equipped with both a direct optical camera named after its detector ``SITe3", and the near-IR imaging camera called ``RetroCam". These data are accompanied with additional imaging taken with the 
 Ir\'en\'ee du Pont 2.5 m telescope. Optical du Pont images were obtained with the direct CCD camera known as ``Tek 5", while near-IR images were taken with the Wide-field 
 InfraRed camera called ``WIRC"; see \citet{hamuy06} for details regarding these instruments.
 
An in-depth description of CSP observational procedures, 
 data reduction techniques, and the computation of definitive photometry
in the {\em natural} photometric system  is described in \citet[][C10]{contreras10}. 
The only difference with respect to the photometry published by 
 C10 and here, is that no host galaxy template subtraction has been performed on the science images of SNe~2005ip and 2006jd. 
 This is simply because these SNe are still visible in deep images taken with the du Pont telescope. However, in the future when useful template images are obtained, an updated version of the photometry will be placed on the CSP webpage.\footnote{Electronic files of SNe~2005ip and 2006jd in {\tt ASCII} format are available on the CSP webpage
{\tt http://obs.carnegiescience.edu/CSP/data .}}

 SN photometry is computed differentially with respect to a local sequence of stars. Absolute photometry of the optical and near-IR local sequence
of SN~2005ip was determined relative to photometric standard stars observed over the course of 6 and (a minimum of) 9 photometric nights, respectively.
The position of each local sequence star and their calibrated magnitudes
are listed in Table~\ref{table1}. 
The averaged optical magnitudes are on the 
\citet{smith02} $u'g'r'i'$ and \citet{landolt92} $BV$ photometric systems, while
the near-IR magnitudes are on the \citet{persson98} $JHK_s$ photometric system.
$Y$-band photometry of the local sequence is calibrated relative
to the ($J-K{_s}$) color relation given in  \citet[][see their Appendix C, equation C2]{hamuy06}.

Limited follow up imaging of SN 2005ip was conducted from space with the UltraViolet Optical Telescope (UVOT) aboard the {\em Swift} satellite (Roming et al. 2005).
Four epochs of $[uvw2][uvm2][uvw1]$-band photometry were obtained covering the flux evolution from day 146 to 1154, while an additional epoch on day 2012 was taken with 
the $[uvw1]$ filter.\footnote{The central wavelengths and FWHM values of the UVOT filters are as follows: $[uvw2]$ ($\lambda_c = 1928$~\AA; FWHM $= 657$ \AA), $[uvm2]$ ($\lambda_c = 2246$~\AA; FWHM $= 498$ \AA), and $[uvw1]$ ($\lambda_c = 2600$~\AA; FWHM $= 693$ \AA).}  
 Aperture photometry was computed from the UVOT images following the method described by Brown et al. (2009), but  updated to the zero-points of \citet{breeveld11}. 
As this SN (and SN~2006jd, see below) has not faded completely in several followup observations, we utilize a background region with a flux level similar to that near the SN position rather than utilizing the standard galaxy-subtraction technique.

 The final UV, optical and near-IR  light curves  of SN~2005ip 
are plotted in Figure~\ref{fig2}. The corresponding 
optical and near-IR photometry  in the natural system are 
listed in Tables~\ref{table2} and \ref{table3}, while the UVOT photometry 
is given in Table~\ref{table4}.
 
 In addition to the photometric follow up, 13 epochs of 
low-resolution visual-wavelength spectroscopy of SN~2005ip covering the first five and a half months 
(23 November 2005 -- 23 April 2006) of evolution was obtained with an assortment of telescopes at LCO and the La Silla Observatory.
Two very late spectra were also obtained 1844 and 2352 days past discovery 
with the 2.5 m Isaac Newton Telescope and the 2.5 m Nordic Optical Telescope, from which we are able to measure 
emission features (e.g. H$\alpha$) associated with post-shock gas.
Each spectrum was reduced in the standard manner using 
 IRAF\footnote{IRAF is distributed by the National Optical Astronomy Observatories, which are operated by the Association of Universities for Research in Astronomy, Inc., under cooperative agreement with the National Science Foundation.}
scripts following the techniques described in \citet{hamuy06}. 
When necessary the fluxing of the spectra was adjusted to match what was 
obtained from the broad-band photometry.
The journal of spectroscopic observations is given in Table~\ref{specjor},
 and 8 of the 15 epochs of spectra are plotted in Figure~\ref{fig3}. 
 
\subsection{Supernova 2006jd}

SN~2006jd was discovered on 12.5 October 2006 UT in 
the SBb galaxy UGC~4179 by 
the Lick Observatory Supernova Search \citep[LOSS;][]{prasad06}.
With J2000.0 coordinates of $\alpha =$08$^{\rm h}$02$^{\rm m}$07${\fs}$43 and
$\delta$ $= +$00$\arcdeg$48$\arcmin$31$\farcs$5, SN~2006jd is located
22$\farcs$0 E and 1$\farcs$3 S from the center of the host galaxy.
A $V$-band image of the galaxy with the SN indicated is shown in 
Figure~\ref{fig4}.
The previous LOSS non-detection on 
21.2 April 2006 UT precludes a precise estimation of the explosion epoch. 
In the remainder of the paper the observations of SN~2006jd will be
 presented with respect to date of discovery, i.e. JD $=$ 2454021.04.
 Initially classified as a Type~IIb event \citep{blondin06}, it was later 
 re-designated by \citet{immler07} as a Type~IIn SN. 
The Galactic color excess in the direction of UGC~4179 is listed in 
NED to be $E(B-V)_{\rm MW} = 0.054$ mag. 
Under close inspection of our spectroscopic sequence 
we find no indication of \ion{Na}{1} D absorption at the redshift of the 
host galaxy, suggesting the SN suffered little to no host extinction. 
In what follows the total color excess is assumed to be only due to Galactic reddening. 
Adopting a standard reddening law characterized by an  $R_{V} = 3.1$ then implies
a visual extinction in the direction of SN~2006jd to be $A_{V} = 0.18$ mag.
 NED lists a recession velocity for UGC~4179 of 5563 $\pm$ 5 km~s$^{-1}$, which when corrected for the 3-component Virgo, GA and Shapley infall model, and assuming
 the standard five-year WMAP cosmological parameters, corresponds to a luminosity distance of 83.8 $\pm$ 1.5 Mpc. 
 
Our photometric observations commenced a week after discovery and extend over 
a duration of nearly four and a half years. 
The local sequence of stars used to compute photometry was calibrated with respect to standard field observations obtained on 11 nights in the optical and, 
depending on the specific star, between 3 and 22 nights in the near-IR. 
The local sequence is given in Table~\ref{table1}.
Final optical and near-IR photometry of SN~2006jd is given in Tables~\ref{table5} and 
\ref{table6}, respectively. The corresponding light curves are plotted in Figure~\ref{fig5}.

Limited follow up imaging of SN~2006jd was also performed with {\em Swift}. 
Seven epochs of $[uvw2][uvm2][uvw1]$-band photometry were obtained covering the flux evolution from day 401 to 790.
The final UVOT photometry is listed in Table~\ref{table4}, and plotted in 
Figure~\ref{fig5}. 
 
We also obtained an epoch of mid-IR imaging with IRAC (InfraRed Array Camera) aboard {\em Spitzer}.
Two channels of imaging were taken at 3.6~$\mu$m and 4.5~$\mu$m 
on 22 June 2011 or 1713.8 days past discovery. 
PSF photometry of the SN implies 
flux values of 1.53$\pm$0.01$\times$10$^{-17}$ erg cm$^{-2}$ s$^{-1}$ \AA$^{-1}$ at 3.6~$\mu$m and 1.59$\pm$0.01$\times$10$^{-17}$ erg cm$^{-2}$ s$^{-1}$ \AA$^{-1}$ at
 4.5~$\mu$m. These values are $\sim$ 30\% of those measured previously in images taken on
day $\sim$ 1150, and are qualitatively in agreement with the slow decline of the $K_s$-band light curve (see below). 
In what follows the {\em Swift} and {\em Spitzer} observations are combined with 
 optical and near-IR photometry, and measurements of the X-ray and
 radio flux to construct a nearly complete SED at day $\sim$ 1638.

Seventeen epochs of low-resolution, long-slit visual-wavelength spectroscopy of SN~2006jd were obtained
with an assortment of telescopes at LCO, and the La Silla and Mauna Kea Observatories 
(see Table~\ref{specjor}). These spectra were also reduced in a standard manner, and their fluxing was adjusted to match that derived from the broad-band photometry.
The spectroscopic sequence covers all phases of the flux evolution, beginning 
2 weeks from discovery to 39 months later. To highlight the rich structure 
contained within these spectra a subset of the full spectroscopic series
is shown in Figure~\ref{fig6}.
 
 \section{RESULTS}
\subsection{Host Galaxy Metallicity}

Here metallicities in the vicinity of SNe~2005ip and 2006jd are derived 
using emission line diagnostics.
To do so for SN~2005ip we extracted from the 2-D spectrum obtained on
day 1844 emission from an \ion{H}{2} region which lies close to the position of the SN.
Inspection of the 1-D spectrum revealed H$\alpha$ and [\ion{N}{2}] $\lambda\lambda$6548, 6583 emission lines from which 
Gaussian fits provide flux measurements 
that imply a local metallicity of 12$+$log(O/H)$=$8.73 $\pm$ 0.18 dex
calibrated on the N2 scale of \citet{pettini04}.
 
The metallicity of the environment of SN~2006jd was also measured in 
a similar fashion. By extracting from the 2-D spectrum taken on day 423
 a series of \ion{H}{2} regions in close proximity to the SN, 
the resultant emission lines imply a N2 calibrated metallicity of 
12$+$log(O/H)$=$8.28 $\pm$ 0.18 dex. For comparison, using the 
 [\ion{N}{2}]/[\ion{O}{2}] diagnostic \citep{kewley02} the series of 
 {H}{2} regions provide metallicity estimates in 
the range of 8.60 $<$ 12$+$log(O/H) $<$ 8.75 dex. 
 
 In summary we find the local environments of SN~2005ip to be
 comparable with the known solar metallicity of $\sim$ 8.69 dex \citep{asplund09},
 while depending on the specific emission line diagnostic, the environment 
 of SN~2006jd lies within the range between that of the LMC to solar. 

\subsection{Light Curves}
\label{lightcurves}

\subsubsection{Supernova 2005ip}

\begin{figure}[t]
\plotone{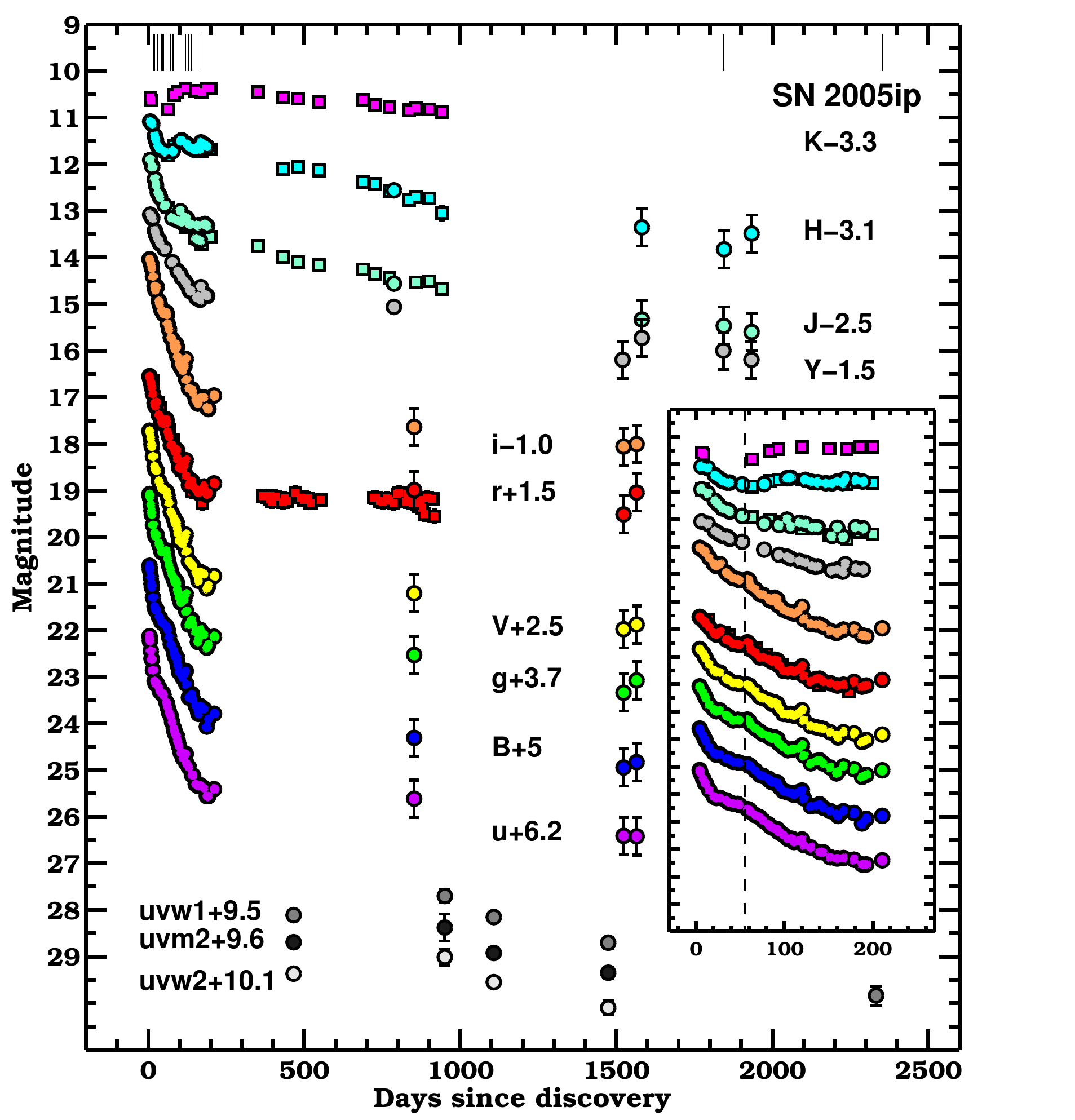}
\figcaption[]{UV, optical and near-IR light curves of SN~2005ip (filled dots), 
and published unfiltered \citep{smith09} and near-IR
\citep{fox09} photometry (filled squares). 
Contained within the inset is the photometric evolution over the first $\sim$ 200 days,
 with the vertical dashed line marking the time when the optical light curves 
 begin to drop at an increased rate. Simultaneously, the $H$- and $K_s$-band light curves
 show an excess in emission.
 These transitions are related to dust condensation (see text). Vertical lines at the top of the figure indicate the epochs in which spectra where obtained (see Table~\ref{specjor}). 
 \label{fig2}} 
\end{figure}

\begin{figure}[t]
\plotone{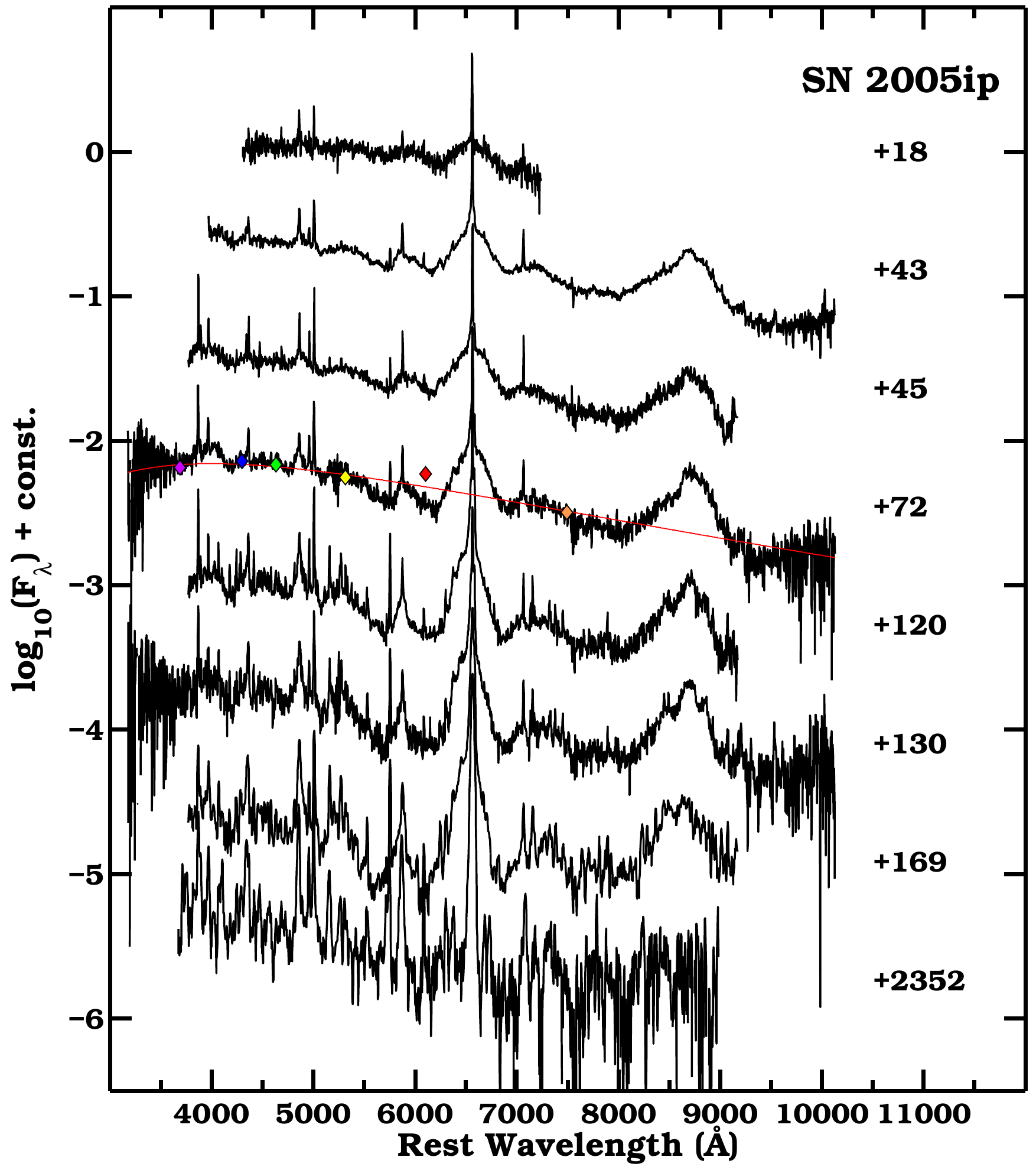}
\figcaption[]{A subset of our spectroscopic sequence of SN~2005ip from 18 to 2352 days past discovery. Over-plotted the day 72 spectrum is the hot $BB_h$ component derived from fits to the $uBgVi$ broad-band flux points (see Section~\ref{uvoir}).  \label{fig3}} 
\end{figure}

\begin{figure}[t]
\plotone{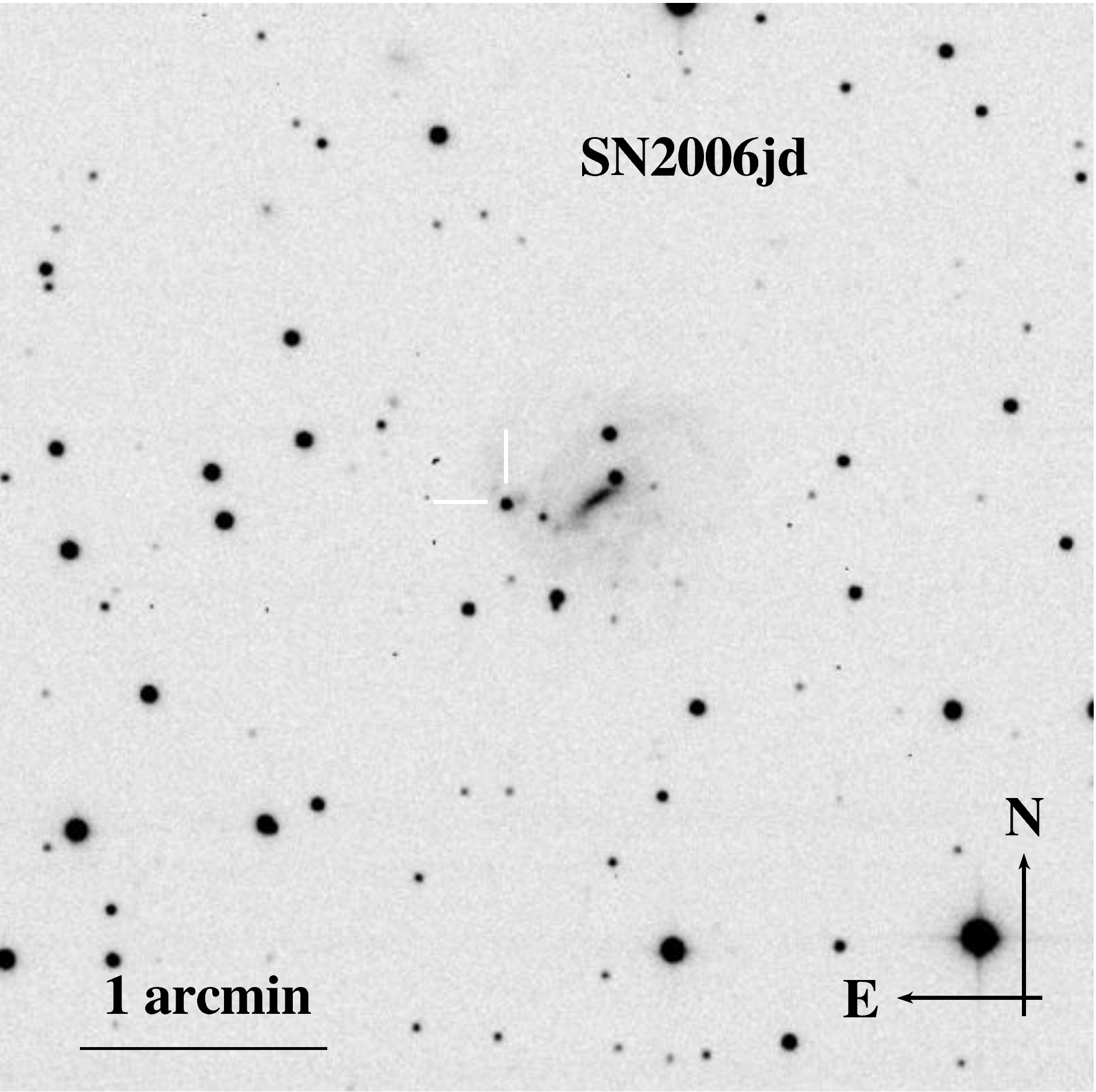}
\figcaption[]{Swope $V$-band image of UGC~4179 with SN~2006jd indicated. \label{fig4}} 
\end{figure}

\begin{figure}[t]
\plotone{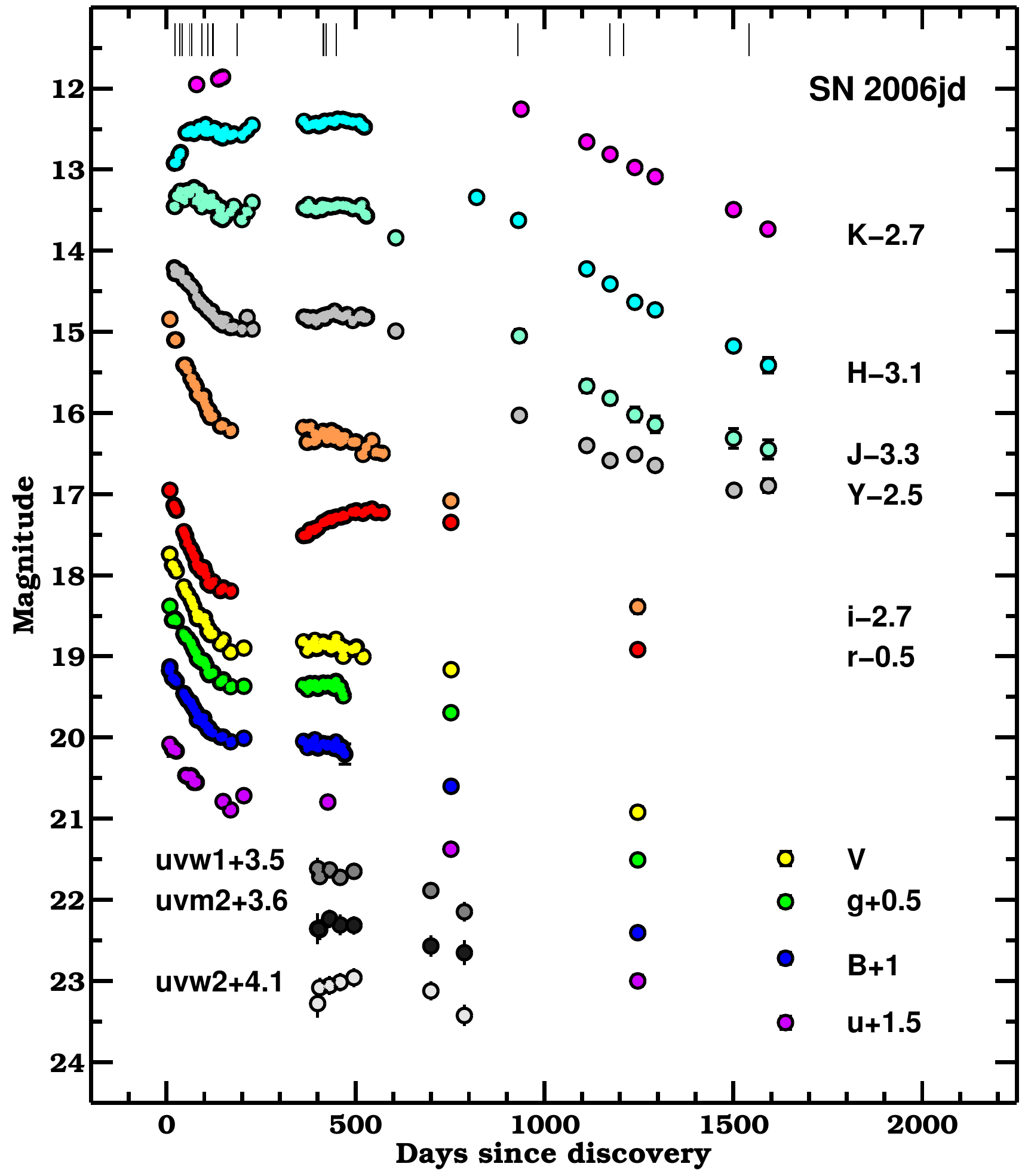}\figcaption[]{UV, optical and near-IR light curves of SN~2006jd.
Vertical lines at the top of the figure mark the epochs in which spectroscopy has been obtained. \label{fig5}} 
\end{figure}

\begin{figure}[t]
\plotone{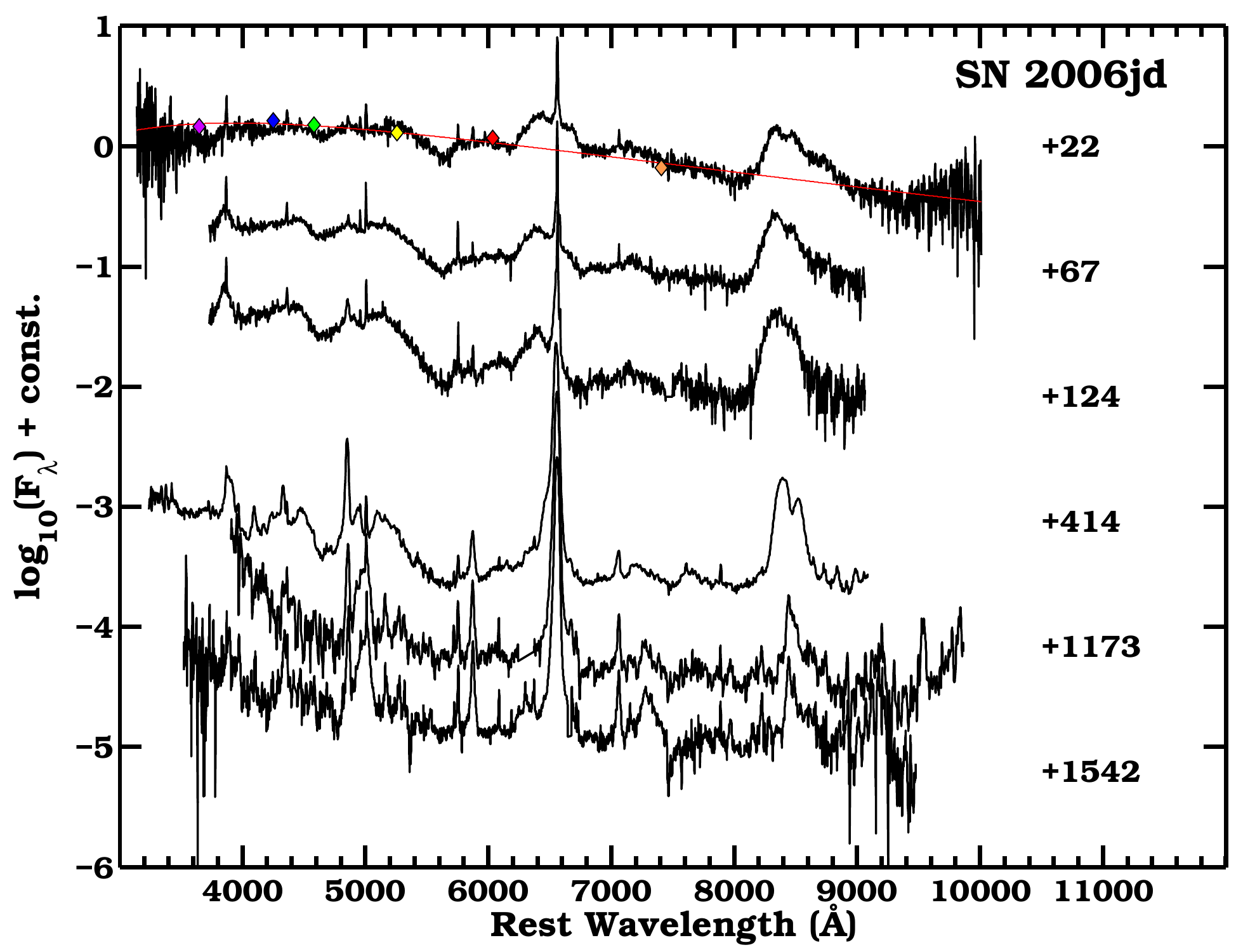}
\figcaption[]{A subset of our spectroscopic sequence of SN~2006jd from 22 to 1542 days past discovery. Over-plotted the day 22 spectrum is the hot $BB_h$ component derived from fits to the $uBgVi$ broad-band flux points. \label{fig6}} 
\end{figure}

The majority of our optical and near-IR photometry plotted in Figure~\ref{fig2} 
follow the initial $\sim$ 200 days of evolution, while several additional epochs 
extend to day 1565 (optical) and day 1934 (near-IR). In addition to our own photometry, included as filled squares in Figure~\ref{fig2} are the unfiltered \citep{smith09} and near-IR \citep{fox09} light curves taken from the literature. 
In general there is excellent agreement between the datasets.
 
Initially, the overall flux evolution is characterized by a rapid, linear decline 
(in magnitude) over the first $\sim$ 55 days.
Subsequently, the slope of the decline further increases in the optical bands, while
simultaneously, the $H$- and $K_s$-band light curves show an increase of emission. 
This is highlighted in the inset of 
Figure~\ref{fig2}, where the vertical dashed line indicates the epoch (day 55) 
when these transitions occur.
As discussed in Section~\ref{dust}, these phenomena are believed to result from the {\em in situ} formation of dust within a cold, dense shell (CDS) of gas that lies between the forward and reverse shock produced by SN--CSM interaction.
If the newly formed dust was distributed homogeneously in a spherical shell one would expect the bluest optical bands to be attenuated more significantly then those in the red,
which is what we observe when comparing the slopes of the light curves derived before dust formed to what is measured post dust formation.

 From day 100 to 150 the $u$- to 
$J$-band light curves continue to decline in brightness, at which point a break occurs post day 150, followed by a long-duration plateau phase. 
This is best illustrated by the $r$-band light curve, which reveals the plateau phase 
extending beyond day 1500.
Contrarily, from the onset of dust formation the $H$ and $K_s$ light curves 
brighten $\approx$ 0.2 mag over a 5 week period, and subsequently,
settle onto a slow 
linear decline phase lasting over the duration of our observations.  

Finally, the UV light curves appear to follow, similar to the redder wavelengths, 
a plateau phase characterized by nearly constant emission between $\sim$ day 500 to day 1000, whereupon
they appear to turn over and slowly decline in flux over time.  

\subsubsection{Supernova 2006jd}

The  light curves of SN~2006jd displayed in Figure~\ref{fig5} reveal a
rich evolution that in some ways resembles SN~2005ip. 
Over the first $\sim$ 150 days 
the $ugriBVY$-band light curves decline in a linear fashion with slopes
of $\lesssim$ 0.01 mag per day.
Subsequently, an abrupt break occurs marking the onset of 
a plateau phase which lasts $\gtrsim$ 500 days. 
Most interestingly, 
over this time the $r$-band light curve reveals an unprecedented
evolution characterized by an excess of flux that peaks on 
day 544 at $\approx$ 1.0 mag above its brightness measured during the start of the plateau phase. As shown later, this excess of flux is attributed to a significant 
increase in H$\alpha$ emission (see Figure~\ref{halpha}) as the SN blast wave interacts 
with a dense shell of material (see Section~\ref{insight}). 
Following the plateau phase that ends some $\sim$ 750 days past discovery, 
the optical light curves evolve along a slow, linear ($\sim$ 0.001 mag per day) decline phase through the end of our observations which conclude on day 1638. 
  
On the other hand, from the commencement of our observations the near-IR $JHK_s$ light curves exhibit an increase in flux which peaks on day $\sim$ 40.
Subsequently, an extended period of constant emission ensues, and then $>$ 500 days later the light curves 
 settle along a slow, linear decline ($\sim$ 0.002 mag per day), which we follow 
 until day 1593.  

Also plotted in Figure~\ref{fig5} are the {\em Swift} UVOT $[uvw2][uvm2][uvw1]$-band light curves.
Beginning on day 401, the UVOT observations cover 7 epochs extending over a duration of 389 days, and indicate that the UV emission, like the optical and near-IR light emission, was also relatively constant during this time period.

\subsection{SEDs and Quasi-Bolometric Light Curves}
\label{uvoir}
In order to estimate the physical parameters, i.e. temperature, radius and luminosity of the 
underlying emitting regions, the available broad-band observations are used to construct SEDs. 
$BB$ fits to the resulting optical and IR SEDs then provide an avenue 
to constrain these parameters, as well as to place lower limits on the bolometric 
luminosity, and as we discuss later, provide dust mass estimates. 

\begin{figure}[t]
\epsscale{1.2}
\plottwo{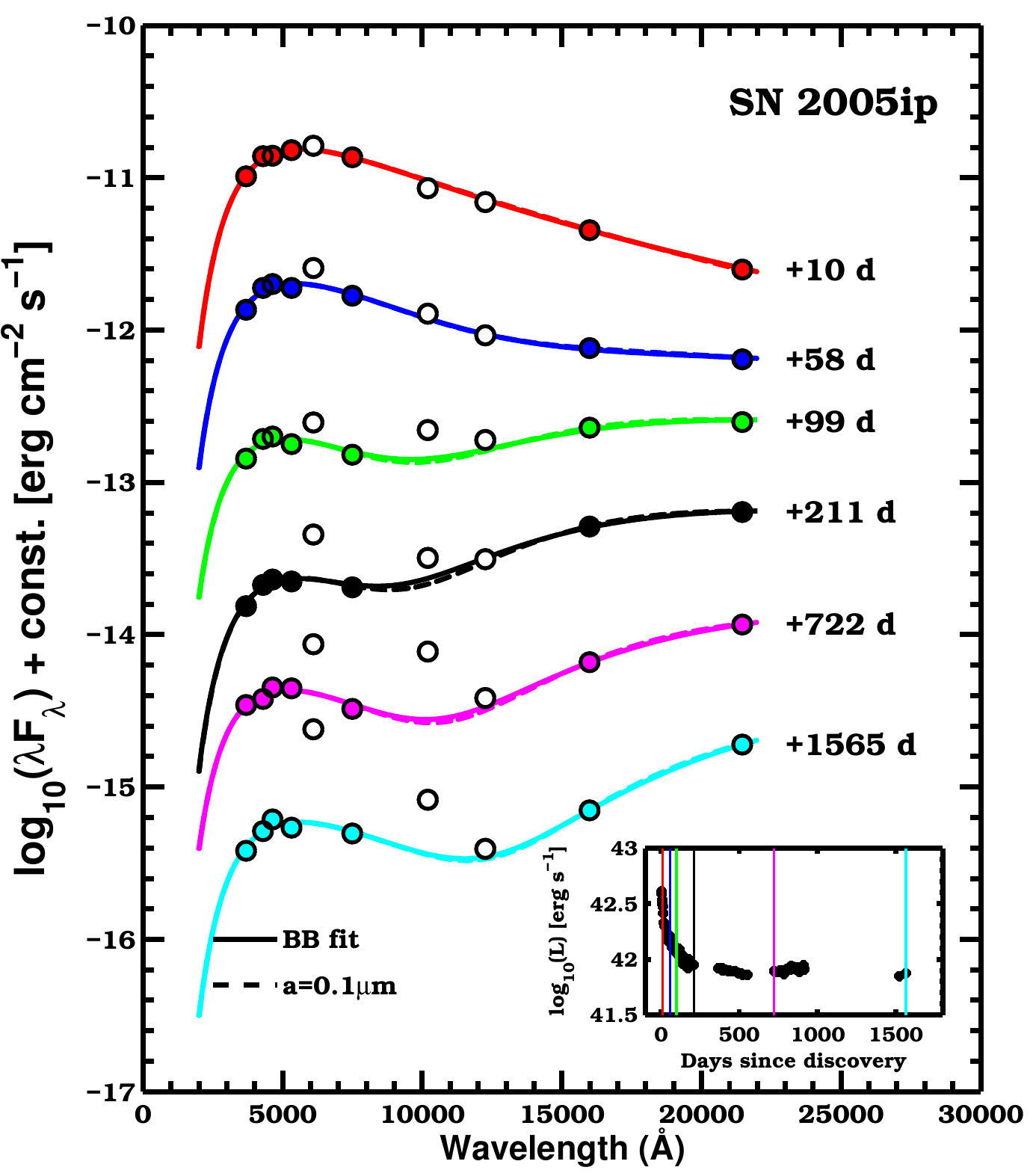}{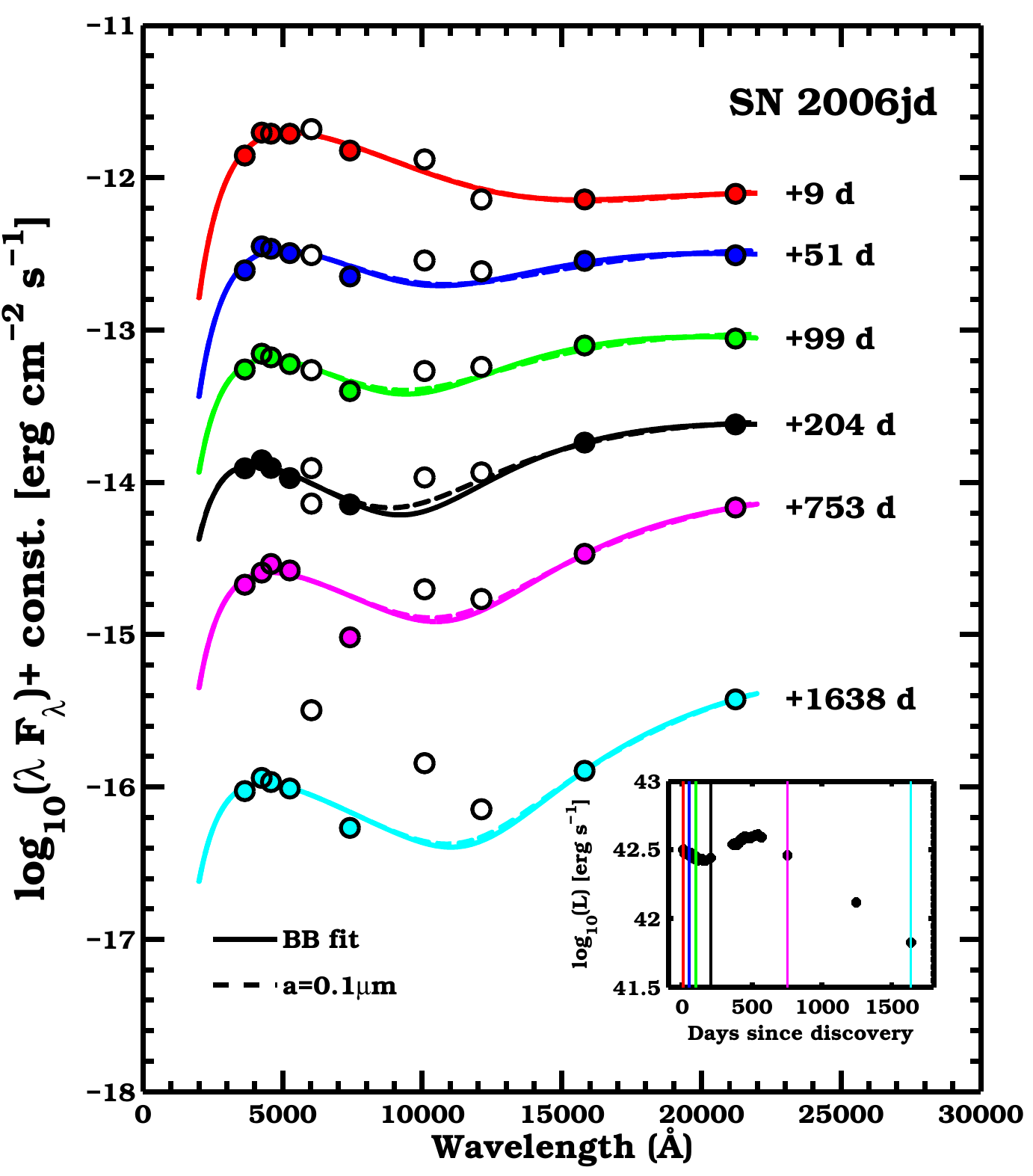}
\figcaption[]{SEDs of SN~2005ip {\em (left)} and SN~2006jd {\em (right)} at six epochs
constructed from broad-band optical ($uBgVri$) and near-IR ($YJHK_s$) photometry.
 Insets indicate the phase of the plotted SEDs with respect to the quasi-bolometric light curve. 
 White filled circles correspond to the $rYJ$ flux points which are excluded 
 from the $BB$ fits due to strong emission-line contamination. 
Solid lines correspond to two-component $BB_{h,w}$ fits to the optical and near-IR flux points, while the dashed lines correspond to modified dust mass $BB_{h,w}$ functions. 
Note that the Rayleigh-Jeans tail of the $BB_{w}$ component is not included in the figure.
\label{fig7}} 
\end{figure}

To construct comprehensive SEDs the near-IR light curves were first interpolated 
following the method of \citet[][see their Section 3]{pastorello09}
so magnitude estimates could be made for those epochs in which 
optical photometry was obtained. Next the optical/near-IR photometry was corrected 
for reddening using the $E(B-V)_{tot}$ color excess values discussed in Section~2, and the standard \citet{cardelli89} extinction curve characterized by $R_V = 3.1$.
The extinction-corrected optical and near-IR photometry was then converted to flux at the effective wavelength of each bandpass. 
The resulting SEDs of SNe~2005ip and 2006jd at a selection of six epochs covering each main phase of flux evolution are plotted in Figure~\ref{fig7}. The inset in each panel of Figure~\ref{fig7} indicates the corresponding phases with regard to the light curve. 

We next proceed to fit each SED simultaneously with a two-component $BB$ function 
in the form of

\begin{equation}
F_\lambda(T_{h,w},R_{h,w})= \frac{2\pi h c^2}{D^2\lambda^5}\left(\frac{R_h^2}{e^{(hc/\lambda k T_h)}-1} + \frac{R_w^2}{e^{(hc/\lambda k T_w)}-1}\right).
\label{eqn1}
\end{equation}

\noindent Here $c$, $h$ and $k$ are the standard symbols for their respective fundamental constants, $T_{h,w}$ and $R_{h,w}$ correspond to the temperature
 and radius of each Planck $BB_{h,w}$ component, and $D$ is the distance to the underlying SN. 
 The first term in Equation~\ref{eqn1} corresponds to a `hot' $BB_h$ component and is derived 
from fits to the $uBgVi$-band flux points, while the second term corresponds to a `warm' $BB_w$ component
derived from fits to the $HK_s$-band flux points. 
These components represent different emitting regions, with the former 
being coupled to the SN ejecta, and the latter to mostly warm, thermal dust emission. 
To obtain reasonably accurate $BB_{h,w}$ fits to the 
 SEDs it proved necessary to exclude the 
 $rYJ$-band flux points 
 as their associated bandpasses include prevalent emission lines of H$\alpha$, \ion{He}{1} $\lambda$10380 and P$\beta$ $\lambda$12830, respectively. 
 The resulting best-fit combined $BB_{h,w}$ functions are over-plotted on the SEDs in 
 Figure~\ref{fig7} as solid lines. 
 Note that for presentation clarity the Rayleigh-Jeans tail of the $BB_{w}$ component is not included in the figure. 

\begin{figure}[t]
\epsscale{1.0}
\plotone{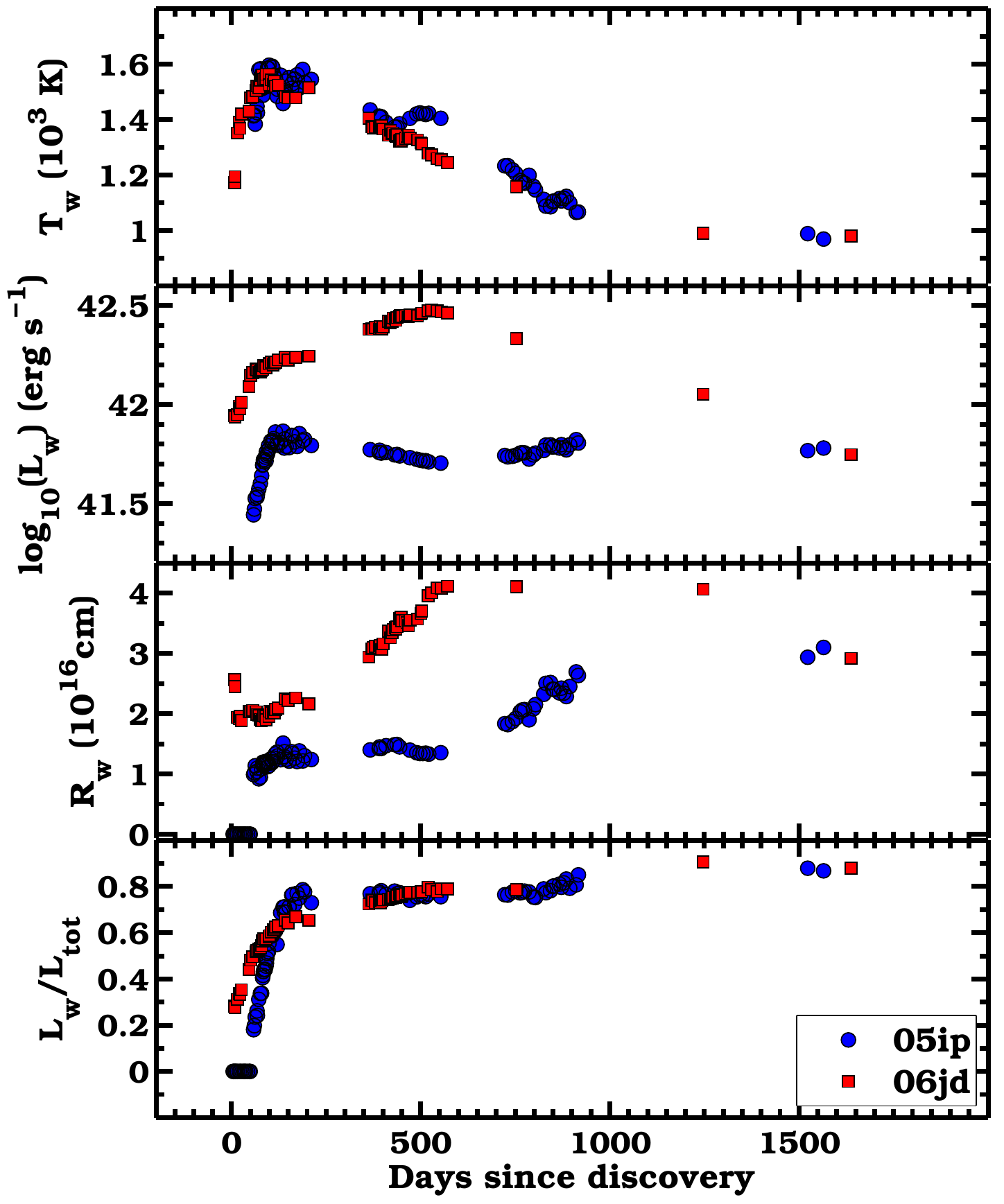}
\figcaption[]{Evolution of the physical parameters: temperature, luminosity and radius derived from the ``warm" $BB_{w}$ component fit to the $H$- and $K_s$-band photometry of SNe~2005ip (filled blue dots) and 2006jd (filled red squares). 
Also shown in the bottom panel is the ratio of the warm $BB_{w}$ luminosity to the 
total luminosity. \label{fig8}} 
\end{figure}

The resulting values of $T_{h,w}$, $L_{h,w}$ and $R_{h,w}$ computed from the 
$BB_{h,w}$ fits 
shown in Figure~\ref{fig7} are given in Table~\ref{table8}, as well as 
 the ratio of $L_{h}$ to $L_{h}$ $+$ $L_{w}$. The corresponding values 
 of $T_{h,w}$, $L_{h,w}$ and $R_{h,w}$ computed from fits to SEDs constructed from 
 our full series of observations are plotted in Figure~\ref{fig8}. Also shown in the bottom panel is the ratio of $L_{w}$ to $L_{h}$ $+$ $L_{w}$.
 Under close inspection, Figure~\ref{fig7} reveals that the SED of both objects exhibit similar characteristics at each of their main evolutionary phases. 
 At the earliest epochs the vast majority of flux is emitted at optical wavelengths.
 However, by day $\gtrsim$ 100 the optical portion of the SED accounts for only half 
 of the total optical$+$near-IR emission, and over time continues to decline, while simultaneously 
 the near-IR emission exhibits an excess that by day $\sim$ 200 dominates the 
 optical$+$near-IR spectrum (see last column of Table~\ref{table8}). 
 The prevalent increase and long-duration of the near-IR excess 
 signifies enhanced thermal emission from warm dust \citep{fox09}.
Interestingly, both objects also have nearly identical values of $T_{w}$ $\sim$ 1500 K at the time 
when newly formed dust is thought to condense, while 
this warm dust component appears to be more luminous 
and extends to further distances in SN~2006jd (see Section~\ref{whereisthedust} for 
further discussion).

\begin{figure}[t]
\epsscale{0.9}
\plotone{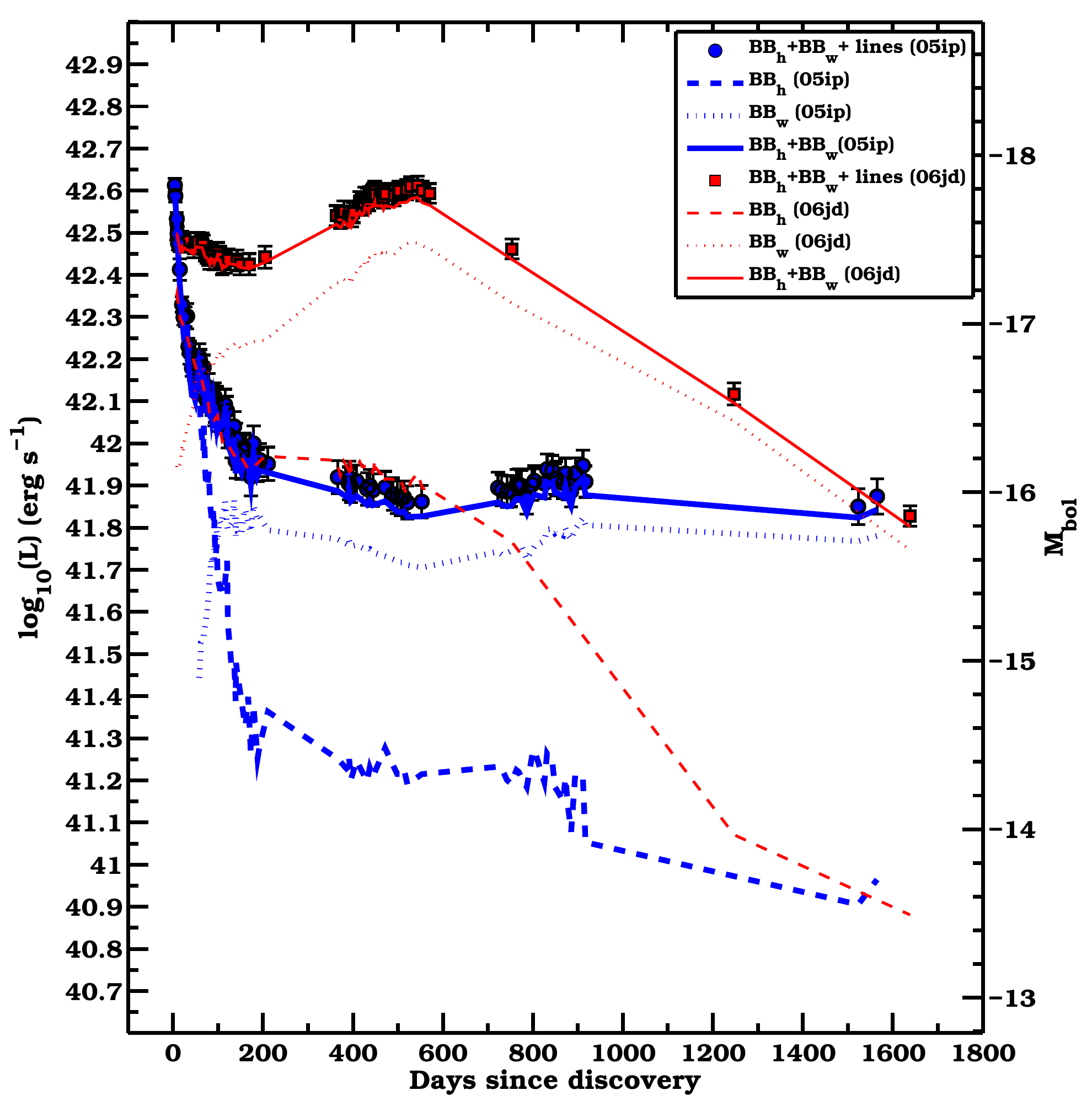}
\figcaption[]{ Quasi-bolometric light curves of SNe~2005ip and 2006jd plotted as points. 
Also included are the $BB_h$ (dashed lines) and $BB_w$ (dotted lines) components, and the summation of the two (solid lines). For clarity the three curves corresponding to SN~2005ip are drawn extra thick.\label{fig9}} 
\end{figure}

 To compute quasi-bolometric light curves of SNe~2005ip and 2006jd
 we take the approach of summing the luminosities derived from 
 the two-component $BB_{h,w}$ fits with the flux contained within the emission lines covered by the $rYJ$-bands (i.e. the flux contained under the dashed line plotted 
 in Figure~\ref{fullSED}, see below). 
The resulting quasi-bolometric light curves are shown in Figure~\ref{fig9} as points.
 Also 
included in this figure are the individual  $BB_h$ (dashed lines) and $BB_w$ (dotted lines) components,
and the summation of these two components (solid lines). 
We note that our quasi-bolometric light curves provide only a lower limit to the 
total flux as they neglect emission associated with wavelength regimes not 
sampled by our photometry, including emission from X-ray to UV wavelengths 
and from mid-IR to radio wavelengths. 
Nevertheless, the  Rayleigh-Jeans tail associated with the $BB_w$ component
{\em does} account for a portion of thermal emission that lies beyond the $K_s$ band,
 and indeed this flux is a dominant source of the quasi-bolometric light curves
 from day $\sim$ 100 and beyond (see dotted lines). 
Note that the $BB_w$ component provides a major part of the total optical$+$near-IR SED
from the onset of the plateau phase and beyond. This long-lasting emission implies that the quasi-bolometric light curve is powered more by SN--CSM interaction rather than radioactive decay.

Comparing the bolometric light curves of SNe~2005ip and 2006jd, 
we find that at the earliest epochs they exhibit similar peak luminosity of roughly 
$\sim$ 3.2$\times$10$^{\rm 42}$ erg~s$^{-1}$, which is equivalent to an absolute bolometric magnitude of $\sim$ $-$17.6.
Later as each object evolves to the mid-points of their respective plateau phase,
 Figure~\ref{fig9} indicates SN~2006jd clearly outshines SN~2005ip, 
reaching a maximum difference in luminosity of $\sim$ 0.7 dex or about 
$\sim$ 1.7 mag (see also middle panel of Figure~\ref{fig8}).
This excess of IR flux observed in SN~2006jd suggests that it has a larger amount of dust. 
To see if this is indeed the case, we proceed to estimate
the dust mass associated with these two SNe. 
 
 \subsection{Dust Emission and Mass Estimates}
\label{dust}

Near-IR observations of SNe~2005ip and 2006jd provide compelling 
evidence for the presence of a warm ($T_w$ $\sim$ 1500 K) dust component.
The ever increasing near-IR emission at early phases supports the notion of 
warm dust condensation. This emission  dominates the  SED within 
$\sim$ 100 days past discovery and continues to do so for the duration of hundreds of days. 
The long-duration of the near-IR excess is best explained by a dust heating mechanism
associated with radiative shocks formed at the interface of the SN--CSM interaction 
\citep[see][for a detailed discussion]{fox09}. 
Interestingly, the turn-on of the near-IR excess coincides 
in time with an ever increasing attenuation of the red wing of the H$\alpha$ profile (see below), 
and in the case of SN~2005ip, a steepening decline in the optical band light curves 
(see Figure~\ref{fig2}).
Such phenomena are usually attributed to dust condensation within 
 either the fast SN ejecta or in the CDS 
 of post-shock gas lying between the interface of the forward and reverse shocks. 
 Dust forming in either of these regions preferentially scatters and absorbs photons 
emitted from the far side of the ejecta and thereby causes the attenuation of the red wing
 of prevalent emission profiles, and can lead to steeper 
 decline rates for the blue light curves. 
 Alternatively, the increased attenuation of the red side of the line profile
 might, in some cases, be related to an optical depth effect (see Section~\ref{whereisthedust} 
and \citealt{smith12}).
 
\begin{figure}[t]
\plottwo{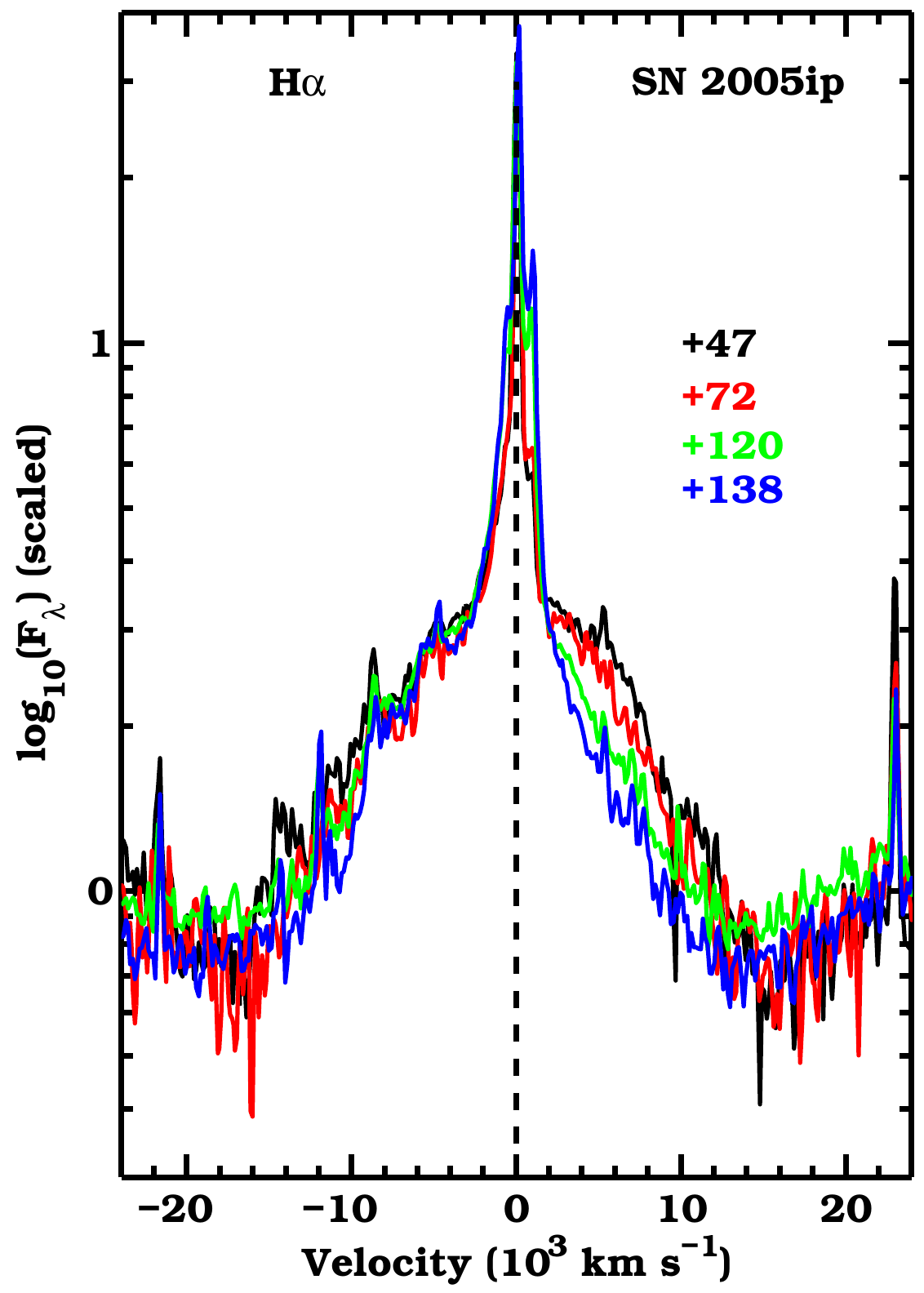}{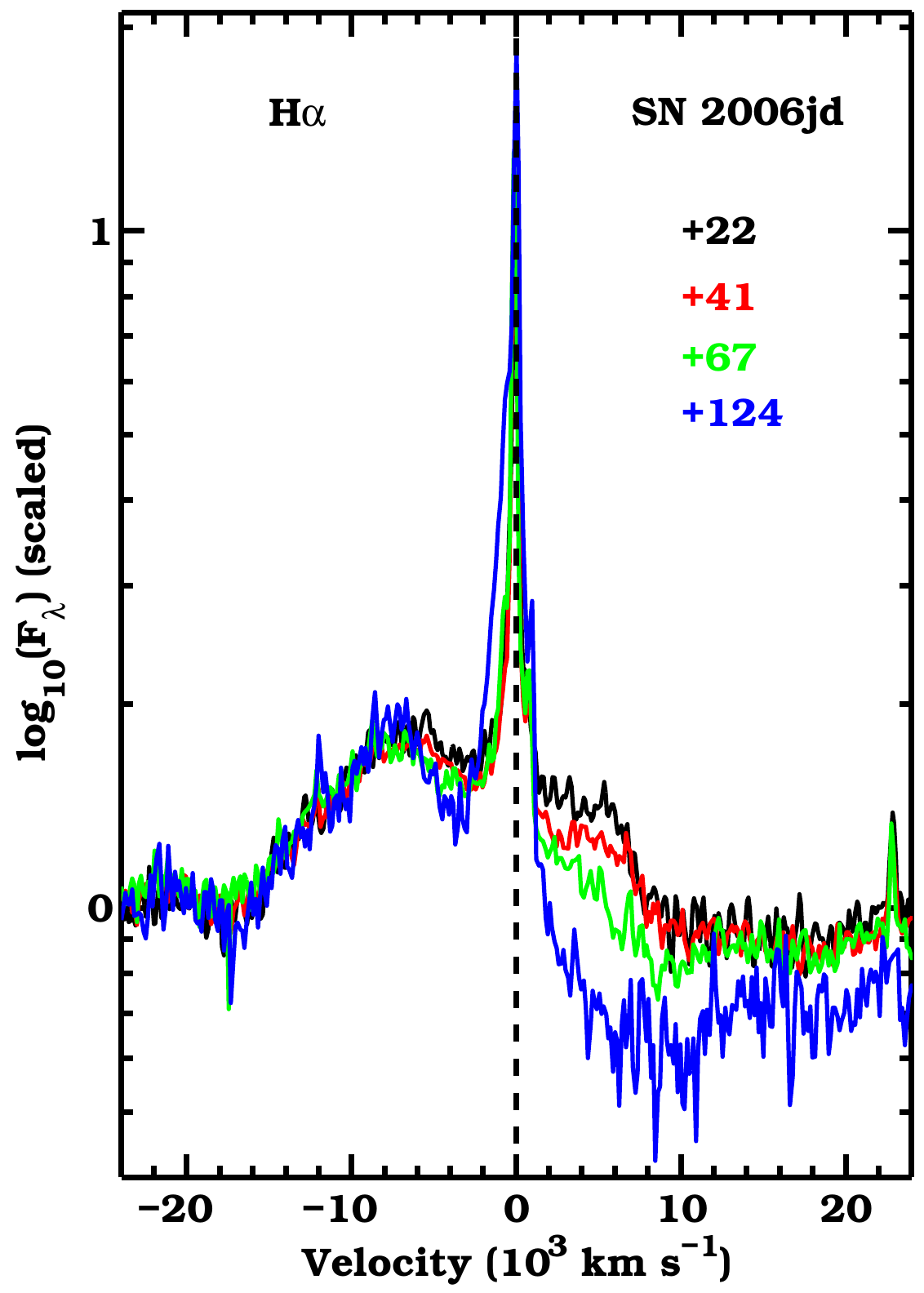}
\figcaption[]{H$\alpha$ line-profile evolution of SNe~2005ip and 2006jd covering 
the initial four months of the linear decline phase.
 The spectra have been scaled and superposed on top of each other. 
 Note in SN~2006jd the broad component is asymmetric, showing an excess of flux in the blue, while the suppression of the red wing over time is clearly more extreme than in SN~2005ip.\label{fig10}} 
\end{figure}

 In Figure~\ref{fig10} the evolution of the H$\alpha$ emission profile of both SNe 2005ip and 2006jd over the first four months of observations is plotted. 
 The ever increasing attenuation of the red side of H$\alpha$ clearly occurred in 
 both objects. 
 This behavior was already noted in SN~2005ip \citep{smith09}, however 
 for SN~2006jd, 
 \citet{fox11} argued there was no evidence for such a phenomenon.\footnote{\citet{fox11} 
based this claim on the comparison of spectra obtained on day 395 and 564. 
Given that the effects of newly formed dust are greatest {\em at early times} 
it is of no great surprise they did not observe this behavior. 
This highlights the necessity of early phase spectroscopy to identify 
signatures of newly formed dust.}
Clearly, SN~2006jd also exhibits an increased attenuation of red flux at early times.
 
 In addition to the possibility that warm dust formed at early times, \citet{fox10,fox11} determined 
from {\em Spitzer} mid-IR imaging that both SNe~2005ip and 2006jd 
 exhibit thermal emission associated with  a cold dust component ($T_c$ $\sim$ 600 K).
 These authors attribute this emission to pre-existing dust, which is illuminated by photons that ultimately originate from 
 radiative shocks produced from the SN--CSM interaction.

We now endeavor to estimate the dust mass associated with the warm 
near-IR and cold mid-IR dust components.
 To estimate the mass ($M_{d}$) of each of these components in 
 SNe~2005ip and 2006jd we follow the prescription 
 of \citet[][and references therein]{fox10}, which entails fitting the SEDs with the following expression
 
\begin{equation}
\label{equation2}
F(\lambda)=\frac{M_d B_\lambda(T_d)\kappa_\lambda(a)}{D^2}.
\end{equation}

\noindent This formalism is valid for small optical depths, and allows us to 
directly constrain $M_{d}$.
Here $B_\lambda(T_d)$ is the Planck $BB$ function, 
$\kappa_\lambda(a)$ is the dust mass absorption coefficient for a dust particle 
with radius, $a$, and $D$ is the distance to the emitting source.
In the following we adopt the $\kappa_\lambda(a)$ functions of 
\citet[][see their Figure 4]{fox10} for a 
dust composition consisting of graphite grains with radii 
$a$ $=$ 0.01~$\mu$m, 0.1~$\mu$m, and 1~$\mu$m.
Based on the absence of the 9.6 $\mu$m feature in their {\em Spitzer} observations,
\citet{fox10}  excluded silicates in their analysis of SN~2005ip. To be consistent, 
in the follow we  compute $M_d$  for both SN assuming that their 
dust components are composed of only graphite.  
Equation~\ref{equation2} is commonly used for a wide variety of astrophysical applications, and is based on the assumption that the emitting
dust is: (1) dominated by one grain size and (2) emits at a single temperature. 
   
Fitting Equation~\ref{equation2} to the series of SEDs plotted in Figure~\ref{fig7}, 
with, $M_d$ and $T_d$ as free parameters,
warm dust mass estimates are obtained for graphite grains with 
 radii of $a$ $=$ 0.01 $\mu$m, 0.1 $\mu$m, and 1 $\mu$m.
 The resulting values for SNe~2005ip and 2006jd are listed in Table~\ref{table9}, and the corresponding best-fit function of Equation~\ref{equation2} for the grain size of radius $a = 0.1~\mu$m is over-plotted on the SEDs of Figure~\ref{fig7} as dashed lines. 
 Note that the solid $BB$ fits included in this figure correspond to a graphite grain size of 
 radius $a = 1.0$~$\mu$m.
 The results of the various fits give increased dust masses as a function of phase, and 
 are strongly sensitive to the adopted grain size. 
 For the range of dust grain sizes and assuming a 
 4$\pi$ covering factor,
 we obtain limits on $M_d$ ranging from 0.2--6.0$\times$10$^{-4}$$M_{\sun}$ 
 and 0.7--9.8$\times$10$^{-4}$$M_{\sun}$ for SNe~2005ip and 2006jd, respectively. 

\begin{figure}[t]
\epsscale{1.0}
\plottwo{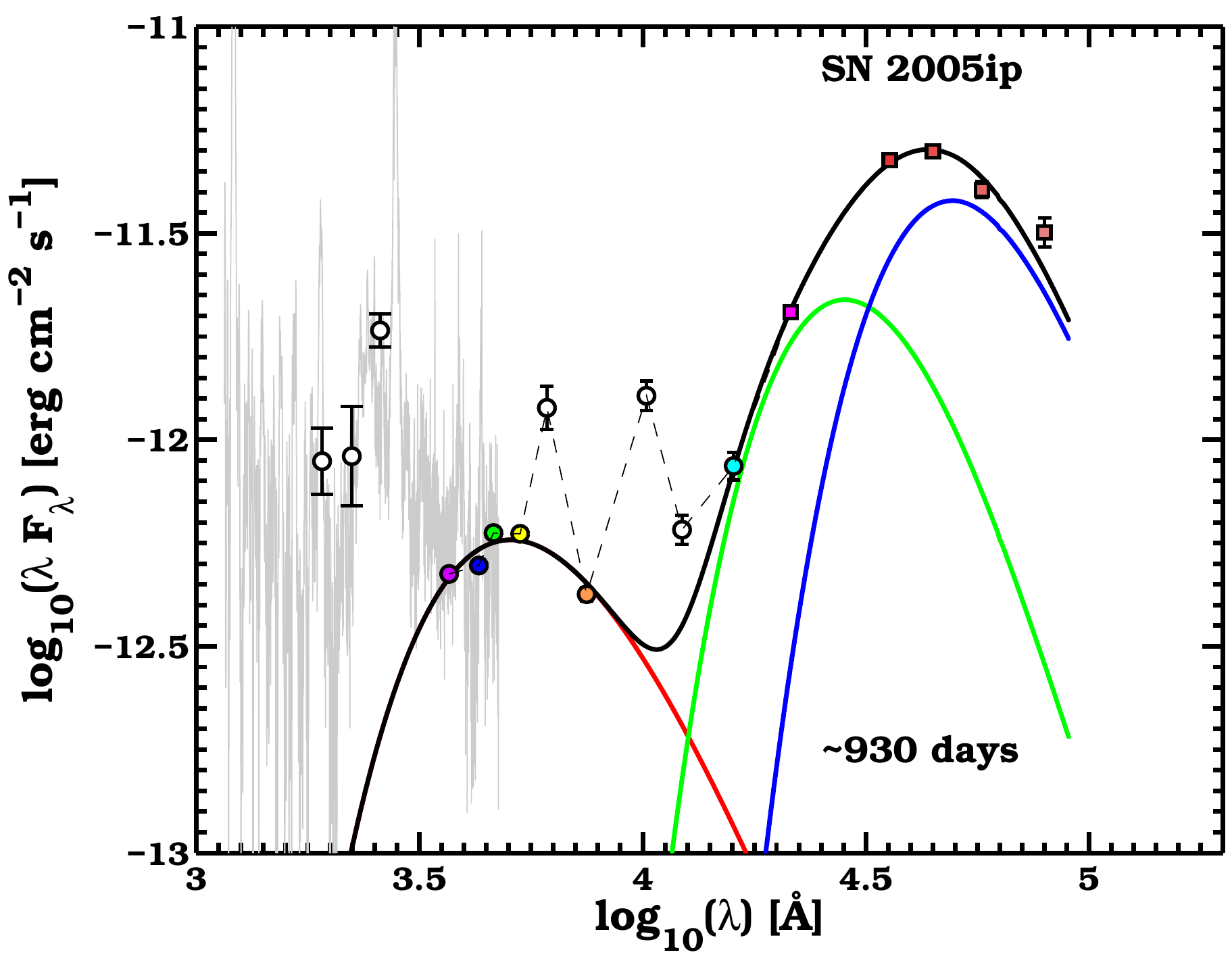}{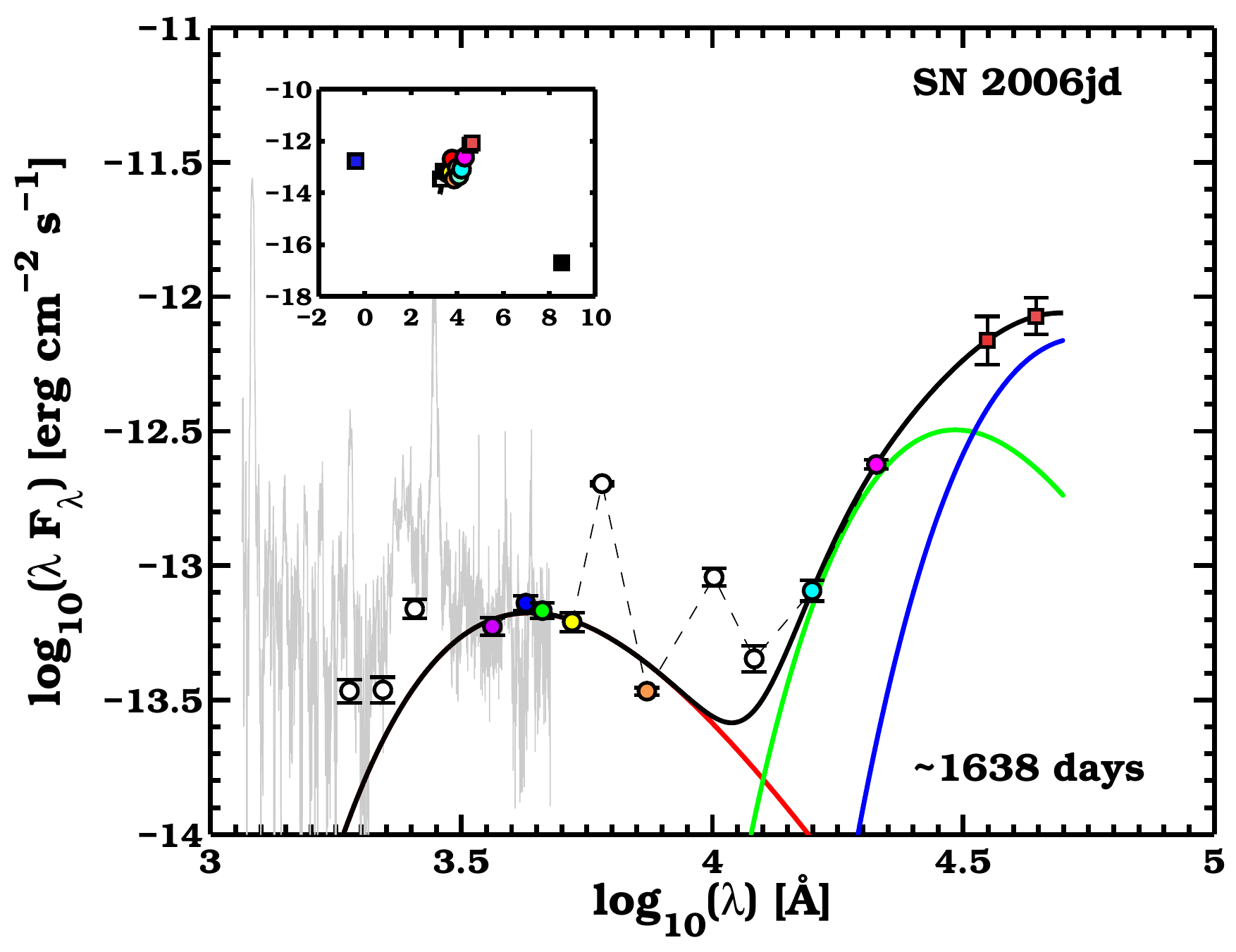}
\figcaption[]{Full SED of SN~2005ip ({\em left}) on day $\sim$ 930 and SN~2006jd 
({\em right}) on day $\sim$ 1638. 
The SEDs of both objects extends from the UV to the mid-IR. 
The inset contained within the plot of SN~2006jd also includes 
X-ray and radio measurements. 
White filled circles correspond to the UV- and $rYJ$-band flux points which were 
excluded from the  $BB$ fits. 
Also plotted in each panel is a late phase ($\sim$ 1000 days past explosion) 
HST spectrum of the 1988Z-like SN~1995N 
\citep{fransson02} scaled to match the ground-based $u$- and $B$-band flux points.
 Black solid lines correspond to the combined three-component $BB$ fit, while
 the red, green and blue lines correspond to the individual hot, warm and cold components.  \label{fullSED}} 
\end{figure}

To place limits on $M_d$ associated with the cold dust component, we 
turn to {\em Spitzer} mid-IR observations. 
 Plotted in Figure~\ref{fullSED} is the full SED of SN~2005ip (left panel)
at day $\sim$ 930 and SN~2006jd (right panel) at day $\sim$ 1638.
Here the mid-IR flux points for SN~2005ip are taken from \citet{fox10}, while 
for SN~2006jd we interpolated between the mid-IR flux values presented in 
\citet{fox10} and the values derived from our recent {\em Spitzer} observations.
In this way we avoid the uncertainty incurred from large extrapolation of 
optical and near-IR fluxes to the date of our latest {\em Spitzer} observation. 
Also plotted in each panel is a late phase  {\em Hubble Space Telescope} (HST)
 spectrum of the 1988Z-like SN~1995N \citep{fransson02}, scaled to match the ground-based 
 $u$- and $B$-band flux points. The HST spectrum suggests that the  
 UV excess exhibited by the UVOT observations
 is associated with  numerous emission lines including, amongst others, 
 [\ion{O}{3}], [\ion{Si}{3}], \ion{Fe}{2}, \ion{Mg}{2}, [\ion{C}{3}], and [\ion{N}{3}]
 \citep[see][Fig. 7]{fransson02}.
 Finally, for completeness the inset contained within the plot of SN~2006jd also 
included reported X-ray and radio measurements \citep{chandra12}. 


 Over-plotted as a black line in each panel of Figure~\ref{fullSED} is the sum of a 
 simultaneous three-component $BB$ fit corresponding to a hot, warm and cold component. 
 In both cases  the hot $BB_{h}$ component fit 
 is made to the $uBgVi$ flux points, while the 
warm $BB_{w}$ component is fit to the $HK_s$ flux points, and the cold $BB_{c}$ component is fit to the mid-IR flux points.
  As mentioned in Section~\ref{uvoir} the $rYJ$ flux points are not used to constrain the best $BB$ fits due to strong emission lines contained within these passbands.
 Following \citet{fox11}, we fit the near- and mid-IR portion of the SED with distinct $BB$ components. 
 This indicates the presence of  two discrete dust emitting regions, which are characterized with warm $BB_{w}$ and cold $BB_{c}$ temperatures of roughly T$_w \sim 900$ K and T$_c \sim$ 500--700 K, respectively.

Plotted as blue lines in Figure~\ref{fullSED} are the 
best-fit cold $BB_{c}$ components for dust grain size of 0.1 $\mu$m, 
and the corresponding 
temperatures, radii and dust masses are listed in Table~\ref{table10}.
These fits suggest cold masses of $\sim$ 0.01 $M_{\sun}$ and $\sim$ 0.02 
$M_{\sun}$ for SN~2005ip and 2006jd, respectively.
As in the case of fitting the warm dust component, our computed values of   
$M_d$ for the cold dust are quite sensitive on 
the composition and size of the dust grains.  
Finally we note that from the full SEDs we are able to estimate that the flux associated with cold dust component at these late epochs represents 
$\sim$ 1\% of the total UVOIR flux.

\subsection{Optical Spectroscopy}
\label{spectroscopy}

\subsubsection{Supernova 2005ip}

The spectroscopic sequence of SN~2005ip shown in Figure~\ref{fig3} follows 
the evolution leading up to the plateau phase, while the last spectrum 
probes the emission at very late times. A detailed description of the spectral evolution of SN~2005ip is presented by \citet{smith09}; here we summarize the overall appearance, and use these observations as a comparison to SN~2006jd. 
 The early phase spectra show a significant multi-component H$\alpha$ 
emission feature and a broad \ion{Ca}{2} 
 $\lambda\lambda$8498, 8542, 8662 bump, superposed 
on a pseudo-continuum exhibiting an excess of flux 
bluewards of $\sim$ 5700 \AA.  
H$\alpha$ dominates the spectrum, and at early phases the 
Balmer decrement, as measured from the sum of all components is $\approx$ 25. 
Narrow emission features attributed to 
[\ion{O}{3}] $\lambda\lambda$4959, 5007,
\ion{He}{1} $\lambda\lambda$5876, 6678, 7065, 7281,
[\ion{Ar}{3}] $\lambda$7136 and 
[\ion{O}{2}] $\lambda$7325 
are observed both at early- and late-times. 
A remarkable feature displayed in the spectra of SN~2005ip are the high-ionization coronal emission lines, e.g. [\ion{Ar}{14}], [\ion{Fe}{14}] and [\ion{Ar}{10}], which are discernible at the earliest epochs, and over time, grow in strength relative to the continuum flux \citep[see][for a detailed analysis of these features]{smith09}. Over the duration of its evolution the majority of emission features remain as narrow lines, however by day 169, 
\ion{He}{1} $\lambda\lambda$5876, 7065 show intermediate-width components, 
characterized by FWHM velocities of 
$\sim$ 3500 km s$^{-1}$ and 2500 km s$^{-1}$, respectively.

 A detailed discussion on the evolution of H$\alpha$ in both SN~2005ip and 2006jd is provided below in Section~\ref{Halphafwhm}.

\subsubsection{Supernova 2006jd}
\label{6jdspectra}

\begin{figure}[t]
\plotone{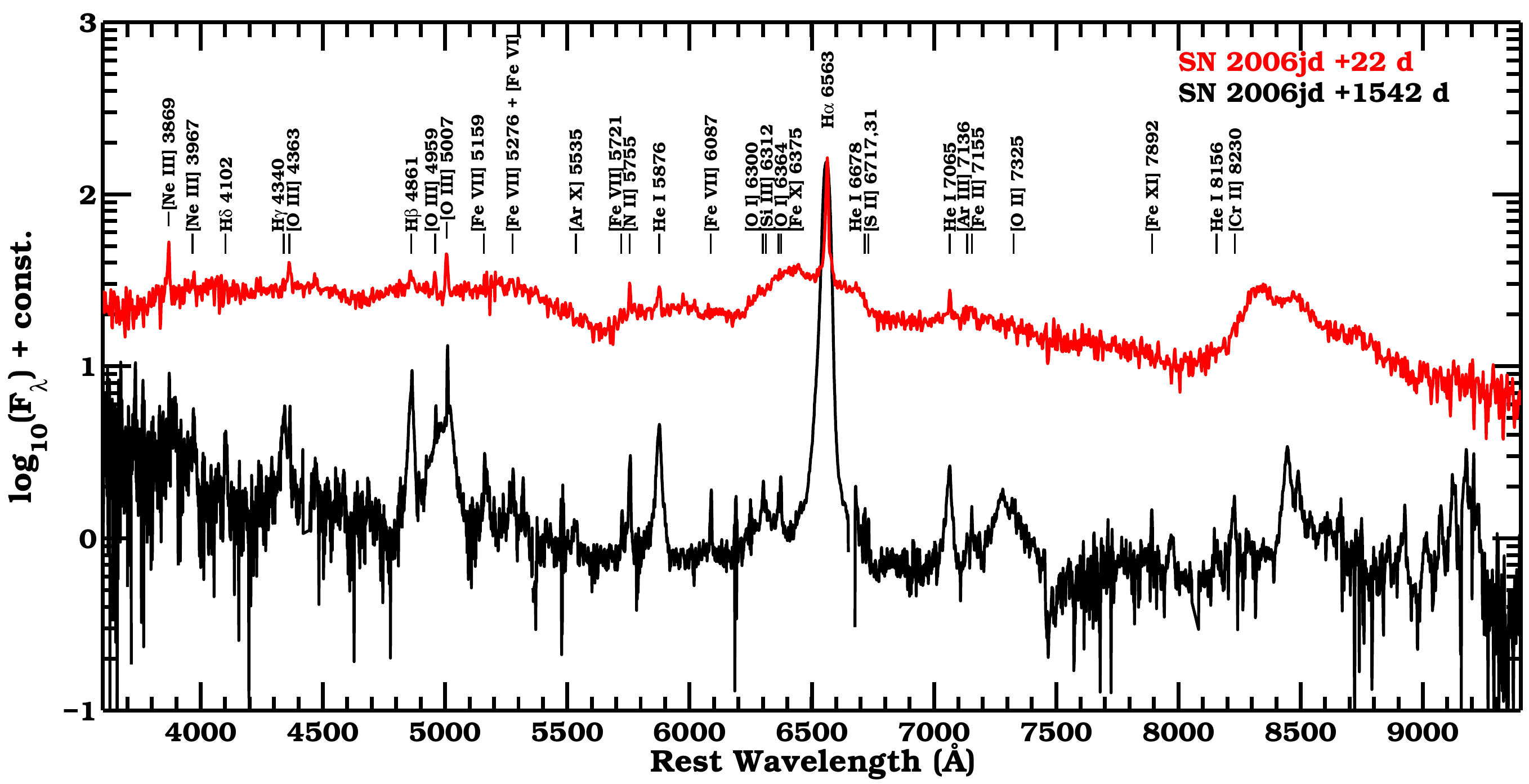}
\figcaption[]{Comparison of the day 22 and day 1542 spectra of SN~2006jd. 
Ions responsible for many of the features are indicated. 
 Features of [\ion{O}{3}] $\lambda\lambda$4959, 5007, \ion{He}{1} $\lambda\lambda$5876, 7065 and H$\alpha$ all exhibit prominent intermediate-width components in the day 1542 spectrum, which are 
 associated with post-shock gas. 
 \label{6jdspeccomp}} 
\end{figure}

The optical spectroscopic sequence of SN~2006jd shown in Figure~\ref{fig6} 
covers 1520 days of evolution.
The appearance of these spectra resembles those of SN~1988Z and SN~2005ip  
(see Section~\ref{88Zcomparison}). 
At early epochs the spectrum of SN~2006jd contains
broad H$\alpha$ and \ion{Ca}{2} $\lambda\lambda$8498, 8542, 8662 emission features,
as well as an excess of continuum flux bluewards of $\sim$ 5700~\AA.
 The H$\alpha$ feature is composed of an intermediate-width component situated on top of a broad, asymmetric component, which shows more flux in the blue wing. Compared to SN~2005ip, the Balmer decrement shown by SN~2006jd, as measured from the sum of components is even more pronounced, 
 ranging from $\approx$ 55 to $\approx$ 40 between the first and last spectra.
 In addition narrow, weaker lines associated with, amongst others, 
[\ion{Ne}{3}] $\lambda\lambda$3869, 3967, [\ion{O}{3}] $\lambda\lambda$4363, 4959, 5007, [\ion{N}{2}] $\lambda$5755 and \ion{He}{1} $\lambda\lambda$4471, 5876, 6678, 7065 are also present. 
Over time these lines tend to increase in relative strength compared to the continuum, while a number of 
conspicuous lines from various other ions appear. 
This is illustrated in Figure~\ref{6jdspeccomp} where the first 
(day 22) and last (day 1542) spectra of SN~2006jd are compared, with
 ions associated with the most prevalent narrow lines labeled, and their corresponding 
 line fluxes are given in Table~\ref{table11}.
 
\begin{figure}[t]
\plotone{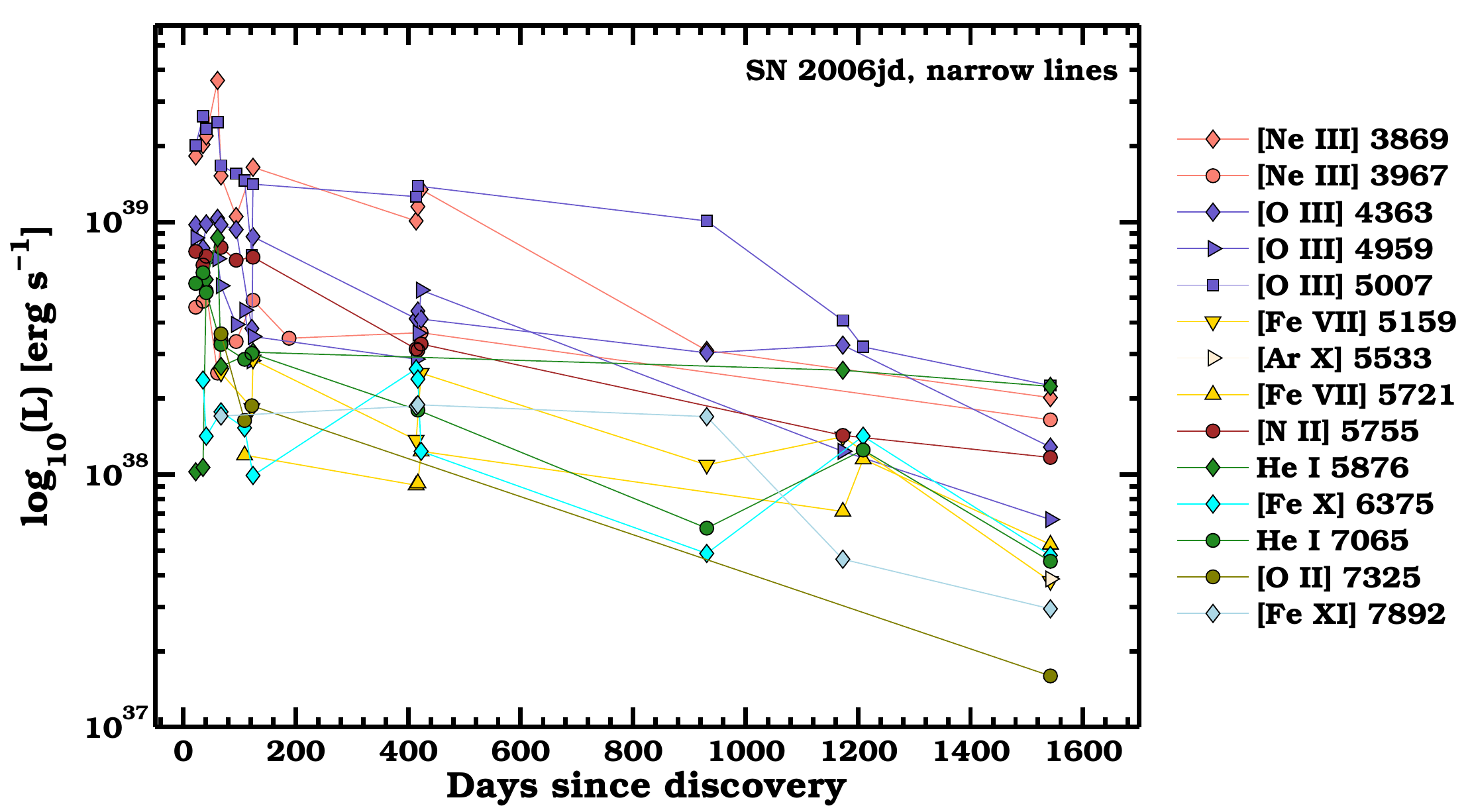}
\figcaption[]{Luminosity evolution of forbidden and
\ion{He}{1} $\lambda\lambda$5876, 7065 emission lines measured from the spectroscopic sequence of SN~2006jd.
 \label{narrowlines}} 
\end{figure}

 From this comparison emission features associated with [\ion{O}{3}] $\lambda\lambda$4959, 5007 and \ion{He}{1} $\lambda\lambda$5876, 7065 are seen to gain an intermediate-width component.
 This component is typically associated with either post-shock gas 
 or unshocked ejecta gas photoionized by shock radiation \citep[e.g.][]{fransson02},
and here is observed to ``turn on" by day 414, and by day 1542 it exhibits FWHM velocities between $\sim$ 1000--1200 km s$^{-1}$. 
 A close inspection of the day 1542 spectrum of SN~2006jd also reveals various medium- and high-ionization coronal lines, e.g. [\ion{Fe}{7}] $\lambda\lambda$5159, 5276, 5721, 6087, [\ion{Fe}{10}] $\lambda$6375, [\ion{Ar}{10}] $\lambda$5535, 
 [\ion{Fe}{10}] $\lambda$6375 and [\ion{Fe}{11}] $\lambda$7892.
Hints of these coronal features are evident as small notches in the earliest spectra (see Figure~\ref{6jdspeccomp}), but by late phases, they too have grown considerably in strength relative to the continuum flux. 
With an ionization potential of 442.3 eV, [\ion{Ar}{10}] $\lambda$5535 is 
the highest excitation coronal line identified in SN~2006jd.
The overall luminosities of the coronal lines identified in SN~2006jd remain relatively constant over the first $\sim$ 1000 days of evolution, but by the time of the last spectrum, their luminosities appear to drop by a factor of ten. This is shown in 
Figure~\ref{narrowlines} which shows the luminosity evolution for the 
coronal lines, and a number of the prominent forbidden and \ion{He}{1} lines.

Provided with the line fluxes given in Table~\ref{table11}, 
we searched for various line diagnostics to constrain physical parameters associated with the underlying emitting gas. Unfortunately, most of the interesting line ratios are blended with lines of \ion{Fe}{2}, however the [\ion{O}{3}] $\lambda$5007/$\lambda$4363 ratio does appear
to be free of contamination. 
Assuming negligible host galaxy reddening, the observed [\ion{O}{3}] $\lambda$5007/$\lambda$4363 ratios measured from the series of spectra imply number densities of $>$ 10$^{6}$~cm$^{-3}$ in the narrow line region. 
The fact that these features are observed 
also implies a minimum gas temperature of 1.0$\times$10$^{4}$~K.
The implications of this quite high gas density are discussed in relation to the mass-loss history of the progenitor in Section~\ref{discussion}.

\subsubsection{Evolution of H${\alpha}$ line profiles}
\label{Halphafwhm}

\begin{figure}[t]
\plottwo{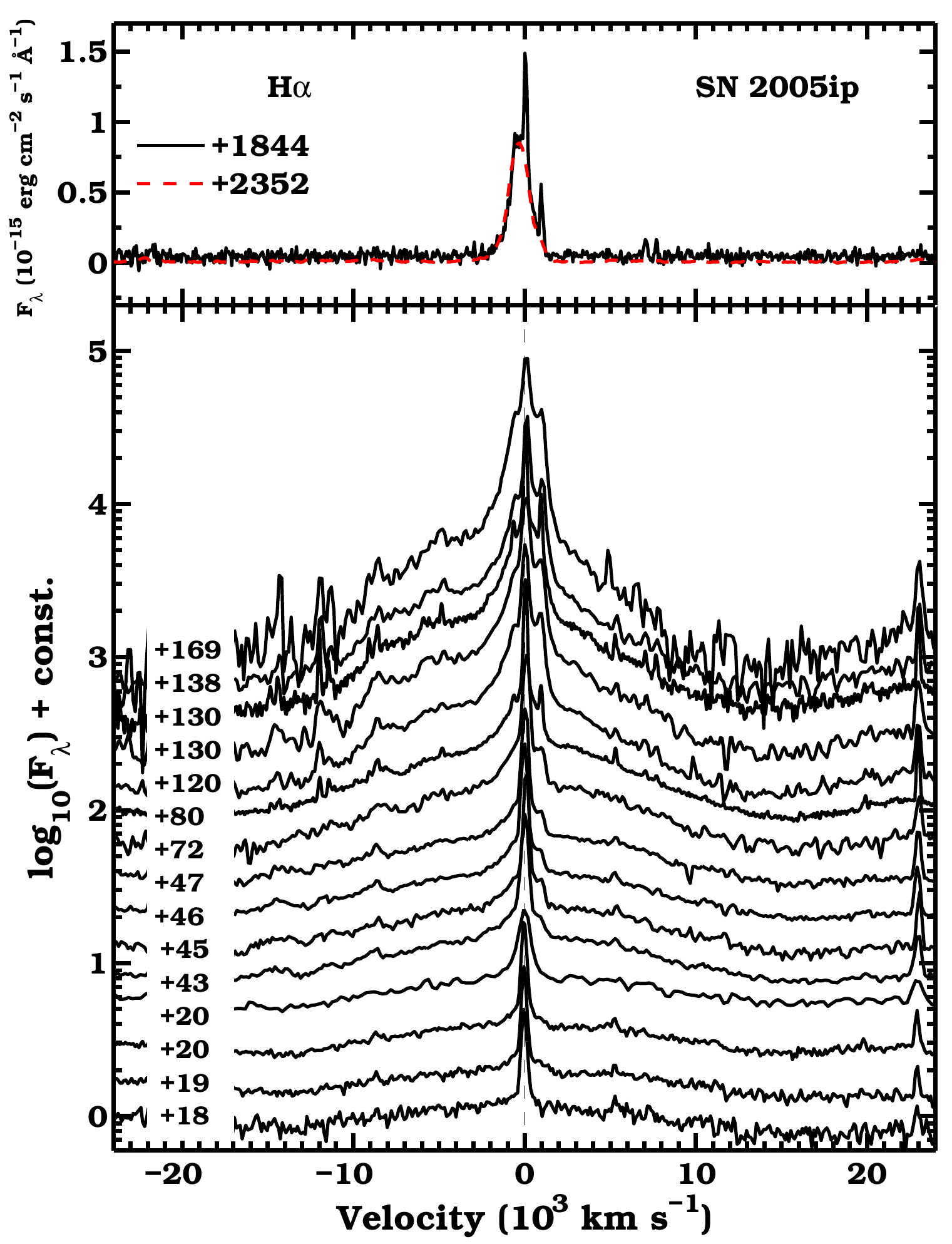}{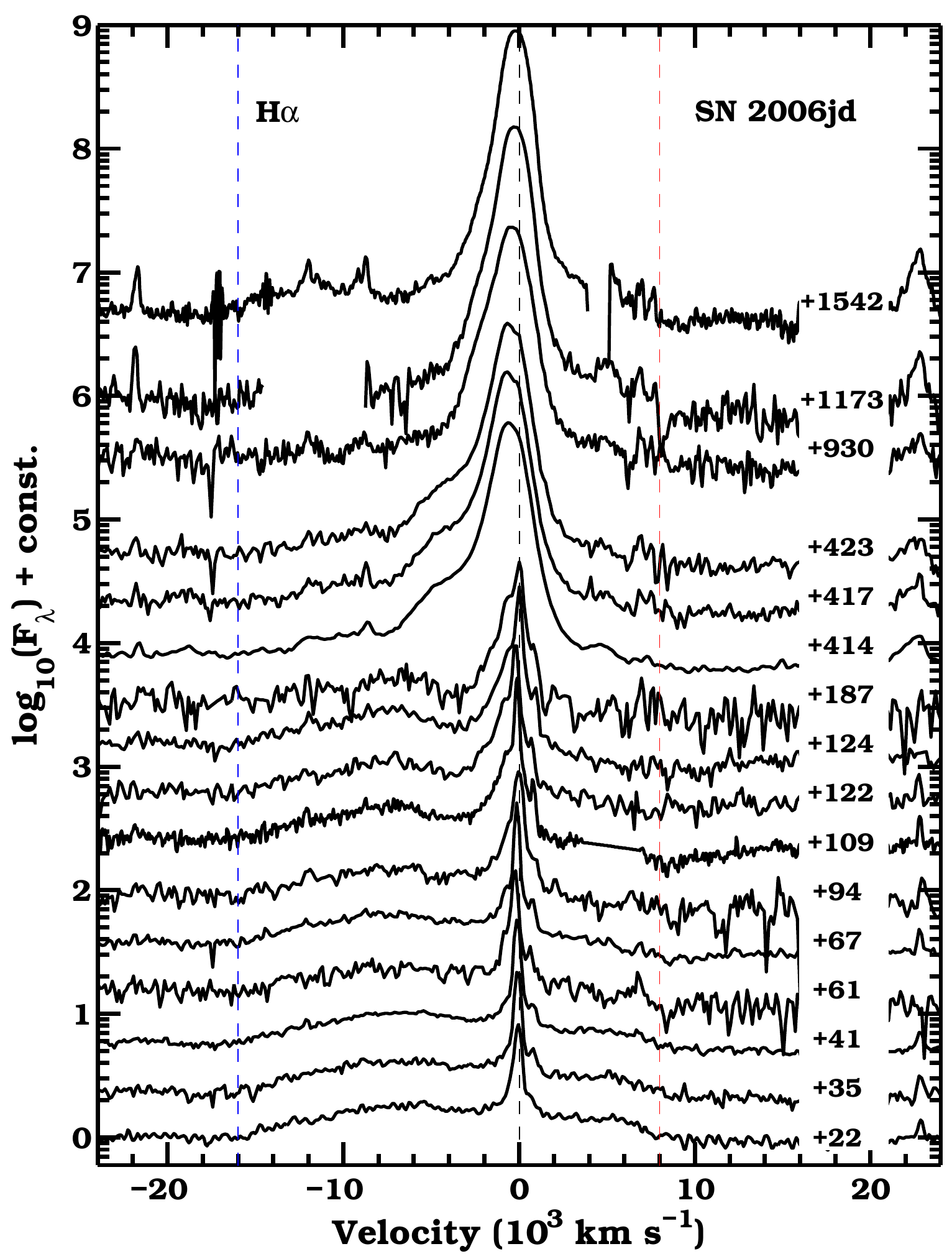}
\figcaption[]{H$\alpha$ line profile evolution
as observed from our spectroscopic time-series of 
 SN~2005ip {\em (left panel)} and SN~2006jd {\em (right panel)}. 
The epoch of each spectrum with respect to the date of discovery is indicated.
In the case of SN~2005ip the two late phase spectra are plotted on top of one another, 
while in the plot of SN~2006jd, vertical lines  highlight the 
asymmetric nature of the broad component. 
\label{halpha}} 
\end{figure}

\begin{figure}[t]
\plotone{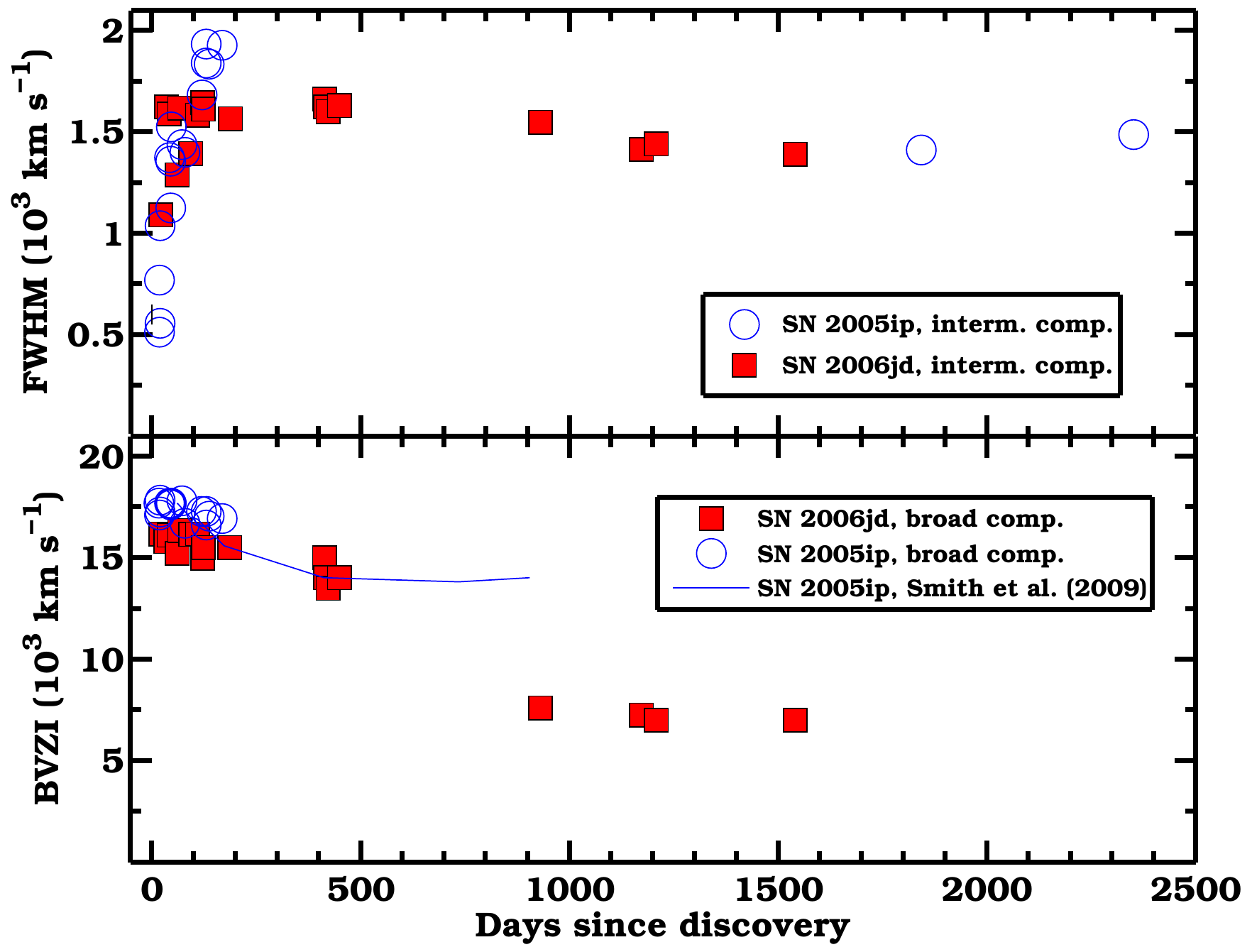}
\figcaption[]{Evolution of the width of the H$\alpha$ profile over time. 
Top panel displays the FWHM velocity of SNe~2005ip and 2006jd obtained from Gaussian fits to the intermediate-width component.
Bottom panel shows the evolution of 
BVZI (blue-velocity at zero intensity) of the broad H$\alpha$ component, including the values measured from \citet{smith09} for SN~2005ip out to $\sim$~900 days. 
 \label{fwhm}} 
\end{figure}

Through careful examination of the H$\alpha$ emission profile and its time-evolution 
one can gain insight on the dynamics of the SN--CSM interaction, a more complete understanding of the shock physics, and clues concerning the geometry of the CSM. 
Presented in Figure~\ref{halpha} (left panel SN~2005ip and right panel SN~2006jd) 
are the full time-series of spectra zoomed in on the position of H$\alpha$.  
Close inspection reveals both broad- and intermediate-width components, and 
in addition, the early phase spectra also exhibit a hint of a narrow component, which
 is only partially resolved.  
Most interestingly, a close comparison of the broad component 
between the two objects reveals 
 this feature to be highly asymmetric in SN~2006jd from the onset of our observations
(see also Figure~\ref{fig10}), showing more flux in the blue wing. 
Possible explanations of this peculiarity could be (i) occultation of the red side of the 
SN by an optically thick ejecta (see Section~\ref{whereisthedust}), 
(ii) an asymmetric explosion, or (iii) fast gas moving through 
a clumpy ejecta \citep[e.g.][]{fransson02}. 
As a result of this asymmetry, traditional Gaussian fits to the broad-component render unreliable FWHM velocity measurements. We therefore adopted the approach of measuring the blue velocity 
at zero intensity (BVZI), which is equivalent to the maximum extent of the broad component, and depending on the geometry of the CSM, can be considered a proxy for the velocity of the forward shock or the
fastest component of the expanding SN ejecta. 
With respect to the intermediate-width component, Gaussian fits provide a robust measure of the average velocity of the post-shock gas.
The resulting FWHM velocity of the intermediate-width component (top panel) and BVZI of the broad component (bottom panel) are plotted in Figure~\ref{fwhm}. 

As revealed by Figure~\ref{fwhm}, at early times the BVZI of the broad component 
reaches $\sim$~18,000~km~s$^{-1}$ and $\sim$~16,000~km~s$^{-1}$ in 
SN~2005ip and SN~2006jd, respectively. 
Although it is difficult to accurately estimate the BVZI of SN~2006jd beyond 500 days, 
under close scrutiny it is observed to evolve from $\sim$ 8000 km~s$^{-1}$ to $\sim$ 7000 km~s$^{-1}$ between day 930 to 1540.
As our spectroscopic coverage of SN~2005ip is limited, we turn to the measurement of \citet{smith09}, which indicate BVZI reached $\gtrsim$~15,000~km~s$^{-1}$ by day 400, and remained at this velocity beyond day 905 (see solid blue line in Figure~\ref{fwhm}). We note that the last several of these BVZI measurements appear to be based on extrapolation and therefore should probably be considered as an upper limit.  
 
 Turning to the time-evolution of the intermediate-width component, 
Figure~\ref{fwhm} shows that over the first $\sim$ 150 days the FWHM velocity in both objects reaches maximum values of $\sim$ 1900 km~s$^{-1}$ and $\sim$ 1700 km~s$^{-1}$ 
in SNe~2005ip and 2006jd, respectively. 
 Subsequently, the FWHM velocity remains relatively constant dropping down to only $\sim$ 1500 km~s$^{-1}$ over the duration of our observations.
 
 Returning to the time-series plotted in Figure~\ref{halpha} 
 (see also below, Figure~\ref{xrays}), unlike SN~2005ip, the flux contained within the intermediate-width component of SN~2006jd is seen to dramatically increase
 from day 187 to 414.
 This period coincides with the onset of the plateau phase 
seen in the light curves, and is driving the excess of flux observed in the $r$-band light curve. 
In Section~\ref{whereisthedust} the BVZI and FWHM velocity measurements are used to 
indicate the location of the warm and cold dust components.

\section{DISCUSSION}
\label{discussion}

\subsection{Comparison of 1988Z-like SNe~IIn}
\label{88Zcomparison}

Objects within the 1988Z-class exhibit strikingly similar 
characteristics, however as we have already shown, subtle 
differences do exist between SNe. 
These differences can be used to decipher the nature of 
the CSM and provide insight on the progenitors. 
We now proceed to compare and contrast the main observational 
characteristics of SNe~2005ip and 2006jd.

\begin{figure}[t]
\epsscale{0.8}
\plotone{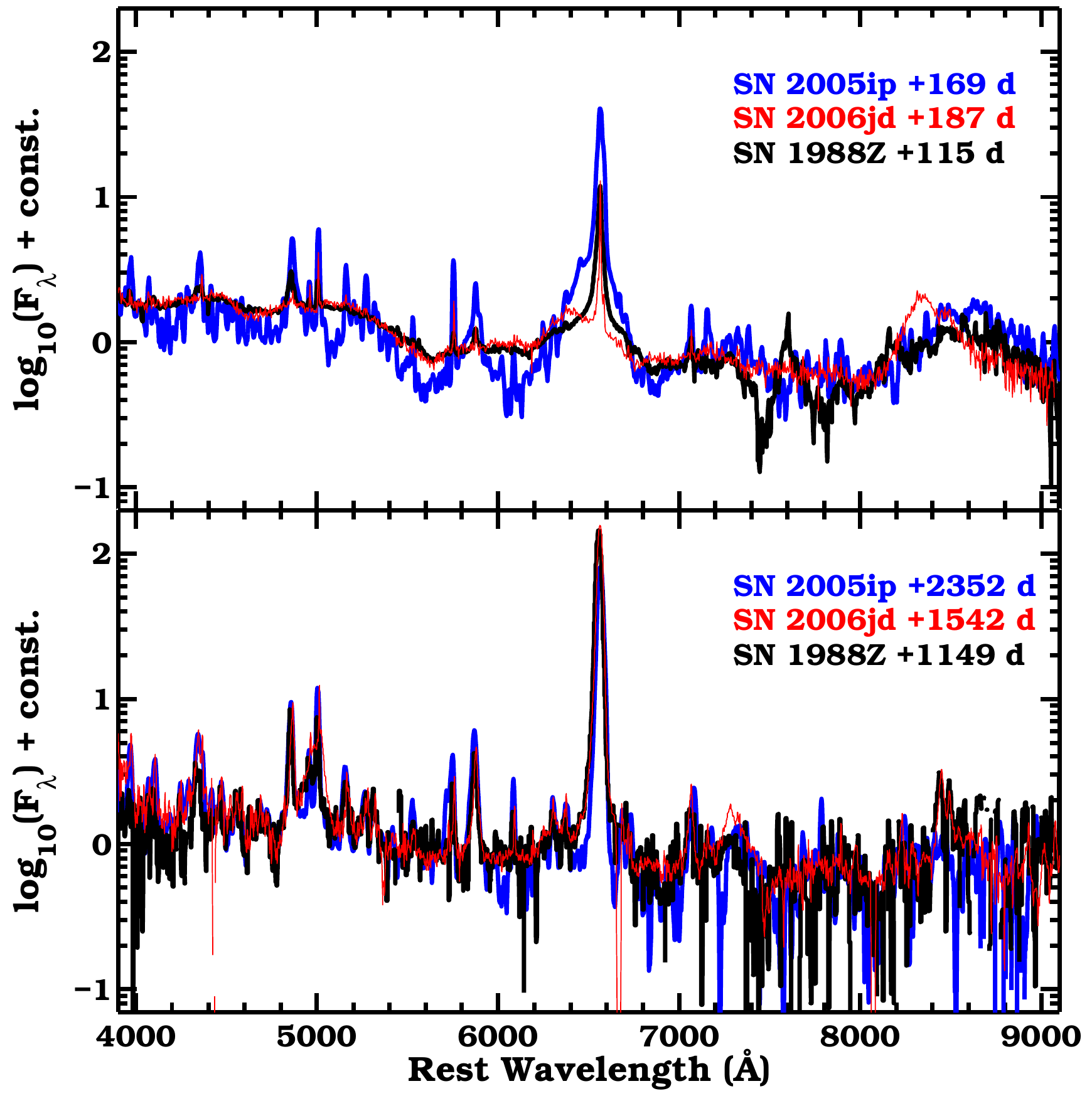}
\figcaption[]{Comparison of early (top) and late (bottom) phase spectra of SNe~1988Z, 2005ip and 2006jd. 
Date provided next to the label of each SN spectrum refers to epoch past discovery.\label{speccomp}} 
\end{figure}

Turning to their spectroscopic appearance, plotted in Figure~\ref{speccomp}
are early phase (top panel) and 
late phase (bottom panel) spectra of SNe~1988Z \citep{turatto93}, 2005ip, and 2006jd. 
Clearly, the overall appearance of these objects is remarkably similar at both early- and 
late-times. At early epochs each spectrum is dominated by a broad H$\alpha$ 
component exhibiting similar BVZI and FWHM values, and as 
noted earlier, an excess of blue continuum flux which is powered by numerous 
\ion{Fe}{2} emission lines \citep[e.g., see][]{foley07}. 
At late times the FWHM velocity of the intermediate-width component 
and the relative strength of the H$\alpha$ feature are nearly identical. 
 Close similarities also appear between the relative strength and FWHM velocity of
 the \ion{He}{1}, [\ion{O}{3}] $\lambda\lambda$4363, 4959, 5007, the \ion{Ca}{2} near-IR triplet, 
 and many of the coronal emission lines. 
 
\begin{figure}[t]
\epsscale{0.8}
\plotone{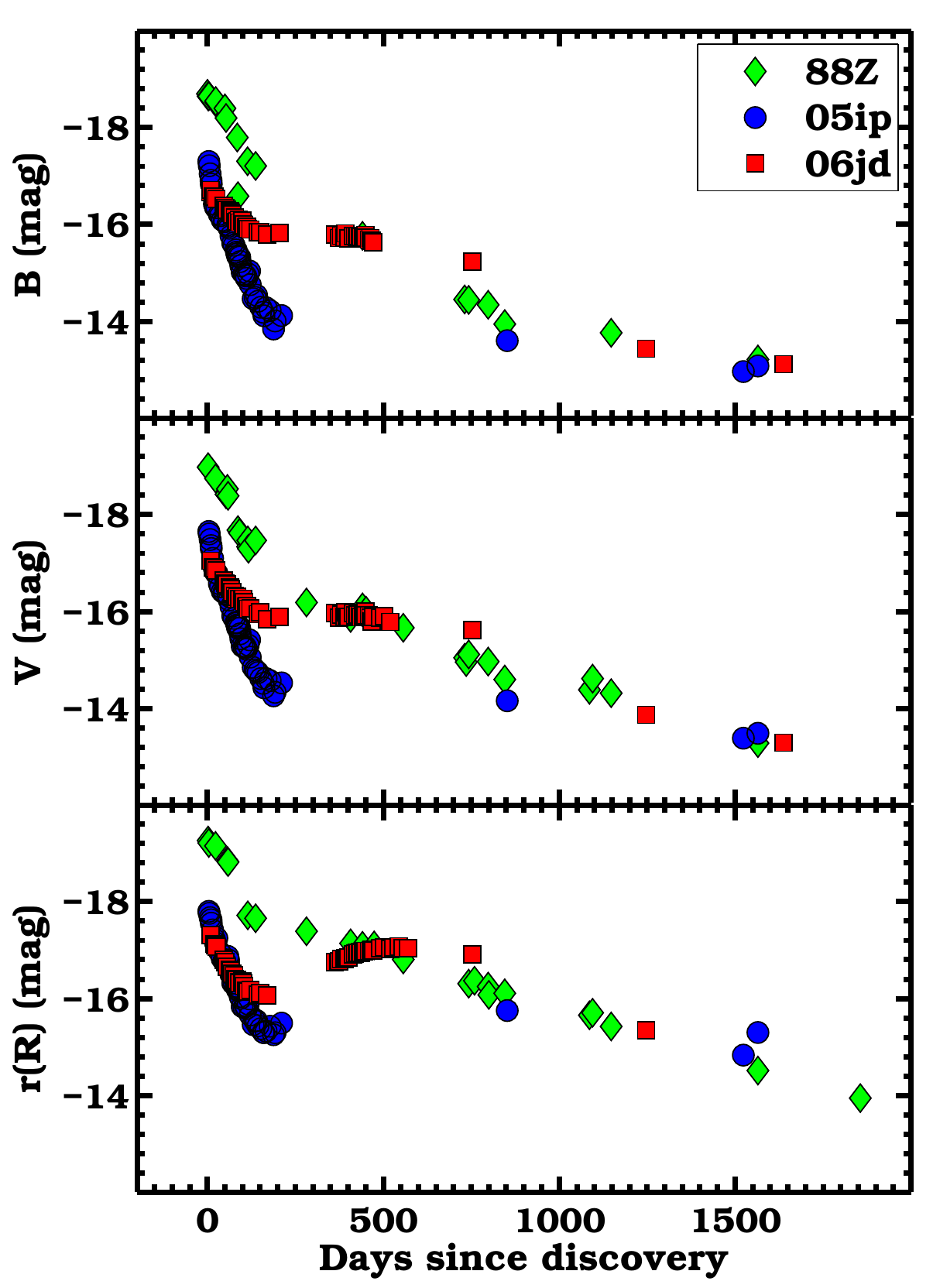}
\figcaption[]{ Comparison of absolute $B$-, $V$-, and $r$-band light curves of 
SNe~1988Z, 2005ip and 2006jd.\label{abslcs}} 
\end{figure}

 In Figure~\ref{abslcs} we compare the absolute $B$-, $V$- and $r/R$-band light curves of 
SNe~1988Z \citep{turatto93}, 2005ip and 2006jd. This comparison provides
evidence of additional
observational diversity amongst members of this class of SNe~IIn. 
With the caveat in mind that the explosion epochs of these SNe are not well constrained, at early epochs 
 SN~1988Z is found to be $\sim$ 1.5 mag brighter than 
 SNe~2005ip and 2006jd.
The photometric evolution over the first $\sim$ 200 days is nearly identical 
in each SN. Beyond this point SN~1988Z exhibits an inflection point followed by a slow, linear decline, while SNe~2005ip and 2006jd exhibit a phase characterized by constant emission. In the case of SN~2005ip, however, its brightness during this plateau phases 
is $\sim$ $-$14 mag, while in SN~2006jd the absolute magnitude is, depending on the specific bandpass, between $-$16 to $-$17 mag. 
Interestingly, as observed in Figure~\ref{abslcs}, by day $\sim$ 1500 all three objects exhibit nearly identical magnitudes in the $B$- and $V$-band light curves of 
 $\sim$ $-$13 mag. 
 
 \subsection{Insight on the progenitor of SN~2006jd}
\label{insight}
 
\begin{figure}[t]
\epsscale{1.1}
\plotone{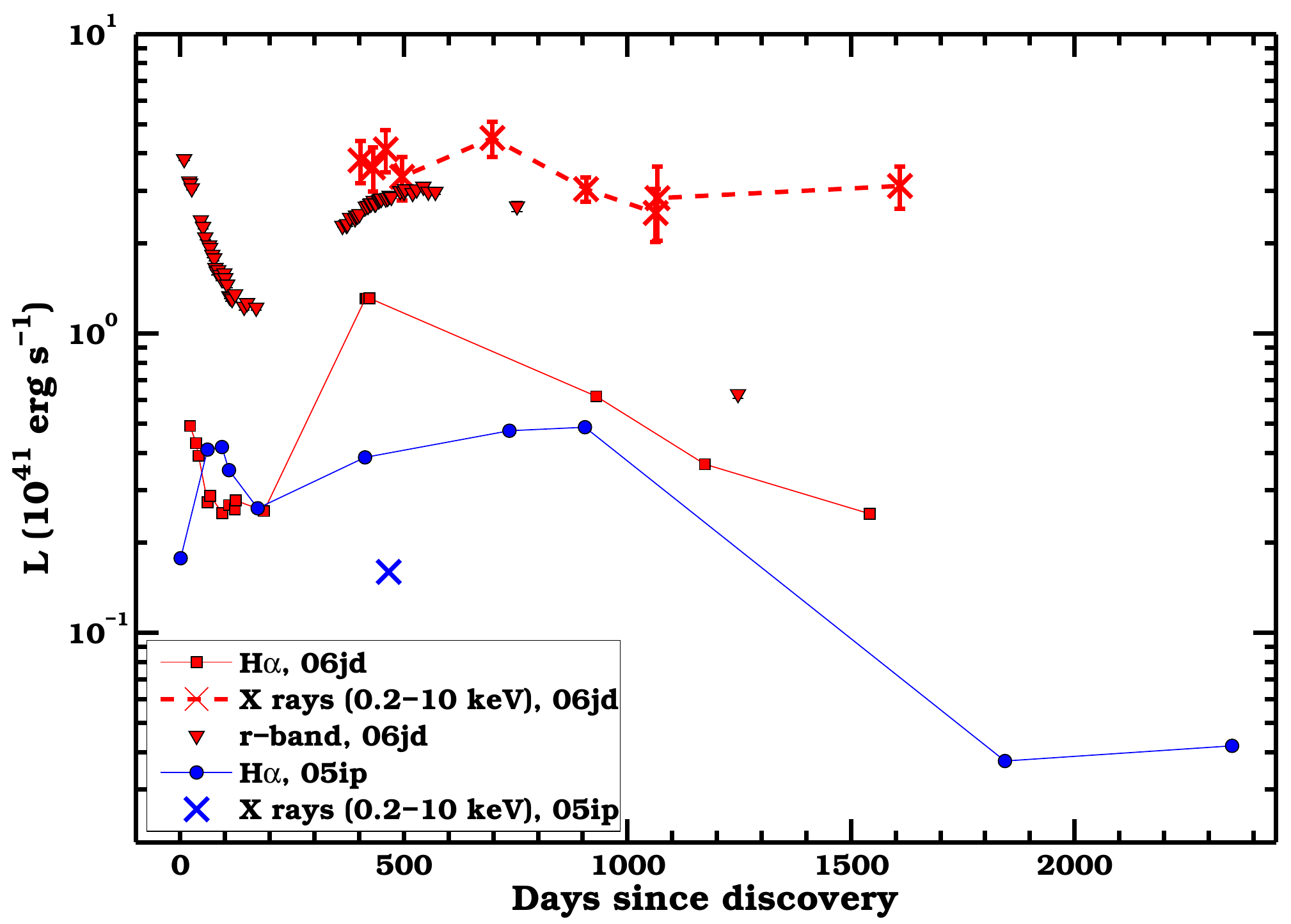}
\figcaption[]{
Comparison of H$\alpha$ and reported X-ray luminosities of SNe~2005ip and 2006jd. 
Also included is the absolute $r$-band light curve of SN~2006jd. 
\label{xrays}} 
\end{figure}

 Figure~\ref{xrays} presents a comparison of the soft X-ray 
 (0.2$-$10 keV) and H$\alpha$ luminosities of 
 SNe~2005ip and 2006jd. The X-ray measurement of SN~2005ip 
 is from
 \citet{immlerpooley07}, and  those of SN~2006jd were recently reported 
 by \citet{chandra12}. 
  Also included in the figure is the absolute $r$-band light curve of SN~2006jd. This comparison reveals that during the plateau phase
 when SN~2006jd clearly outshone SN~2005ip at optical/near-IR wavelengths,
 the X-ray (and H$\alpha$) luminosity of SN~2006jd roughly exceeds SN~2005ip 
 by an order of a magnitude. Interestingly, the X-ray luminosity of SN~2006jd
 appears to remain relatively constant for at least $\sim$ 600 days. 
 
 A detailed discussion of the X-ray luminosity is presented 
 by \citet{chandra12}, nevertheless here we speculate that the flat evolution
 observed in the X-ray light curve of SN~2006jd may result from the SN blast wave interacting with a dense, 
 torus-like shell of material located equatorially around the SN,
 akin to what is observed in $\eta$ Carinae \citep{smith06}.
 As the SN blast wave interacts with this shell of material radiative shocks produce 
 copious amounts of X-rays, which in turn, gives rise to: 
 (i) the plateau phase of the optical light curves, (ii) the significant 
strengthening over time of the
 intermediate-width component of H$\alpha$, hence the $r$-band excess,
 (iii) the various emission lines that exhibit intermediate-width components and,
 (iv) the coronal lines which are formed from photo-ionization of the pre-SN wind. 
In this model the broad H$\alpha$ component is then 
formed from the SN blast wave interacting with the 
rarified progenitor wind located 
 in the polar directions of the progenitor
star, perpendicular to the plane of the dense shell \citep[e.g.][]{blondin96}. 
 The equatorial shell likely originates 
 from a pre-SN LBV eruption. 
 Such a shell can explain both the $>$ 10$^{6}$ cm$^{3}$ gas density 
  measured from the [\ion{O}{3}] $\lambda$5007/$\lambda$4363 ratio, and a
  portion of the near-IR emission corresponding to warm dust, 
 which we estimate in Section~\ref{dust} to be located 
 $\sim$ 2.5$\times$10$^{16}$ cm from the progenitor star.
  
 The diversity of the X-ray luminosities exhibited in Figure~\ref{xrays} 
 is likely due to
differences in the density and/or geometry of the CSM which is shocked
by the SN blast wave. When CS interaction dominates 
radioactivity and the shock wave is radiative the total X-ray luminosity 
depends on the CS density, shock radius, $R$, and velocity, $V$, as $L \propto
\rho_{CSM} R^2 V^3$. For a steady, spherically symmetric wind $L
\propto \dot M V^3 \approx $ constant. In the case of an anisotropic or clumpy
CSM the latter relation is not necessarily valid, and the luminosity
depends on the exact density distribution.
Most of the luminosity produced from the shock interaction
is emitted as X rays. A fraction of this energy is emitted
 inwards
towards the SN ejecta and is reprocessed into optical/UV lines
\citep{chevalier94}, while part of the outgoing X-ray flux
may be absorbed by the CSM, and there re-radiated as narrow optical/UV
lines, or escape. 
There is therefore a direct relation 
 between the CS density and both the X-ray flux and the optical/UV flux from the SN,
 and this  applies to the high density CSM  measured in SN~2006jd.

The higher levels of optical emission observed in SN~2006jd over SN~2005ip 
at later phases  (see Figure~\ref{abslcs}) is best
 explained by a higher density of CS material that forms the intermediate-width 
velocity (post-shock gas) component.
Furthermore, the `bump' exhibited by the $r$-band light curve of SN~2006jd is 
more naturally explained with a shell like geometry of dense material, 
as is seen in LBVs like $\eta$ Carinae, and less  aligned with expectations 
from a steady- or clumpy-wind scenario. 

Within this context, the lower X-ray luminosity and gas densities
\citep{smith09} observed in SN~2005ip,
along with the appearance of high-ionization coronal lines observed at the earliest times, suggest that the mass-loss mechanism of
its progenitor was less extreme. As advocated by 
\citet{smith09}, it appears more likely that SN~2005ip exploded while enshrouded in a
relatively spherical, clumpy-wind environment.

\subsection{Location of the warm and cold dust components}
\label{whereisthedust}

Armed with the warm and cold $BB$ 
radii ($R_w$ and $R_c$) computed in Section~\ref{dust}, in combination with velocity estimates of the forward and reverse shocks obtained from measurements made in 
Section~\ref{Halphafwhm}, we endeavor to 
constrain the location of both dust components in SN~2005ip and SN~2006jd.
Plotted in each panel of Figure~\ref{dustposition} 
is the location of $R_w$ and $R_c$. 
Here $R_c$ is given for a range of covering factor (see below), where
values below 4$\pi$ account for deviations from spherical symmetry. 
Also plotted in Figure~\ref{dustposition} is 
the position of the forward shock determined from the BVZI value of the broad 
H$\alpha$ component, and
the location of the reverse shock as dictated by 
the BVZI of 
intermediate-width component, which has been computed by taking
 three sigma of the measured 
FWHM velocity shown in Figure~\ref{fwhm}.
Finally, plotted in Figure~\ref{dustposition} 
are dust evaporation radii, $R_{evap}$, for three grain sizes. 
These curves represent a cavity around the SN 
in which all dust is evaporated by the radiation field 
produced from the initial shock break-out. 
To estimate $R_{evap}$ we make use of the following expression \citep[e.g.][]{fox10}

\begin{equation}
\label{equation4}
 R_{evap} = \left ( \frac{L_{max}<Q>}{16\pi \sigma T^{4}_{evap}} \right )^{1/2} ,
\end{equation}

\noindent which is valid under the assumption that the dust is optically thin. 
Here $L_{max}$ is the maximum bolometric luminosity of the SN, 
$< Q>$ is the ratio of the Planck averaged dust absorption efficiency to the dust emission efficiency, $\sigma$ is the Stefan-Boltzmann constant, and $T_{evap}$ is the evaporation temperature ($\sim 1900$ K for graphite).
The parameter $<Q>$ depends upon the evaporation temperature of 
the dust, the grain size of the dust, and the temperature of the SN. For 
our calculations of 
 $R_{evap}$ we adopted values of $T_{evap}$ $=$ 1900 K and $T_{SN}$ $=$ 7000 K.
 This value of $T_{SN}$ was determined from the $BB_h$ fits presented in Section~\ref{uvoir} (see Table~\ref{table8}).
 Our earliest observations of SNe 2005ip and 2006jd indicate 
$L_{max}$ $\sim$ 3.2$\times$10$^{42}$ erg~s$^{-1}$ or 8.3$\times$10$^{8}$ $L_{\odot}$, however it is probably somewhat higher before 
the discovery dates.
For a peak luminosity of 
10$^{9}$ $L_{\sun}$
 and a graphite composition with grain radii of $a = 1.0~\mu$m, 0.1 $\mu$m and 0.01 $\mu$m, we obtain as shown in Figure~\ref{dustposition}, 
values of $R_{evap}$ of $\sim$ 1.5$\times$10$^{16}$ cm, $\sim$ 3.0$\times$10$^{16}$ cm and $\sim$ 5.0$\times$10$^{16}$ cm, respectively. 
Unfortunately, $R_{evap}$ also depends sensitively on the grain composition,
 for instance in the case of a silicate dust composition, $R_{evap}$ could be a factor of 
 $\sim 5$ higher \citep{dwek85}. Therefore, 1.5$\times$10$^{16}$ cm should be considered
 a minimum radius for dust evaporation. 

Figure~\ref{dustposition} suggests that the warm dust component in 
SN~2005ip  formed well within the CDS during the first several months after the SN explosion. 
However,  as revealed from the right panel of Figure~\ref{dustposition}, 
the warm dust component of SN~2006jd appears to precede the forward shock and $R_{evap}$ for 
grain size of $a = 0.1$ and 1~$\mu$m for the initial 200 days of evolution, 
whereupon the forward shock surpasses the warm dust $R_w$.

This finding raises some perplexing questions. First of all, how does the warm dust 
component in SN~2006jd find itself to be located beyond the forward shock during the 
first $\sim$ 200 days, and how does this dust component survive once it is 
swept up by the SN blast wave? 
On day 9 the warm dust $R_w$ is computed to be $\sim$ 2.5$\times$10$^{16}$ cm, 
while the shock radius only reaches  $\sim$ 1.6$\times$10$^{15}$ cm. 
If a portion of the warm dust actually lies beyond the forward shock, 
this would imply that it is pre-existing and was ejected by the progenitor star 
prior to exploding in a shell at a minimum distance of $\sim$ 2.5$\times$10$^{16}$ cm.
However, at the earliest phases this pre-existing warm dust component could be evaporated 
from the initial SN outburst.

A solution of this problem could be that the CSM of the SN is highly anisotropic, e.g., 
a bipolar structure with a disk like dense equatorial region, as was discussed in
Section~\ref{insight}.
The dust would then be pre-existing in the CSM and heated by the radiation from the SN. In this case the SN would expand roughly spherically until the equatorial disk is encountered. The interaction between the ejecta and disk would then slow down the equatorial region of the ejecta, while the polar region of the ejecta would expand largely unimpeded.
This would explain the larger radius of the dust emitting region compared to the blast wave at early times, 
and the opposite situation at late times. As was discussed in the previous sub-section, this picture 
is consistent with the evolution of the H$\alpha$ line, which changes in both shape and intensity 
at an epoch corresponding to 
$\sim R_w/V_{max} \sim 170 \cdot  (R_w/1.5\times 10^{16} \ {\rm cm}) (V_{max}/10^4 \ {\rm km \ s}^{-1})^{-1}$ days, 
which signals the encounter of the ejecta with CS shell.

An equatorial shell may also explain the evolution of the H$\alpha$  line shape. 
At 22 days the line extends to $\sim 16,000~{\rm km \ s}^{-1}$ on the blue side and 
$\sim 8000~{\rm km \ s}^{-1}$ on the red (see right panel of Figure~\ref{halpha}). 
This asymmetry may be a result of  occultation of the receding side of the ejecta by 
the opaque photosphere. If the photosphere has a radius, $R_p$, and the maximum radius of the ejecta is, 
$R_{max} (= V_{max} t)$, the extent on the red side in the case of homologous expansion will be 
$V_{red} = V_{max} (1 - (R_p/R_{max} )^2)^{1/2}$. 
For the above values we get $R_p/R_{max}  \sim 0.87$ at 22 days.

As long as the radius of the ejecta is smaller that the inner radius of the shell, the H$\alpha$ emission from the receding ejecta will be seen as a red wing. 
However, once the radius of the ejecta becomes larger than the shell, the red wing of the  H$\alpha$  line will only be seen if the optical depth of the dust in the shell is low. 
If this is not the case the extent of the red wing will rapidly decrease, while the blue wing will remain largely unaffected.  
This is consistent with what is seen from 41 days to 124 days 
as illustrated in the right panel of Figure~\ref{fig10}. 
If  any of the  pre-existing dust survives the initial radiation field associated with the SN 
outburst, then the increased asymmetry observed in the H$\alpha$ line profile could be 
 caused by obscuration due to dust absorption. 
 This scenario does not exclude the possibility that dust condensation occurs
 within the CDS at late epochs, but does not require it either. 
 It does, however, solve the problem with the location of the  warm dust relative to the SN blast wave.  At the same time it also explains the early indication of dust emission seen, which is a problematic  to explain solely by  newly formed dust.

The  intermediate H$\alpha$ component is seen to increase in both luminosity and width during the first few hundred days 
(see Figure~\ref{halpha}).  
As mentioned previously, the dramatic increased prominence of the intermediate 
component is  driven by the SN shock wave
moving  into the dense CS shell, 
with a velocity of $V_{disk}  \sim V_{ej} (\rho_{ej}/\rho_{disk})^{1/2} \ll V_{ej} $. 
A reverse shock into the ejecta would then have a velocity of $V_{ej} -V_{disk} \approx V_{ej}$ 
in the observer frame. 
At the time of this interaction we expect to observe an increase in both the luminosity of the intermediate component from the 
shock going into the disk and of the ejecta component from the reverse shock into the ejecta, 
and possibly also  X ray-ionized ejecta in the plane of the shell.

 Alternatively, the last non-detection date of SN 2006jd in KAIT search images 
occurred $\sim 170$ days before discovery, implying that there is a possibility that 
the explosion epoch could be significantly earlier than the discovery date. 
If the SN did occur some $\sim 5$ months prior to discovery then the forward shock radius 
at $\sim 200$ days would be comparable to $R_w$. 
This would then suggest that the warm dust formed in the vicinity of the forward shock. 
However, if the progenitor exploded significantly prior to discovery, this would imply  
that SN 2006jd was substantially brighter than SNe 1988Z and 2005ip. Given the nature of 
the SED at the time of our first observations, we, however, believe the explosion of 
SN~2006jd did not occur long before the time of discovery.

 Turning to the location of the cold, pre-existing dust component, as demonstrated in  
Figure 19, $R_c$ varies significantly as a function of the adopting covering factor (CF). 
For spherical symmetry CF $= 4\pi$, however, in the case of SNe IIn it is not obvious that such 
a case is appropriate. 
In fact, observations suggest that the majority of LBVs are surrounded  
by shells of dense material, and also exhibit enhanced brightness regions that are aligned to a
 preferred axis of symmetry \citep{nota95}. 
For example, the well-studied LBV-candidate HD~167625 displays strong evidence for a pre-existing 
cold ($T_{BB} \sim 130$ K) dust component which is distributed within an equatorially enhanced torus 
inclined $60^o$ with respect to the observer \citep{hutsemekers94,ohara03}. 
The measured geometry of this  shell implies a CF $= 2\pi$.
Given the indications described earlier that the CSM surround SN 2006jd resides in  
a dense equatorial torus, an appropriate CF would place the cold dust well beyond the
  location implied by spherical symmetry, and as indicated in Figure 19, also just beyond the fast forward shock plowing 
through CSM in the polar direction. In the case of SN~2005ip where the CS environment  
appears to be more consistent with a clumpy wind, the cold dust component is also likely  
described by a CF that deviates significantly from spherical symmetry \citep[see][]{fox10}. 
We again stress that a spherically symmetric CF provides lower limits on $R_w$ and $R_c$, 
and our ability to constrain these parameters is dependent on a large number of underlying assumptions.

\section{CONCLUSIONS}
\label{conclusion}

The observational properties of SN~2006jd and SN~2005ip have a number of 
similarities with the 1988Z-like subclass of SN~IIn. However differences are also evident,  most notably between their light curve morphologies, X-ray luminosities,  and the ionization potentials of the coronal lines and the time-scale in which these features appear. 
These differences are most likely traced back to the mass-loss history of the progenitors. With densities on the order of $>$ 10$^{6}$ cm$^{-3}$, and an X-ray luminosity an order of a magnitude higher than in SN~2005ip, the progenitor of SN~2006jd likely experienced a significant mass-loss event during its late-state evolution akin to the great 19th century eruption of $\eta$ Carinae. In the case of SN~2005ip, its CSM was comparably less dense and probably clumpy \citep{smith09}, explaining both the lower X-ray luminosity and the early turn-on of high-ionization coronal lines. 
Within this context one could then expect a continuum of SN~IIn luminosities to trace 
the mass-loss history, with the most luminous objects like SN~2006gy exploding in the midst of, or soon after, an LBV-like outburst \citep[e.g.][]{moriya12}; while the more moderately luminous objects, like SN~2006jd, find themselves exploding in the centuries or eons after suffering an eruptive LBV outburst. Given the differences between SN~2006jd and SN~2005ip, it appears that the mass-loss mechanism of the latter may be more aligned with a 
super-wind or perhaps mass-ejections of a less extreme nature.

We calculate that both SN~2005ip and SN~2006jd are associated with similar amounts of  warm ($\sim$ 10$^{-4}$ M$_{\sun}$) and  cold ($\sim$ 10$^{-2}$ M$_{\sun}$) dust.
Given the uncertainties involved in the derivations, it is encouraging to note that these 
estimates are consistent with the dust masses computed for other SNe~IIn 
\citep[e.g.][]{gerardy02,pozzo04,meikle07,andrews11,gall11}.

\acknowledgments 

This material is based upon work supported by NSF under 
grants AST--0306969, AST--0908886, AST--0607438, and AST--1008343. 
The Oskar Klein Centre is funded by the Swedish Research Council. 
C. F. acknowledges support from the Swedish Research Council and the Swedish 
National Space Board. 
O.D. Fox would like to thank the NASA Postdoctoral Program fellowship for support. 
J. P. Anderson and M. Hamuy acknowledge support by CONICYT through FONDECYT grant 3110142, and the Millennium Center for Supernova Science (P10-064-F).
The Dark Cosmology Centre is funded by the Danish National Research Foundation. 
The Isaac Newton Telescope is operated on the island of La Palma by the Isaac Newton Group in the Spanish Observatorio del Roque de los Muchachos of the Instituto de 
Astrof\'isica de Canarias. Also based on observations made with the Nordic Optical Telescope, operated
on the island of La Palma jointly by Denmark, Finland, Iceland,
Norway, and Sweden, in the Spanish Observatorio del Roque de los
Muchachos of the Instituto de Astrofisica de Canarias.

\clearpage

\begin{deluxetable} {ccccccccccccc}
\rotate
\tabletypesize{\tiny}
\tablenum{1}
\tablecolumns{13}
\tablewidth{0pt}
\tablecaption{Local Sequences of SNe~2005ip and 2006jd in the Standard System\label{table1}}
\tablehead{
\colhead{STAR ID} &
\colhead{R.A.}  &
\colhead{Decl.}  &
\colhead{$u'$}   &
\colhead{$g'$}   &
\colhead{$r'$}   &
\colhead{$i'$}   &
\colhead{$B$}   &
\colhead{$V$}  &
\colhead{$Y$}   &
\colhead{$J$}   &
\colhead{$H$}  &
\colhead{$K_s$}} 
\startdata
\multicolumn{13}{c}{\bf SN 2005ip}\\
01	& 09:32:15.63	& $+$08:24:43.82& $ \cdots $	& $ \cdots $	& $ \cdots $	& $ \cdots $	& $ \cdots $	& $ \cdots $	&  11.890(008)&  11.684(007)	&  11.428(007)	& $ \cdots $	\\
02	& 09:32:13.24	& $+$08:27:15.47&  15.560(033)	& $ \cdots $	& $ \cdots $	& $ \cdots $	&  14.658(007)	&  14.048(014)	&  13.183(008)&  12.956(006)	&  12.654(006)	& $ \cdots $	\\
03	& 09:32:20.99	& $+$08:26:13.89&  17.726(018)	&  15.625(010)	&  14.843(007)	&  14.531(006)	&  16.134(012)	&  15.176(009)	&  13.716(009)&  13.368(008)	&  12.874(007)	& $ \cdots $	\\
04	& 09:32:14.03	& $+$08:26:12.68&  17.276(015)	&  15.550(010)	&  14.921(007)	&  14.708(006)	&  15.999(008)	&  15.179(009)	&  13.996(007)&  13.699(006)	&  13.282(007)	& $ \cdots $	\\
05	& 09:32:16.12	& $+$08:30:07.69&  18.475(036)	&  15.955(011)	&  15.019(007)	&  14.695(007)	&  16.519(011)	&  15.434(006)	&  13.876(010)&  13.519(009)	&  13.003(007)	& $ \cdots $	\\
06	& 09:32:15.71	& $+$08:24:18.79&  17.918(014)	&  16.076(008)	&  15.416(006)	&  15.191(006)	&  16.538(007)	&  15.693(008)	&  14.457(008)&  14.145(007)	&  13.706(007)	& $ \cdots $	\\
07	& 09:32:24.28	& $+$08:30:15.54&  18.271(034)	&  16.445(009)	&  15.742(007)	&  15.460(007)	&  16.913(014)	&  16.044(007)	&  14.704(014)&  14.388(012)	&  13.887(011)	& $ \cdots $	\\
08	& 09:32:16.41	& $+$08:24:38.99&  17.864(025)	&  16.663(012)	&  16.234(008)	&  16.092(006)	&  17.002(007)	&  16.396(007)	&  15.489(009)&  15.242(008)	&  14.926(012)	& $ \cdots $	\\
09	& 09:32:20.49	& $+$08:26:11.93&  19.223(038)	&  17.095(011)	&  16.326(006)	&  16.055(006)	&  17.630(008)	&  16.652(009)	& $ \cdots $	& $ \cdots $	& $ \cdots $	& $ \cdots $	\\
10	& 09:32:16.60	& $+$08:29:19.12&  18.394(029)	&  16.964(010)	&  16.462(006)	&  16.294(006)	&  17.344(007)	&  16.650(006)	&  15.664(011)&  15.390(010)	&  15.059(016)	& $ \cdots $	\\
11	& 09:31:57.50	& $+$08:26:07.65&  19.668(101)	&  17.409(016)	&  16.462(006)	&  16.096(006)	&  17.952(009)	&  16.890(006)	& $ \cdots $	& $ \cdots $	& $ \cdots $	& $ \cdots $	\\
12	& 09:31:54.46	& $+$08:28:01.39&  17.907(024)	&  16.917(009)	&  16.593(008)	&  16.465(008)	&  17.204(007)	&  16.715(006)	& $ \cdots $	& $ \cdots $	& $ \cdots $	& $ \cdots $	\\
13	& 09:31:57.32	& $+$08:26:57.25&  18.633(024)	&  17.190(011)	&  16.642(008)	&  16.426(006)	&  17.601(008)	&  16.858(008)	& $ \cdots $	& $ \cdots $	& $ \cdots $	& $ \cdots $	\\
14	& 09:32:23.92	& $+$08:30:04.83&  19.141(047)	&  17.428(009)	&  16.807(007)	&  16.605(007)	&  17.872(012)	&  17.063(016)	& $ \cdots $	& $ \cdots $	& $ \cdots $	& $ \cdots $	\\
15	& 09:32:17.64	& $+$08:25:39.73&  18.914(029)	&  17.645(011)	&  17.159(010)	&  16.989(007)	&  18.004(009)	&  17.338(007)	&  16.352(013)&  16.075(013)	&  15.705(018)	& $ \cdots $	\\
16	& 09:32:14.88	& $+$08:26:35.63& $ \cdots $	&  18.832(019)	&  17.453(012)	&  16.601(010)	&  19.546(030)	&  18.054(016)	&  15.396(010)&  14.955(006)	&  14.331(007)	& $ \cdots $	\\
17	& 09:32:20.10	& $+$08:25:58.43& $ \cdots $	&  18.714(015)	&  17.767(008)	&  17.394(007)	&  19.276(024)	&  18.175(009)	&  16.518(010)&  16.150(013)	&  15.569(013)	& $ \cdots $	\\
18	& 09:32:03.16	& $+$08:25:19.58&  20.256(208)	&  18.493(008)	&  17.803(007)	&  17.555(007)	&  18.936(025)	&  18.080(008)	&  16.801(014)&  16.469(014)	&  16.017(026)	& $ \cdots $	\\
19	& 09:32:07.43	& $+$08:30:15.63& $ \cdots $	&  19.397(018)	&  18.018(010)	&  16.644(011)	&  20.265(055)	&  18.596(014)	&  15.132(012)&  14.629(008)	&  14.044(009)	& $ \cdots $	\\
20	& 09:32:12.68	& $+$08:29:08.69& $ \cdots $	&  19.550(020)	&  18.313(008)	&  17.788(008)	&  20.170(062)	&  18.889(018)	&  16.808(019)&  16.375(022)	&  15.692(017)	& $ \cdots $	\\
21	& 09:32:17.38	& $+$08:24:07.50& $ \cdots $	&  18.873(015)	&  18.437(012)	&  18.297(014)	&  19.147(032)	&  18.606(012)	& $ \cdots $	& $ \cdots $	& $ \cdots $	& $ \cdots $	\\
22	& 09:31:59.77	& $+$08:24:46.23&  20.242(099)	&  18.989(011)	&  18.402(014)	&  18.001(011)	&  19.381(025)	&  18.599(016)	& $ \cdots $	& $ \cdots $	& $ \cdots $	& $ \cdots $	\\
23	& 09:31:57.95	& $+$08:28:48.65&  20.104(157)	&  19.326(019)	&  18.848(018)	&  18.668(035)	&  19.670(033)	&  19.049(018)	& $ \cdots $	& $ \cdots $	& $ \cdots $	& $ \cdots $	\\
24	& 09:31:57.94	& $+$08:29:35.05& $ \cdots $	&  20.545(055)	&  19.156(021)	&  18.329(012)	&  21.531(186)	&  19.796(039)	& $ \cdots $	& $ \cdots $	& $ \cdots $	& $ \cdots $	\\
25	& 09:32:10.29	& $+$08:28:28.85& $ \cdots $	&  20.651(060)	&  19.333(021)	&  18.588(024)	&  21.290(246)	&  19.983(043)	& $ \cdots $	& $ \cdots $	& $ \cdots $	& $ \cdots $	\\
26	& 09:32:17.25	& $+$08:22:57.90& $ \cdots $	& $ \cdots $	& $ \cdots $	& $ \cdots $	& $ \cdots $	& $ \cdots $	&  12.277(012)&  11.966(010)	&  11.536(009)	& $ \cdots $	\\
27	& 09:32:19.74	& $+$08:22:25.79& $ \cdots $	& $ \cdots $	& $ \cdots $	& $ \cdots $	& $ \cdots $	& $ \cdots $	&  14.229(014)&  13.923(011)	&  13.500(010)	& $ \cdots $	\\
28	& 09:32:23.51	& $+$08:22:31.37& $ \cdots $	& $ \cdots $	& $ \cdots $	& $ \cdots $	& $ \cdots $	& $ \cdots $	&  15.265(012)&  14.833(015)	&  14.221(011)	& $ \cdots $	\\
29	& 09:32:07.26	& $+$08:22:24.42& $ \cdots $	& $ \cdots $	& $ \cdots $	& $ \cdots $	& $ \cdots $	& $ \cdots $	&  15.405(010)&  14.960(010)	&  14.461(013)	& $ \cdots $	\\
30	& 09:32:26.86	& $+$08:27:18.68& $ \cdots $	& $ \cdots $	& $ \cdots $	& $ \cdots $	& $ \cdots $	& $ \cdots $	&  15.486(010)&  15.085(009)	&  14.448(009)	& $ \cdots $	\\
31	& 09:32:20.35	& $+$08:21:14.18& $ \cdots $	& $ \cdots $	& $ \cdots $	& $ \cdots $	& $ \cdots $	& $ \cdots $	&  15.492(012)&  15.213(013)	&  14.860(014)	& $ \cdots $	\\
32	& 09:32:29.01	& $+$08:22:24.85& $ \cdots $	& $ \cdots $	& $ \cdots $	& $ \cdots $	& $ \cdots $	& $ \cdots $	&  15.809(011)&  15.465(014)	&  15.075(011)	& $ \cdots $	\\
33	& 09:32:18.96	& $+$08:26:16.51& $ \cdots $	& $ \cdots $	& $ \cdots $	& $ \cdots $	& $ \cdots $	& $ \cdots $	&  16.385(010)&  15.809(010)	&  15.322(009)	& $ \cdots $	\\
34	& 09:32:32.65	& $+$08:20:48.73& $ \cdots $	& $ \cdots $	& $ \cdots $	& $ \cdots $	& $ \cdots $	& $ \cdots $	&  16.370(016)&  16.077(014)	&  15.765(025)	& $ \cdots $	\\
35	& 09:32:32.86	& $+$08:25:49.37& $ \cdots $	& $ \cdots $	& $ \cdots $	& $ \cdots $	& $ \cdots $	& $ \cdots $	&  16.632(014)&  16.097(013)	&  15.593(012)	& $ \cdots $	\\
36	& 09:32:21.46	& $+$08:26:46.25& $ \cdots $	& $ \cdots $	& $ \cdots $	& $ \cdots $	& $ \cdots $	& $ \cdots $	&  17.111(020)&  16.589(017)	&  15.988(017)	& $ \cdots $	\\
37	& 09:32:03.40	& $+$08:24:08.96& $ \cdots $	& $ \cdots $	& $ \cdots $	& $ \cdots $	& $ \cdots $	& $ \cdots $	&  17.503(025)&  16.996(021)	&  16.415(028)	& $ \cdots $	\\
38	& 09:32:16.02	& $+$08:22:14.84& $ \cdots $	& $ \cdots $	& $ \cdots $	& $ \cdots $	& $ \cdots $	& $ \cdots $	&  17.485(027)&  17.254(039)	&  16.867(056)	& $ \cdots $	\\
39	& 09:32:04.01	& $+$08:26:45.42& $ \cdots $	& $ \cdots $	& $ \cdots $	& $ \cdots $	& $ \cdots $	& $ \cdots $	&  17.781(035)&  17.359(037)	&  16.956(055)	& $ \cdots $	\\
40	& 09:32:31.39	& $+$08:30:59.44& $ \cdots $	& $ \cdots $	& $ \cdots $	& $ \cdots $	& $ \cdots $	& $ \cdots $	&  16.086(020)&  15.729(009)	&  15.162(024)	& $ \cdots $	\\
41	& 09:32:11.66	& $+$08:24:34.09& $ \cdots $	& $ \cdots $	& $ \cdots $	& $ \cdots $	& $ \cdots $	& $ \cdots $	&  16.369(009)&  15.850(010)	&  15.168(009)	& $ \cdots $	\\
42	& 09:32:05.43	& $+$08:21:30.85& $ \cdots $	& $ \cdots $	& $ \cdots $	& $ \cdots $	& $ \cdots $	& $ \cdots $	&  16.601(014)&  16.084(018)	&  15.394(021)	& $ \cdots $	\\
\multicolumn{13}{c}{\bf SN 2006jd}\\
01	& 08:02:19.48	& $+$00:48:13.64	& $ \cdots $	& $ \cdots $	& $ \cdots $	& $ \cdots $	& $ \cdots $	& $ \cdots $	&  12.037(004)	&  11.842(006)	&  11.595(005)	& $ \cdots $	\\
02	& 08:02:16.88	& $+$00:46:15.74	&  15.796(016)	& $ \cdots $	& $ \cdots $	& $ \cdots $	& $ \cdots $	& $ \cdots $	& $ \cdots $	& $ \cdots $	& $ \cdots $	& $ \cdots $	\\
03	& 08:02:14.35	& $+$00:47:14.17	&  16.288(009)	& $ \cdots $	& $ \cdots $	& $ \cdots $	&  15.227(008)	& $ \cdots $	& $ \cdots $	& $ \cdots $	& $ \cdots $	& $ \cdots $	\\
04	& 08:02:04.82	& $+$00:46:44.54	&  15.806(016)	& $ \cdots $	& $ \cdots $	& $ \cdots $	&  14.802(011)	& $ \cdots $	& $ \cdots $	& $ \cdots $	& $ \cdots $	& $ \cdots $	\\
05	& 08:02:04.61	& $+$00:50:32.68	& $ \cdots $	& $ \cdots $	& $ \cdots $	& $ \cdots $	& $ \cdots $	& $ \cdots $	&  11.513(006)	&  11.147(010)	&  10.656(010)	& $ \cdots $	\\
06	& 08:02:04.17	& $+$00:45:01.01	&  18.042(017)	&  15.829(014)	&  14.980(013)	&  14.694(011)	&  16.345(005)	&  15.340(010)	& $ \cdots $	& $ \cdots $	& $ \cdots $	& $ \cdots $	\\
07	& 08:01:54.05	& $+$00:46:03.90	&  16.605(009)	&  15.524(015)	&  15.199(011)	&  15.095(012)	&  15.798(005)	&  15.309(006)	& $ \cdots $	& $ \cdots $	& $ \cdots $	& $ \cdots $	\\
08	& 08:02:20.33	& $+$00:45:01.33	&  17.113(010)	&  15.853(012)	&  15.382(006)	&  15.245(012)	&  16.218(005)	&  15.558(005)	& $ \cdots $	& $ \cdots $	& $ \cdots $	& $ \cdots $	\\
09	& 08:01:56.97	& $+$00:47:22.42	&  17.067(010)	&  15.824(011)	&  15.404(008)	&  15.272(007)	&  16.172(005)	&  15.559(005)	& $ \cdots $	& $ \cdots $	& $ \cdots $	& $ \cdots $	\\
10	& 08:01:57.06	& $+$00:49:16.39	&  17.567(017)	&  15.952(009)	&  15.372(008)	&  15.176(008)	&  16.384(006)	&  15.596(006)	& $ \cdots $	& $ \cdots $	& $ \cdots $	& $ \cdots $	\\
11	& 08:02:16.35	& $+$00:51:28.98	&  17.038(013)	&  15.849(009)	&  15.429(007)	&  15.279(007)	&  16.197(005)	&  15.581(005)	& $ \cdots $	& $ \cdots $	& $ \cdots $	& $ \cdots $	\\
12	& 08:02:13.17	& $+$00:47:46.72	&  18.961(037)	&  16.495(011)	&  15.489(009)	&  15.134(008)	&  17.078(005)	&  15.940(006)	&  14.275(005)	&  13.903(006)	&  13.365(004)	& $ \cdots $	\\
13	& 08:02:14.43	& $+$00:48:21.49	&  17.764(014)	&  16.146(010)	&  15.516(006)	&  15.278(011)	&  16.594(005)	&  15.764(005)	& $ \cdots $	& $ \cdots $	& $ \cdots $	& $ \cdots $	\\
14	& 08:02:00.31	& $+$00:47:41.50	&  17.071(011)	&  15.900(009)	&  15.520(006)	&  15.402(006)	&  16.222(005)	&  15.656(005)	& $ \cdots $	& $ \cdots $	& $ \cdots $	& $ \cdots $	\\
15	& 08:02:08.48	& $+$00:50:00.31	&  17.098(015)	&  16.003(008)	&  15.678(006)	&  15.584(008)	&  16.301(005)	&  15.794(004)	& $ \cdots $	& $ \cdots $	& $ \cdots $	& $ \cdots $	\\
16	& 08:01:59.72	& $+$00:44:21.48	&  18.363(087)	&  16.440(013)	&  15.721(007)	&  15.461(007)	&  16.942(007)	&  16.000(006)	& $ \cdots $	& $ \cdots $	& $ \cdots $	& $ \cdots $	\\
17	& 08:01:52.81	& $+$00:51:27.65	&  17.090(010)	&  16.019(006)	&  15.739(005)	&  15.648(005)	&  16.288(005)	&  15.829(005)	& $ \cdots $	& $ \cdots $	& $ \cdots $	& $ \cdots $	\\
18	& 08:02:19.20	& $+$00:50:48.30	&  17.744(017)	&  16.334(006)	&  15.834(005)	&  15.666(005)	&  16.731(005)	&  16.027(004)	& $ \cdots $	& $ \cdots $	& $ \cdots $	& $ \cdots $	\\
19	& 08:01:55.16	& $+$00:44:40.06	&  17.514(023)	&  16.346(006)	&  15.934(005)	&  15.794(005)	&  16.678(005)	&  16.087(005)	& $ \cdots $	& $ \cdots $	& $ \cdots $	& $ \cdots $	\\
20	& 08:02:12.15	& $+$00:48:40.00	&  17.767(015)	&  16.406(008)	&  15.875(005)	&  15.675(006)	&  16.809(005)	&  16.079(004)	&  14.979(004)	&  14.714(007)	&  14.350(005)	& $ \cdots $	\\
21	& 08:02:10.86	& $+$00:49:11.78	&  18.235(039)	&  16.872(006)	&  16.303(005)	&  16.074(005)	&  17.280(005)	&  16.531(004)	&  15.369(005)	&  15.083(007)	&  14.687(005)	& $ \cdots $	\\
22	& 08:01:53.31	& $+$00:47:23.35	&  18.221(150)	&  16.916(008)	&  16.360(005)	&  16.156(005)	&  17.316(005)	&  16.576(004)	& $ \cdots $	& $ \cdots $	& $ \cdots $	& $ \cdots $	\\
23	& 08:02:02.96	& $+$00:46:22.33	&  17.935(029)	&  16.808(008)	&  16.411(005)	&  16.268(005)	&  17.121(005)	&  16.556(004)	& $ \cdots $	& $ \cdots $	& $ \cdots $	& $ \cdots $	\\
24	& 08:02:12.92	& $+$00:46:00.44	&  18.478(025)	&  17.313(006)	&  16.852(005)	&  16.689(005)	&  17.678(006)	&  17.030(004)	& $ \cdots $	& $ \cdots $	& $ \cdots $	& $ \cdots $	\\
25	& 08:01:59.37	& $+$00:48:54.47	&  18.519(045)	&  17.398(006)	&  16.934(005)	&  16.754(006)	&  17.747(006)	&  17.114(005)	& $ \cdots $	& $ \cdots $	& $ \cdots $	& $ \cdots $	\\
26	& 08:01:58.49	& $+$00:48:21.20	&  19.670(070)	&  17.903(007)	&  17.173(006)	&  16.897(005)	&  18.395(006)	&  17.493(005)	& $ \cdots $	& $ \cdots $	& $ \cdots $	& $ \cdots $	\\
27	& 08:02:29.36	& $+$00:45:35.32	& $ \cdots $	& $ \cdots $	& $ \cdots $	& $ \cdots $	& $ \cdots $	& $ \cdots $	&  13.363(006)	&  13.079(009)	&  12.736(009)	& $ \cdots $	\\
28	& 08:02:26.29	& $+$00:46:11.42	& $ \cdots $	& $ \cdots $	& $ \cdots $	& $ \cdots $	& $ \cdots $	& $ \cdots $	&  14.052(004)	&  13.721(006)	&  13.248(005)	& $ \cdots $	\\
29	& 08:02:24.21	& $+$00:49:14.59	& $ \cdots $	& $ \cdots $	& $ \cdots $	& $ \cdots $	& $ \cdots $	& $ \cdots $	&  14.313(005)	&  14.111(006)	&  13.845(006)	& $ \cdots $	\\
30	& 08:02:28.22	& $+$00:48:16.92	& $ \cdots $	& $ \cdots $	& $ \cdots $	& $ \cdots $	& $ \cdots $	& $ \cdots $	&  14.683(005)	&  14.304(007)	&  13.753(006)	& $ \cdots $	\\
31	& 08:02:16.12	& $+$00:46:29.78	& $ \cdots $	& $ \cdots $	& $ \cdots $	& $ \cdots $	& $ \cdots $	& $ \cdots $	&  15.224(006)	&  14.880(006)	&  14.396(006)	& $ \cdots $	\\
32	& 08:02:02.04	& $+$00:48:41.18	& $ \cdots $	& $ \cdots $	& $ \cdots $	& $ \cdots $	& $ \cdots $	& $ \cdots $	&  15.811(006)	&  15.414(007)	&  14.801(005)	& $ \cdots $	\\
33	& 08:02:01.87	& $+$00:48:09.72	& $ \cdots $	& $ \cdots $	& $ \cdots $	& $ \cdots $	& $ \cdots $	& $ \cdots $	&  16.285(007)	&  16.018(008)	&  15.649(010)	& $ \cdots $	\\
34	& 08:02:22.32	& $+$00:48:52.06	& $ \cdots $	& $ \cdots $	& $ \cdots $	& $ \cdots $	& $ \cdots $	& $ \cdots $	&  16.339(007)	&  16.009(008)	&  15.507(007)	& $ \cdots $	\\
35	& 08:02:04.97	& $+$00:47:21.19	& $ \cdots $	& $ \cdots $	& $ \cdots $	& $ \cdots $	& $ \cdots $	& $ \cdots $	&  16.976(007)	&  16.654(011)	&  16.254(016)	& $ \cdots $	\\
36	& 08:02:07.43	& $+$00:45:39.02	& $ \cdots $	& $ \cdots $	& $ \cdots $	& $ \cdots $	& $ \cdots $	& $ \cdots $	&  17.032(011)	&  16.598(010)	&  16.029(014)	& $ \cdots $	\\
37	& 08:02:14.59	& $+$00:49:38.28	& $ \cdots $	& $ \cdots $	& $ \cdots $	& $ \cdots $	& $ \cdots $	& $ \cdots $	&  17.147(012)	&  16.701(019)	&  16.094(011)	& $ \cdots $	\\
38	& 08:02:17.31	& $+$00:49:58.44	& $ \cdots $	& $ \cdots $	& $ \cdots $	& $ \cdots $	& $ \cdots $	& $ \cdots $	&  17.278(011)	&  17.052(017)	&  16.769(020)	& $ \cdots $	\\
39	& 08:02:16.27	& $+$00:48:33.66	& $ \cdots $	& $ \cdots $	& $ \cdots $	& $ \cdots $	& $ \cdots $	& $ \cdots $	&  17.236(013)	&  16.871(013)	&  16.330(013)	& $ \cdots $	\\
40	& 08:02:02.16	& $+$00:46:44.98	& $ \cdots $	& $ \cdots $	& $ \cdots $	& $ \cdots $	& $ \cdots $	& $ \cdots $	&  17.207(012)	&  17.026(014)	&  16.759(025)	& $ \cdots $	\\
41	& 08:02:14.48	& $+$00:45:47.88	& $ \cdots $	& $ \cdots $	& $ \cdots $	& $ \cdots $	& $ \cdots $	& $ \cdots $	&  17.567(021)	&  17.113(017)	&  16.590(020)	& $ \cdots $	\\
42	& 08:02:17.91	& $+$00:45:51.66	& $ \cdots $	& $ \cdots $	& $ \cdots $	& $ \cdots $	& $ \cdots $	& $ \cdots $	&  17.752(014)	&  17.334(016)	&  16.841(026)	& $ \cdots $	\\
43	& 08:02:05.77	& $+$00:48:48.10	& $ \cdots $	& $ \cdots $	& $ \cdots $	& $ \cdots $	& $ \cdots $	& $ \cdots $	&  15.547(007)	&  15.208(006)	&  14.745(004)	&  14.557(020)	\\
44	& 08:02:08.05	& $+$00:48:06.34	& $ \cdots $	& $ \cdots $	& $ \cdots $	& $ \cdots $	& $ \cdots $	& $ \cdots $	&  15.394(006)	&  14.990(006)	&  14.368(006)	&  14.158(020)	\\
45	& 08:02:04.41	& $+$00:47:42.32	& $ \cdots $	& $ \cdots $	& $ \cdots $	& $ \cdots $	& $ \cdots $	& $ \cdots $	&  15.853(006)	&  15.668(008)	&  15.434(007)	& $ \cdots $	\\
46	& 08:02:05.73	& $+$00:48:03.60	& $ \cdots $	& $ \cdots $	& $ \cdots $	& $ \cdots $	& $ \cdots $	& $ \cdots $	&  18.069(015)	&  17.626(014)	&  16.960(026)	&  16.571(046)	\\
47	& 08:02:10.65	& $+$00:48:47.23	& $ \cdots $	& $ \cdots $	& $ \cdots $	& $ \cdots $	& $ \cdots $	& $ \cdots $	&  15.214(006)	&  14.976(006)	&  14.673(005)	&  14.558(020)	\\
48	& 08:02:22.43	& $+$00:47:06.40	& $ \cdots $	& $ \cdots $	& $ \cdots $	& $ \cdots $	& $ \cdots $	& $ \cdots $	&  11.684(006)	&  11.318(008)	&  10.810(007)	& $ \cdots $	\\
49	& 08:02:22.20	& $+$00:46:29.10	& $ \cdots $	& $ \cdots $	& $ \cdots $	& $ \cdots $	& $ \cdots $	& $ \cdots $	&  14.052(005)	&  13.829(006)	&  13.548(004)	& $ \cdots $	\\
50	& 08:02:22.15	& $+$00:45:49.64	& $ \cdots $	& $ \cdots $	& $ \cdots $	& $ \cdots $	& $ \cdots $	& $ \cdots $	&  15.058(004)	&  14.644(007)	&  14.045(005)	& $ \cdots $	\\
51	& 08:02:26.44	& $+$00:47:09.20	& $ \cdots $	& $ \cdots $	& $ \cdots $	& $ \cdots $	& $ \cdots $	& $ \cdots $	&  16.684(009)	&  16.433(009)	&  16.065(013)	& $ \cdots $	\\
\enddata
\tablecomments{One sigma uncertainties given in parentheses in thousandths of a
  magnitude correspond to an rms of the magnitudes obtained on
  photometric nights.}
\end{deluxetable}

\clearpage
\begin{deluxetable}{cccccccc}
\setlength{\tabcolsep}{0.02in}
\tabletypesize{\scriptsize}
\tablecolumns{8}
\tablenum{2}
\tablewidth{0pt}
\tablecaption{Optical Photometry of SN~2005ip in the Natural System\label{table2}}
\tablehead{
\colhead{JD--2450000+} &
\colhead{$u$ }  &
\colhead{$g$ }   &
\colhead{$r$ }   &
\colhead{$i$ }   &
\colhead{$B$ }   &
\colhead{$V$ } &
\colhead{Telescope\tablenotemark{a}} } 
\startdata
3683.86  &  15.925(016)  & 15.375(012)  & 15.044(018)  & 15.034(018)  & 15.613(011)  & 15.215(013) & SWO \\ 
3684.86  &  16.029(017)  & 15.441(009)  & 15.087(011)  & 15.051(012)  & 15.700(007)  & 15.262(010) & SWO \\
3687.86  &  16.240(018)  & 15.603(009)  & 15.172(010)  & 15.130(010)  & 15.864(009)  & 15.378(010) & SWO \\
3689.86  &  16.403(014)  & 15.731(010)  & 15.218(015)  & 15.151(017)  & 16.000(009)  & 15.503(009) & SWO \\
3690.86  &  16.427(018)  & 15.815(007)  & 15.288(009)  & 15.228(011)  & 16.076(007)  & 15.564(007) & SWO \\
3694.86  &  16.655(014)  & 16.039(009)  & 15.426(015)  & 15.407(014)  & 16.295(009)  & 15.766(009) & SWO \\
3699.82  &  16.885(012)  & 16.207(012)  & 15.623(013)  & 15.611(015)  & 16.487(010)  & 15.981(011) & SWO \\
3702.81  &  16.940(012)  & 16.286(013)  & 15.693(020)  & 15.699(019)  & 16.564(010)  & 16.064(016) & SWO \\
3706.81  &  16.933(011)  & 16.252(033)  & 15.599(044)  & 15.630(049)  & 16.559(019)  & 16.074(034) & SWO \\
3713.84  &  17.026(011)  & 16.424(017)  & 15.873(019)  & 15.932(026)  & 16.703(011)  & 16.290(016) & SWO \\
3716.83  &  17.104(011)  & 16.489(020)  & 15.910(024)  & 16.015(027)  & 16.729(014)  & 16.352(018) & SWO \\
3721.78  &  17.143(019)  & 16.617(010)  & 16.024(017)  & 16.133(023)  & 16.822(012)  & 16.451(012) & SWO \\
3726.83  &  17.155(014)  & 16.593(014)  & 16.054(023)  & 16.179(028)  & 16.843(010)  & 16.459(018) & SWO \\
3728.83  &  17.198(013)  & 16.604(016)  & 16.069(021)  & 16.210(032)  & 16.855(011)  & 16.487(019) & SWO \\
3737.79  &  17.343(012)  & 16.585(054)  & 15.973(069)  & 16.174(077)  & 16.911(029)  & 16.516(047) & SWO \\
3739.81  &  17.368(017)  & 16.650(035)  & 16.044(050)  & 16.233(060)  & 16.949(023)  & 16.545(041) & SWO \\
3741.77  &  17.447(017)  & 16.762(015)  & 16.236(017)  & 16.419(023)  & 16.988(011)  & 16.625(014) & SWO \\
3746.78  &  17.564(014)  & 16.908(013)  & 16.345(020)  & 16.545(031)  & 17.145(011)  & 16.765(017) & SWO \\
3751.77  &  17.645(030)  & 17.017(015)  & 16.528(010)  & 16.718(018)  & 17.297(019)  & 16.956(018) & SWO \\
3755.76  &  17.769(033)  & 17.113(016)  & 16.521(025)  & 16.747(045)  & 17.338(021)  & 17.007(024) & SWO \\
3761.72  &  17.899(018)  & 17.180(025)  & 16.589(027)  & 16.864(035)  & 17.433(012)  & 17.084(026) & SWO \\
3762.71  &  17.938(018)  & 17.219(030)  & 16.611(034)  & 16.932(034)  & 17.483(013)  & 17.101(023) & SWO \\
3763.70  &  17.963(022)  & 17.226(022)  & 16.605(028)  & 16.912(035)  & 17.513(016)  & 17.140(023) & SWO \\
3764.72  &  17.992(019)  & 17.259(025)  & 16.646(031)  & 16.937(040)  & 17.519(019)  & 17.166(027) & SWO \\
3765.70  &  18.014(020)  & 17.280(029)  & 16.674(033)  & 16.972(040)  & 17.568(020)  & 17.209(027) & SWO \\
3768.72  &  18.053(018)  & 17.235(048)  & 16.611(056)  & 16.851(068)  & 17.559(031)  & 17.158(047) & SWO \\
3771.69  &  18.076(020)  & 17.288(047)  & 16.633(051)  & 16.920(071)  & 17.581(031)  & 17.209(047) & SWO \\
3772.69  &  18.170(021)  & 17.423(028)  & 16.798(037)  & 17.124(042)  & 17.675(019)  & 17.325(028) & SWO \\
3774.69  &  18.191(028)  & 17.480(019)  & 16.835(028)  & 17.143(037)  & 17.760(016)  & 17.425(023) & SWO \\
3778.68  &  18.284(049)  & 17.588(020)  & 17.008(016)  & 17.283(031)  & 17.892(027)  & 17.579(022) & SWO \\
3783.61  &  18.427(034)  & 17.701(016)  & 17.031(024)  & 17.352(040)  & 17.934(020)  & 17.599(024) & SWO \\
3786.69  &  18.419(023)  & 17.682(030)  & 16.984(040)  & 17.379(049)  & 17.959(019)  & 17.578(034) & SWO \\
3788.67  &  18.428(022)  & 17.670(036)  & 17.001(041)  & 17.401(048)  & 17.971(019)  & 17.607(039) & SWO \\
3790.66  &  18.542(027)  & 17.664(033)  & 17.050(040)  & 17.428(051)  & 18.023(022)  & 17.644(036) & SWO \\
3795.67  &  18.493(026)  & 17.637(055)  & 16.956(061)  & 17.333(088)  & 17.929(040)  & 17.555(060) & SWO \\
3799.63  &  18.451(035)  & 17.524(052)  & 16.849(053)  & 17.178(077)  & 17.871(041)  & 17.447(056) & SWO \\
3801.62  &  18.650(026)  & 17.890(035)  & 17.183(040)  & 17.622(053)  & 18.151(024)  & 17.804(039) & SWO \\
3809.62  &  18.741(083)  & 18.176(033)  & 17.383(019)  & 17.805(043)  & 18.441(045)  & 18.017(035) & SWO \\
3815.62  &  $ \cdots $   & 18.098(042)  & 17.308(049)  & 17.825(069)  & 18.415(024)  & 18.066(042) & SWO \\
3819.59  &  18.906(024)  & 18.071(053)  & 17.294(062)  & 17.829(074)  & 18.370(035)  & 18.095(046) & SWO \\
3823.59  &  18.920(031)  & 18.181(048)  & 17.405(047)  & 17.901(070)  & 18.492(032)  & 18.124(049) & SWO \\
3831.57  &  19.099(040)  & 18.330(036)  & 17.465(045)  & 18.054(070)  & 18.617(031)  & 18.248(041) & SWO \\
3838.58  &  19.137(094)  & 18.548(029)  & 17.540(033)  & 18.135(060)  & 18.717(051)  & 18.343(043) & SWO \\
3840.53  &  19.139(061)  & 18.477(034)  & 17.549(037)  & 18.108(064)  & 18.793(034)  & 18.437(039) & SWO \\
3846.56  &  19.125(030)  & 18.289(054)  & 17.515(042)  & 18.009(075)  & 18.625(042)  & 18.247(047) & SWO \\
3858.49  &  19.186(041)  & 18.392(084)  & 17.415(084)  & 18.004(118)  & 18.686(055)  & 18.291(081) & SWO \\
3867.53  &  19.355(101)  & 18.670(039)  & 17.589(037)  & 18.217(070)  & 19.066(059)  & 18.601(043) & SWO \\
3872.47  &  19.356(036)  & 18.600(066)  & 17.545(064)  & 18.251(081)  & 18.900(043)  & 18.529(065) & SWO \\
3890.50  &  19.208(065)  & 18.435(069)  & 17.347(060)  & 17.960(111)  & 18.788(057)  & 18.333(080) & SWO \\
4531.66  &  19.412(400)  & 18.829(400)  & 17.091(400)  & 18.635(400)  & 19.306(400)  & 18.705(400) & DUP \\
5202.75  &  20.211(400)  & 19.636(400)  & 18.011(400)  & 19.057(400)  & 19.942(400)  & 19.475(400) & DUP \\
5244.67  &  20.222(400)  & 19.373(400)  & 17.543(400)  & 19.002(400)  & 19.828(400)  & 19.375(400) & DUP \\
\hline
\enddata
\tablenotetext{a}{DUP and SWO correspond to the du Pont
 and Swope telescopes, respectively.}
\tablecomments{One sigma uncertainties given in parentheses are in millimag.}
\end{deluxetable}

\clearpage
\begin{deluxetable}{ccccc}
\tabletypesize{\scriptsize}
\tablecolumns{5}
\tablenum{3}
\tablewidth{0pt}
\tablecaption{Near-infrared Photometry of SN~2005ip\label{table3}}
\tablehead{
\colhead{JD--2450000+}  &
\colhead{$Y$}   &
\colhead{$J$}   &
\colhead{$H$}   &
\colhead{Telescope\tablenotemark{a}}   }
\startdata
3685.85  & 14.579(028) & 14.407(023) & 14.179(021) & SWO \\ 
3692.87  & 14.657(033) & 14.546(028) & 14.236(028) & SWO \\ 
3700.86  & 14.926(025) & 14.820(026) & 14.491(028) & SWO \\
3704.84  & 15.007(044) & 14.951(041) & 14.585(041) & SWO \\ 
3709.80  & 15.102(029) & 15.105(031) & 14.701(031) & SWO \\
3714.82  & 15.128(040) & 15.152(037) & 14.761(038) & SWO \\
3717.84  & 15.207(035) & 15.216(040) & 14.765(038) & SWO \\
3731.86  & 15.314(047) & 15.371(048) & 14.826(041) & SWO \\
3756.77  & 15.601(013) & 15.653(015) & 14.816(013) & SWO \\
3773.77  & 15.790(018) & 15.705(016) & 14.690(010) & DUP \\
3782.72  & 15.858(059) & 15.511(053) & 14.602(032) & SWO \\
3786.68  & 15.930(022) & 15.744(021) & 14.582(017) & DUP \\
3796.70  & 16.012(063) & 15.631(057) & $ \cdots $  & SWO \\
3803.62  & 16.077(056) & 15.660(057) & 14.664(041) & SWO \\
3810.68  & 16.187(067) & 15.787(062) & 14.731(034) & SWO \\
3814.66  & 16.222(064) & 15.782(061) & $ \cdots $  & SWO \\
3817.63  & 16.211(060) & 15.787(060) & 14.748(034) & SWO \\
3826.57  & $ \cdots $  & $ \cdots $  & 14.784(036) & SWO \\
3833.60  & 16.361(022) & 16.101(023) & 14.791(016) & DUP \\
3839.56  & 16.325(061) & 15.808(069) & 14.720(039) & SWO \\
3845.56  & 16.405(025) & 16.135(023) & 14.777(018) & SWO \\
3848.55  & 16.137(064) & $ \cdots $  & 14.628(040) & SWO \\
3860.52  & 16.288(058) & 15.780(066) & 14.676(037) & SWO \\
3868.52  & 16.326(077) & 15.823(067) & 14.724(035) & SWO \\
4466.78  & 16.562(028) & 17.059(047) & 15.657(024) & DUP \\
5199.77  & 17.694(400) & $ \cdots $  & $ \cdots $  & DUP \\
5261.57  & 17.224(400) & 17.831(400) & 16.453(400) & DUP \\
5523.84  & 17.498(400) & 17.965(400) & 16.924(400) & DUP \\
5613.61  & 17.695(400) & 18.098(400) & 16.587(400) & DUP \\  
\hline
\enddata
\tablenotetext{a}{DUP and SWO correspond to the du Pont
 and Swope telescopes, respectively.}
\tablecomments{One sigma uncertainties given in parentheses are in millimag.}
\end{deluxetable}

\clearpage
\begin{deluxetable}{ccccc}
\tabletypesize{\scriptsize}
\tablecolumns{5}
\tablenum{4}
\tablewidth{0pt}
\tablecaption{UVOT far-UV Photometry of SN~2005ip and SN~2006jd\label{table4}}
\tablehead{
\colhead{JD--2454000}  &
\colhead{Phase$^{a}$}      &
\colhead{$uvw2$}   &
\colhead{$uvm2$}   &
\colhead{$uvw1$}   }
\startdata
\multicolumn{5}{c}{\bf SN 2005ip}\\
145.51  &  465.9 &18.267(096) & 18.093(092) & 17.612(081)\\
630.13  &  950.5 &17.911(172) & 17.777(292) & 17.200(131)\\
786.71  & 1107.0 &18.451(098) & 18.322(103) & 17.654(081)\\
1153.96 & 1474.3 &19.000(150) & 18.748(123) & 18.202(115)\\
2012.26 & 2332.6 &$\cdots$    &    $\cdots$ & 19.336(206)\\ 
\multicolumn{5}{c}{\bf SN 2006jd}\\
420.77& 399.7&   19.179(156) & 18.756(171) & 18.114(115)\\
426.30& 405.3&   18.983(100) & 18.767(110) & 18.212(084)\\
452.48& 431.4&   18.957(097) & 18.632(096) & 18.132(080)\\
480.63& 459.6&   18.916(096) & 18.709(108) & 18.224(081)\\
517.09& 496.0&   18.860(096) & 18.713(095) & 18.149(081)\\
720.39& 699.3&   19.023(098) & 18.969(113) & 18.386(085)\\
809.52& 788.5&   19.325(116) & 19.053(134) & 18.648(101)\\
\hline
\enddata
\tablenotetext{a}{Days past discovery.}

\tablecomments{One sigma uncertainties given in parentheses are in millimag.}
\end{deluxetable}

\clearpage
\begin{deluxetable}{ccccccccc}
\tabletypesize{\scriptsize}
\tablecolumns{9}
\tablewidth{0pt}
\tablenum{5}
\tablecaption{Spectroscopic Observations\label{specjor}}
\tablehead{
\colhead{Epoch$^{a}$} &
\colhead{Date} &
\colhead{Julian Date} &
\colhead{Telescope} &
\colhead{Instrument} &
\colhead{Range} &
\colhead{Resolution} &
\colhead{\# of exposures} &
\colhead{Integration} \\ 
\colhead{} &
\colhead{} &
\colhead{JD--2450000+} &
\colhead{} &
\colhead{} &
\colhead{(\AA)} &
\colhead{(FWHM \AA)} &
\colhead{} &
\colhead{(s)}}
\startdata
\multicolumn{9}{c}{\bf SN 2005ip}\\
18  &2005 Nov 23  & 3697.8   & du Pont & MS     &  3780 -- 7289   & 7  & 3 & 400 \\
19  &2005 Nov 24  & 3698.8   & du Pont & MS     &  3780 -- 7290   & 7  & 3 & 600 \\
20  &2005 Nov 25  & 3699.8   & du Pont & MS     &  3780 -- 7289   & 7  & 3 & 400 \\
20  &2005 Nov 25  & 3699.8   & CTIO60  & CS     &  3200 -- 9535   & 14 & 3 & 600 \\
43  &2005 Dec 18  & 3722.8   & NTT     & EMMI   &  4000 -- 10200  & 8 & 3 & 300 \\
45  &2005 Dec 20  & 3724.8   & du Pont & WFCCD  &  3800 -- 9235   & 8 & 3 & 400 \\
46  &2005 Dec 21  & 3725.8   & du Pont & WFCCD  &  3800 -- 9208   & 8 & 3 & 500 \\
48  &2005 Dec 23  & 3727.8   & du Pont & WFCCD  &  3800 -- 9235   & 8 & 3 & 500 \\
72  &2006 Jan 16  & 3751.8   & NTT     & EMMI   &  4000 -- 10200  & 8 & 3 & 200 \\
80  &2006 Jan 24  & 3759.8   & Clay    & LDSS   &  3788 -- 9980   & 4 & 3 & 300 \\
120 &2006 Mar 05  & 3799.7   & du Pont & WFCCD  &  3800 -- 9235   & 8 & 3 & 600 \\
130 &2006 Mar 15  & 3809.6   & NTT     & EMMI   &  4000 -- 10200  & 8 & 3 & 300 \\
130 &2006 Mar 15  & 3809.6   & CLAY    & LDSS   &  3785 -- 9969   & 4 & 3 & 400 \\
138 &2006 Mar 23  & 3817.6   & du Pont & WFCCD  &  3800 -- 9235   & 8 & 3 & 600 \\
169 &2006 Apr 23  & 3848.5   & du Pont & WFCCD  &  3800 -- 9235   & 8 & 3 & 900 \\
1844&2010 Nov 10  & 5523.7   & INT     & IDS    &  3871 -- 7756   & 6 & 3 & 1200 \\
2352&2012 Apr 13  & 6031.0   & NOT     & Alfosc &  3675 -- 8985   & 17& 1 & 3600  \\
\hline
\multicolumn{9}{c}{\bf SN 2006jd}\\
22  &2006 Nov 03  & 4042.8  & NTT     & EMMI   & 3200 -- 10200  & 8  & 3 & 200 \\ 
35  &2006 Nov 16  & 4055.8  & du Pont & WFCCD  & 3803 -- 9235   & 8  & 3 & 500 \\
41  &2006 Nov 22  & 4061.8  & du Pont & WFCCD  & 3800 -- 9235   & 8  & 3 & 600 \\
61  &2006 Dec 13  & 4082.0  & du Pont & B\&C   & 3625 -- 9832   & 8  & 3 & 600 \\
67  &2006 Dec 18  & 4087.8  & du Pont & WFCCD  & 3800 -- 9235   & 8  & 3 & 600 \\
94  &2007 Jan 14  & 4114.7  & du Pont & B\&C   & 3625 -- 9835   & 8  & 3 & 1800\\
109 &2007 Jan 29  & 4129.6  & Baade   & IMACS  & 4279 -- 9538   & 4  & 3 & 1200\\
122 &2007 Feb 11  & 4142.7  & du Pont & WFCCD  & 3800 -- 9235   & 8  & 1 & 1200\\
124 &2007 Feb 13  & 4144.7  & du Pont & WFCCD  & 3800 -- 9235   & 8  & 3 & 1200\\
188 &2007 Apr 18  & 4208.5  & du Pont & WFCCD  & 3800 -- 9235   & 8  & 3 & 1200\\
414 &2007 Nov 30  & 4434.8  & 3.6-m   & EFOSC  & 3300 -- 9260   & 13 & 3 & 1200\\
417 &2007 Dec 03  & 4437.8  & du Pont & WFCCD  & 3800 -- 9235   & 8  & 3 & 1200\\
423 &2007 Dec 09  & 4443.8  & du Pont & WFCCD  & 3800 -- 9235   & 8  & 3 & 1200\\
449 &2008 Jan 04  & 4469.7  & 3.6-m   & EFOSC  & 3300 -- 9260   & 13 & 3 & 1200\\
930 &2009 Apr 30  & 4951.5  & Clay    & LDSS   & 3704 -- 9424   & 7  & 3 & 900\\
1173&2009 Dec 28  & 5193.9  & Baade   & IMACS  & 3976 -- 10050  & 6  & 1 & 900\\
1209&2010 Feb 02  & 5229.7  & Baade   & IMACS  & 3976 -- 10050  & 5  & 3 & 1200\\
1542&2011 Jan 01  & 5563.0  & Gemini-North & GMOS & 3588 -- 9653& 8  & 1 & 3000\\
\enddata
\tablenotetext{a}{Days past discovery.}
\end{deluxetable}

\clearpage
\begin{deluxetable}{cccccccc}
\setlength{\tabcolsep}{0.02in}
\tabletypesize{\scriptsize}
\tablecolumns{8}
\tablenum{6}
\tablewidth{0pt}
\tablecaption{Optical Photometry of SN~2006jd in the Natural System\label{table5}}
\tablehead{
\colhead{JD--2450000+} &
\colhead{$u$ }  &
\colhead{$g$ }   &
\colhead{$r$ }   &
\colhead{$i$ }   &
\colhead{$B$ }   &
\colhead{$V$ } &
\colhead{Telescope\tablenotemark{a}} } 
\startdata
4028.89  &  $\cdots$   &  $\cdots$   &  $\cdots$   &  $\cdots$   & 18.177(061)  & 17.741(017)   & SWO \\  
4029.88  & 18.583(024) & 17.881(011) & 17.453(012) & 17.545(027) & 18.130(015)  & 17.745(013)   & SWO \\
4036.88  & 18.646(089) & 18.051(015) &  $\cdots$   &  $\cdots$   & 18.265(024)  & 17.872(014)   & SWO \\
4040.87  & 18.644(038) & 18.051(015) & 17.638(014) &  $\cdots$   & 18.274(014)  & 17.892(014)   & SWO \\
4042.86  &  $\cdots$   & 18.038(012) & 17.658(014) & 17.799(016) &  $\cdots$    &  $\cdots$     & SWO \\
4046.82  & 18.666(061) & 18.061(016) & 17.697(013) & 17.799(016) & 18.309(025)  & 17.947(017)   & SWO \\
4066.86  &  $\cdots$   & 18.226(015) & 17.963(015) & 18.111(032) & 18.457(017)  & 18.146(019)   & SWO \\
4071.76  & 18.967(042) & 18.270(013) & 18.019(016) & 18.110(019) & 18.497(019)  & 18.205(018)   & SWO \\ 
4076.81  & 18.980(069) & 18.280(017) & 18.107(018) & 18.166(020) & 18.538(026)  & 18.233(020)   & SWO \\    
4085.79  & 18.971(028) & 18.333(014) & 18.174(019) & 18.273(022) & 18.566(014)  & 18.299(017)   & SWO \\
4087.83  & 18.983(024) & 18.363(018) & 18.195(021) & 18.277(024) & 18.600(017)  & 18.325(022)   & SWO \\    
4092.81  & 19.053(030) & 18.416(019) & 18.253(023) & 18.341(026) & 18.639(018)  & 18.383(024)   & SWO \\                
4096.77  &  $\cdots$   & 18.446(018) & 18.282(022) & 18.347(022) &  $\cdots$    &  $\cdots$     & SWO \\
4099.77  & 19.051(028) & 18.463(016) & 18.362(019) & 18.382(027) & 18.692(018)  & 18.471(023)   & SWO \\                   
4102.74  &  $\cdots$   & 18.530(023) & 18.393(027) & 18.472(025) & 18.784(035)  & 18.533(031)   & SWO \\       
4105.80  &  $\cdots$   & 18.545(024) & 18.380(025) & 18.485(030) & 18.780(029)  & 18.481(023)   & SWO \\                  
4110.76  &  $\cdots$   & 18.540(015) & 18.423(019) & 18.506(024) & 18.748(016)  & 18.523(017)   & SWO \\                 
4112.77  &  $\cdots$   & 18.568(017) & 18.448(021) & 18.505(025) & 18.802(016)  & 18.537(019)   & SWO \\                
4119.74  &  $\cdots$   & 18.554(015) & 18.407(021) & 18.491(028) & 18.759(024)  & 18.525(021)   & SWO \\
4121.74  &  $\cdots$   & 18.571(017) & 18.447(020) & 18.573(027) & 18.841(017)  & 18.523(021)   & SWO \\
4125.72  &  $\cdots$   & 18.615(020) & 18.501(026) & 18.623(027) & 18.850(017)  & 18.598(022)   & SWO \\
4130.65  &  $\cdots$   & 18.698(017) & 18.599(021) & 18.699(028) & 18.902(022)  & 18.693(024)   & SWO \\                 
4133.71  &  $\cdots$   & 18.724(031) & 18.608(033) & 18.667(036) & 18.884(043)  & 18.669(038)   & SWO \\
4136.74  &  $\cdots$   & 18.720(017) & 18.622(020) & 18.751(032) & 18.931(022)  & 18.730(021)   & SWO \\
4143.69  &  $\cdots$   & 18.704(025) & 18.582(033) & 18.744(033) & 18.946(023)  & 18.725(025)   & SWO \\
4163.67  &  $\cdots$   & 18.818(023) & 18.687(025) & 18.861(028) & 18.995(039)  & 18.841(028)   & SWO \\
4170.61  & 19.288(024) & 18.788(023) & 18.655(030) & 18.857(035) & 18.995(020)  & 18.798(027)   & SWO \\  
4190.56  & 19.391(053) & 18.872(019) & 18.695(021) & 18.916(029) & 19.051(024)  & 18.945(022)   & SWO \\
4225.53  & 19.217(065) & 18.868(022) & $\cdots$    & $\cdots$    & 19.009(031)  & 18.896(024)   & SWO \\
4383.84  &  $\cdots$   & 18.858(020) & 18.011(013) & 18.879(038) & 19.047(018)  & 18.821(024)   & SWO \\
4392.88  &  $\cdots$   & $\cdots$    & 18.004(014) & 19.061(035) & $\cdots$     & $\cdots$      & SWO \\
4394.84  &  $\cdots$   & 18.905(023) & 17.995(019) & 19.061(035) & 19.122(024)  & 18.921(027)   & SWO \\
4400.80  &  $\cdots$   & 18.835(033) & 17.943(014) & 18.877(051) & 19.070(055)  & 18.854(065)   & SWO \\
4411.83  &  $\cdots$   & 18.875(025) & 17.948(017) & 19.051(033) & 19.108(022)  & 18.892(029)   & SWO \\  
4413.86  &  $\cdots$   & 18.835(019) & 17.925(012) & 18.956(050) & 19.027(020)  & 18.804(025)   & SWO \\
4417.77  &  $\cdots$   & 18.876(022) & 17.925(017) & 19.034(035) & $\cdots$     & 18.902(031)   & SWO \\
4421.75  &  $\cdots$   & 18.893(021) & 17.912(016) & 18.989(036) & 19.129(024)  & 18.898(030)   & SWO \\
4435.78  &  $\cdots$   & 18.829(018) & 17.850(012) & 18.925(036) & 19.075(020)  & 18.825(024)   & SWO \\
4440.78  &  $\cdots$   & 18.863(029) & 17.844(022) & 18.984(037) & 19.083(026)  & 18.868(029)   & SWO \\
4445.77  &  $\cdots$   & 18.837(032) & 17.826(021) & 19.023(035) & 19.107(022)  & 18.864(032)   & SWO \\  
4447.74  & 19.295(026) & 18.831(026) & 17.823(015) & 18.941(033) & 19.083(023)  & 18.850(029)   & SWO \\
4453.77  &  $\cdots$   & 18.838(026) & 17.802(018) & 18.986(036) & 19.091(021)  & 18.854(027)   & SWO \\
4455.72  &  $\cdots$   & 18.841(022) & 17.822(011) & 19.005(034) & 19.101(022)  & 18.904(023)   & SWO \\ 
4457.76  &  $\cdots$   & 18.868(024) & 17.824(012) & 18.918(033) & 19.096(031)  & 18.905(029)   & SWO \\             
4464.72  &  $\cdots$   & 18.892(023) & 17.798(020) & 19.025(037) & 19.133(020)  & 18.888(024)   & SWO \\             
4466.76  &  $\cdots$   & 18.878(023) & 17.788(018) & 18.978(037) & 19.131(020)  & 18.875(026)   & SWO \\   
4469.67  &  $\cdots$   & 18.809(020) & 17.784(013) & 18.948(033) & 19.055(017)  & 18.791(024)   & SWO \\
4470.70  &  $\cdots$   & 18.839(023) & 17.784(014) & 18.989(036) & 19.109(022)  & 18.871(030)   & SWO \\           
4480.70  &  $\cdots$   & 18.874(020) & 17.782(015) & 19.059(038) & 19.156(021)  & 18.865(027)   & SWO \\
4484.64  &  $\cdots$   & 18.929(017) & 17.776(011) & 18.991(032) & 19.120(027)  & 18.912(025)   & SWO \\
4488.68  &  $\cdots$   & 18.985(037) & 17.759(014) & 19.037(050) & 19.176(059)  & 18.998(045)   & SWO \\            
4492.74  &  $\cdots$   &  $\cdots$   & 17.768(012) & 18.991(036) & 19.204(126)  & 18.896(022)   & SWO \\
4512.65  &  $\cdots$   &  $\cdots$   & 17.721(012) & 19.058(048) & $\cdots$     & 18.913(064)   & SWO \\                  
4521.65  &  $\cdots$   &  $\cdots$   & 17.725(014) & 19.055(035) & $\cdots$     & 18.904(024)   & SWO \\
4523.60  &  $\cdots$   &  $\cdots$   & 17.705(017) & 19.059(039) & $\cdots$     & 18.885(025)   & SWO \\
4540.64  &  $\cdots$   &  $\cdots$   & 17.736(018) & 19.209(037) & $\cdots$     & 19.003(024)   & DUP \\
4549.54  &  $\cdots$   &  $\cdots$   & 17.706(014) & 19.119(036) & $\cdots$     & $\cdots$      & SWO \\  
4564.51  &  $\cdots$   &  $\cdots$   & 17.684(013) & 19.040(038) & $\cdots$     & $\cdots$      & SWO \\
4575.55  &  $\cdots$   &  $\cdots$   & 17.726(014) & 19.182(037) & $\cdots$     & $\cdots$      & SWO \\
4591.49  &  $\cdots$   &  $\cdots$   & 17.727(021) & 19.196(043) & $\cdots$     & $\cdots$      & SWO \\
4774.31  & 19.876(030) & 19.194(019) & 17.848(044) & 19.783(040) & 19.600(020)  & 19.163(037)   & DUP \\ 
5268.21  & 21.499(067) & 21.009(061) & 19.416(033) & 21.088(085) & 21.405(059)  & 20.921(067)   & DUP \\
5659.29  & 22.013(080) & 21.524(072) & $\cdots$    & $\cdots$    & 21.721(071)  & 21.492(088)   & DUP \\
\hline
\enddata
\tablenotetext{a}{DUP and SWO correspond to the du Pont
 and Swope telescopes, respectively.}
\tablecomments{One sigma uncertainties given in parentheses are in millimag.}
\end{deluxetable}

\clearpage
\begin{deluxetable}{cccccc}
\tabletypesize{\scriptsize}
\tablecolumns{6}
\tablenum{7}
\tablewidth{0pt}
\tablecaption{Near-infrared Photometry of SN~2006jd\label{table6}}
\tablehead{
\colhead{JD--2450000+}  &
\colhead{$Y$}   &
\colhead{$J$}   &
\colhead{$H$}   &
\colhead{$K_s$}   &
\colhead{Telescope\tablenotemark{a}}}
\startdata
4041.84  & 16.711(024) & 16.753(043) & 16.021(039) & $\cdots$    & SWO \\  
4043.80  & 16.779(036) & 16.752(052) & 16.012(053) & $\cdots$    & SWO \\        
4047.83  & 16.738(031) & 16.624(042) & $\cdots$    & $\cdots$    & SWO \\
4053.86  & 16.789(022) & 16.641(036) & 15.925(037) & $\cdots$    & SWO \\
4057.77  & 16.763(033) & 16.567(039) & 15.890(035) & $\cdots$    & SWO \\
4066.85  & 16.854(021) & 16.669(021) & $\cdots$    & $\cdots$    & DUP \\
4072.76  & 16.853(019) & 16.565(027) & 15.645(025) & $\cdots$    & SWO \\
4077.80  & 16.903(019) & 16.558(031) & 15.646(022) & $\cdots$    & SWO \\
4082.84  & 16.922(019) & 16.596(034) & 15.615(025) & $\cdots$    & SWO \\
4086.82  & 16.936(021) & 16.550(027) & 15.612(028) & $\cdots$    & SWO \\
4093.76  & 16.981(036) & 16.523(031) & 15.653(038) & $\cdots$    & SWO \\
4100.75  & 17.073(024) & 16.684(021) & 15.625(021) & $\cdots$    & DUP \\
4107.76  & 17.114(030) & 16.563(027) & 15.577(029) & $\cdots$    & SWO \\
4109.82  & 17.148(026) & 16.621(028) & $\cdots$    & $\cdots$    & SWO \\
4114.74  & 17.149(023) & 16.756(039) & 15.604(029) & $\cdots$    & SWO \\
4120.72  & 17.188(025) & 16.644(024) & 15.569(018) & $\cdots$    & SWO \\
4122.75  & 17.189(022) & 16.688(024) & 15.636(023) & $\cdots$    & SWO \\
4124.73  & 17.192(025) & 16.662(030) & 15.546(025) & $\cdots$    & SWO \\
4126.67  & 17.206(028) & 16.699(033) & 15.604(029) & $\cdots$    & SWO \\
4129.65  & 17.238(025) & 16.696(026) & 15.639(021) & $\cdots$    & SWO \\
4135.65  & 17.257(026) & 16.741(025) & 15.622(024) & $\cdots$    & SWO \\
4138.63  & 17.275(027) & 16.652(034) & 15.633(023) & $\cdots$    & SWO \\
4140.61  & 17.245(030) & 16.727(032) & 15.620(027) & $\cdots$    & SWO \\
4146.64  & 17.308(030) & 16.729(028) & 15.595(027) & $\cdots$    & SWO \\
4150.63  & 17.332(027) & 16.724(031) & 15.602(024) & $\cdots$    & SWO \\
4158.66  & 17.380(021) & 16.887(021) & 15.687(015) & 14.584(021) & DUP \\
4162.61  & 17.345(037) & 16.803(041) & 15.651(035) & $\cdots$    & SWO \\
4164.64  & 17.344(031) & 16.771(031) & 15.669(024) & $\cdots$    & SWO \\
4169.58  & 17.413(021) & 16.915(021) & 15.709(021) & 14.559(021) & DUP \\          
4175.55  & 17.353(029) & 16.881(046) & 15.627(028) & $\cdots$    & SWO \\
4189.56  & 17.446(046) & 16.823(041) & 15.684(042) & $\cdots$    & SWO \\
4198.51  & 17.441(036) & 16.755(052) & 15.660(034) & $\cdots$    & SWO \\
4220.48  & 17.463(047) & 16.915(063) & 15.672(040) & $\cdots$    & SWO \\
4234.47  & 17.322(081) & 16.817(074) & 15.616(054) & $\cdots$    & SWO \\
4247.46  & 17.466(054) & 16.704(056) & 15.551(054) & $\cdots$    & SWO \\
4384.86  & 17.319(030) & 16.775(034) & 15.507(021) & $\cdots$    & SWO \\
4393.85  & 17.325(034) & 16.791(037) & 15.555(021) & $\cdots$    & SWO \\
4396.87  & 17.354(027) & 16.731(030) & 15.559(026) & $\cdots$    & SWO \\
4408.85  & 17.326(034) & 16.783(038) & 15.542(033) & $\cdots$    & SWO \\
4416.83  & 17.372(025) & 16.806(027) & 15.532(020) & $\cdots$    & SWO \\
4424.84  & 17.330(028) & 16.780(032) & 15.560(025) & $\cdots$    & SWO \\
4431.81  & 17.311(032) & 16.792(041) & 15.528(046) & $\cdots$    & SWO \\
4432.75  & 17.326(032) & 16.743(028) & 15.544(038) & $\cdots$    & SWO \\
4437.79  & 17.298(024) & 16.745(028) & 15.519(029) & $\cdots$    & SWO \\
4439.78  & 17.330(021) & 16.768(028) & 15.505(029) & $\cdots$    & SWO \\
4446.86  & 17.288(021) & 16.753(024) & $\cdots$    & $\cdots$    & SWO \\
4454.73  & 17.282(024) & 16.761(023) & 15.501(022) & $\cdots$    & SWO \\
4465.76  & 17.246(019) & 16.753(018) & 15.514(016) & $\cdots$    & SWO \\
4473.78  & 17.289(026) & 16.737(026) & 15.479(015) & $\cdots$    & SWO \\
4487.75  & 17.314(023) & 16.741(030) & 15.482(016) & $\cdots$    & SWO \\
4498.73  & 17.289(020) & 16.751(020) & 15.496(015) & $\cdots$    & SWO \\
4513.70  & 17.359(057) & 16.780(049) & 15.512(021) & $\cdots$    & SWO \\
4530.65  & 17.315(021) & 16.785(024) & 15.511(023) & $\cdots$    & SWO \\                  
4536.62  & 17.292(023) & 16.742(027) & 15.550(021) & $\cdots$    & SWO \\
4544.64  & 17.331(022) & 16.843(024) & 15.574(027) & $\cdots$    & SWO \\                    
4550.63  & 17.320(027) & 16.872(028) & $\cdots$    & $\cdots$    & SWO \\
4627.50  & 17.490(025) & 17.140(040) & $\cdots$    & $\cdots$    & SWO \\
4841.65  & $\cdots$    & $\cdots$    & 16.441(021) & $\cdots$    & DUP \\
4954.84  & 18.527(056) & 18.348(074) & 16.725(051) & 14.955(021) & SWO \\
5133.30  & 18.900(060) & 18.968(079) & 17.324(039) & 15.358(030) & DUP \\                        
5194.76  & 19.085(055) & 19.119(075) & 17.508(030) & 15.512(023) & DUP \\
5260.57  & 19.012(052) & 19.321(092) & 17.734(037) & 15.673(026) & DUP \\         
5314.50  & 19.145(067) & 19.440(106) & 17.829(065) & $\cdots$    & DUP\\
5521.50  & 19.453(062) & 19.611(121) & 18.273(067) & 16.194(025) & DUP \\          
5613.72  & 19.396(083) & 19.749(117) & 18.510(096) & 16.433(043) & DUP \\
\hline
\enddata
\tablenotetext{a}{DUP and SWO correspond to the du Pont
 and Swope telescopes, respectively.}
\tablecomments{One sigma uncertainties given in parentheses are in millimag.}
\end{deluxetable}

\clearpage
\begin{deluxetable}{lccccccc}
\tablecolumns{8}
\tablenum{8}
\tablewidth{0pt}
\tablecaption{Physical Parameters of $BB$ fits\label{table8}}
\tablehead{
\colhead{Epoch$^{a}$}    &
\colhead{T$_w$}          &
\colhead{log$_{10}$(L$_w$)}     &
\colhead{R$_w$}          &
\colhead{T$_h$}          &
\colhead{log$_{10}$(L$_h$)}     &
\colhead{R$_h$}          &
\colhead{$L_{hot}/L_{tot}$}\\
\colhead{(days)}         &
\colhead{(K)}            &
\colhead{(erg s$^{-1}$)} &
\colhead{(10$^{16}$ cm)} &
\colhead{(K)}            &
\colhead{(erg s$^{-1}$)} &
\colhead{(10$^{14}$ cm)} &
\colhead{(percent)}}     
\startdata
\multicolumn{8}{c}{\bf SN 2005ip}\\
10      &$\cdots$&$\cdots$&$\cdots$& 6484 & 42.485 & 15.58  & $\cdots$\\
58      & 1413  & 41.444  & 0.989  & 6704 & 42.100 & 9.353  & 82      \\
99      & 1596  & 41.775  & 1.135  & 7146 & 41.680 & 5.073  & 45      \\
211     & 1544  & 41.794  & 1.240  & 6613 & 41.363 & 4.115  & 27      \\
722     & 1233  & 41.743  & 1.834  & 7123 & 41.233 & 3.053  & 24     \\
1565    & 969   & 41.781  & 3.101  & 6565 & 40.965 & 2.639  & 13      \\
\hline
\multicolumn{8}{c}{\bf SN 2006jd}\\
9       & 1194  & 41.938 & 2.447  & 6999  & 42.355 & 10.51  & 72 \\
51      & 1478  & 42.148 & 2.034  & 7353  & 42.180 & 8.523  & 57 \\
99      & 1562  & 42.210 & 1.955  & 8097  & 42.221 & 6.165  & 42 \\
204     & 1515  & 42.245 & 2.165  & 9307  & 41.969 & 4.176  & 35  \\
753     & 1157  & 42.333 & 4.106  & 8018  & 41.766 & 4.450  & 21  \\
1638    &  980  & 41.747 & 2.917  & 8479  & 40.881 & 1.437  & 12  \\
\hline
\enddata
\tablenotetext{a}{Days past discovery.}
\end{deluxetable}

\clearpage
\begin{deluxetable}{cccc}
\tablecolumns{4}
\tablewidth{0pt}
\tablenum{9}
\tablecaption{Warm Dust Mass Estimates\label{table9}}
\tablehead{
\colhead{Days Past} &
\colhead{M${_d}$$(10^{-4}$$M_{\sun})$} &
\colhead{M${_d}$$(10^{-4}$$M_{\sun})$} &
\colhead{M${_d}$$(10^{-4}$$M_{\sun})$}\\
\colhead{Discovery} &
\colhead{$a=1~\mu$$m$} &
\colhead{$a=0.1~\mu$$m$} &
\colhead{$a=0.01~\mu$$m$}}
\startdata
\multicolumn{4}{c}{\bf SN 2005ip}\\
10  & $\cdots$ & $\cdots$ & $\cdots$ \\
58  & 0.20 & 0.45 & 0.53 \\
99  & 0.26 & 0.62 & 0.71 \\
211 & 0.32 & 0.83 & 0.93 \\
722 & 0.72 & 1.88 & 2.09 \\
1565& 2.05 & 5.41 & 6.00 \\
\hline
\multicolumn{4}{c}{\bf SN 2006jd}\\
9   & 1.02 & 2.88 &3.32 \\
51  & 0.72 & 1.66 &2.04 \\
99  & 0.67 & 1.60 &1.93 \\
204 & 0.80 & 2.40 &2.71 \\
753 & 2.86 & 8.72 &9.84 \\
1638& 1.44 & 4.62 &5.14 \\
\enddata
\end{deluxetable}

\clearpage
\begin{deluxetable}{ccccc}
\tablecolumns{5}
\tablewidth{0pt}
\tablenum{10}
\tablecaption{Pre-Existing Dust Masses and $BB_c$ Physical Parameters\label{table10}}
\tablehead{
\colhead{SN} &
\colhead{Days Past}   & 
\colhead{$M_{d}$} &
\colhead{$R$} &
\colhead{$T$}\\
\colhead{} &
\colhead{Discovery} &
\colhead{($M_{\sun})$} &
\colhead{(10$^{16}$cm)} &
\colhead{(K)}}
\startdata
2005ip    & 930  & 0.01    & 5.14  & 502 \\
2006jd    & 1150 & 0.02    & 8.47  & 726  \\
2006jd    & 1638 & 0.02    & 9.28  & 578  \\
\enddata
\tablecomments{Best-fit values for graphite composition with grain size of 
0.1 $\mu$m.}
\end{deluxetable}

\clearpage
\begin{deluxetable}{lcccc}
\tablecolumns{4}
\scriptsize
\tablewidth{0pt}
\tablenum{11}
\tablecaption{Fluxes of Lines Indicated in Figure~\ref{6jdspeccomp}\label{table11}}
\tablehead{
\colhead{ID}          &
\colhead{$\lambda$}   & 
\colhead{Flux d22}    &
\colhead{Flux d1542}  \\
\colhead{}            &
\colhead{(\AA)}       &
\colhead{(10$^{-16}$ erg s$^{-1}$ cm$^{-2}$)} &
\colhead{(10$^{-16}$ erg s$^{-1}$ cm$^{-2}$)}}
\startdata
$[$\ion{Ne}{3}$]$    &3869     	& 21.70    & 2.40      \\
$[$\ion{Ne}{3}$]$    &3967      & 5.47     & 1.96      \\
H$\delta$            &4102      & $\cdots$ & 1.57      \\
H$\gamma$            &4340     	& $\cdots$ & 1.00      \\
$[$\ion{O}{3}$]$     &4363    	& 11.60    & 1.53      \\
H$\beta$             &4861      & 6.82     & 0.95      \\
$[$\ion{O}{3}$]$     &4959      & 10.30    & 0.79      \\
$[$\ion{O}{3}$]$     &5007      & 23.90    & 2.68      \\
$[$\ion{Fe}{7}$]$    &5159    	& $\cdots$ & 0.45      \\
$[$\ion{Fe}{7}$]$    &5276 $+[$\ion{Fe}{6}$]$ 	& $\cdots$ & 0.38 \\
$[$\ion{Ar}{10}$]$   &5533     	& $\cdots$ & 0.46      \\
$[$\ion{Fe}{7}$]$    &5721      & $\cdots$ & 0.63      \\
$[$\ion{N}{2}$]$     &5755      &9.09      & 1.39      \\
\ion{He}{1}          &5876  	&1.22   & 2.66         \\
$[$\ion{Fe}{7}$]$    &6087  	& $\cdots$ & 0.55      \\
$[$\ion{O}{1}$]$     &6300  	& $\cdots$ & 0.41      \\
$[$\ion{Si}{3}$]$    &6312  	& $\cdots$ & 0.15      \\
$[$\ion{O}{1}$]$     &6364  	& $\cdots$ & 0.27      \\
$[$\ion{Fe}{10}$]$   &6375  	& $\cdots$ & 0.57      \\
H$\alpha$            &6563  	&86.15     & $\cdots$  \\
$[$\ion{S}{2}$]$     &6717/31   & $\cdots$ & 0.31/0.12  \\
\ion{He}{1}          &7065      &6.79      & 0.54       \\
$[$\ion{Ar}{3}$]$    &7136  	& $\cdots$ & 0.30       \\
$[$\ion{Fe}{2}$]$    &7155  	& $\cdots$ & 0.36       \\
$[$\ion{O}{2}$]$     &7325  	& $\cdots$ & 0.19       \\
$[$\ion{Fe}{11}$]$   &7892 	& $\cdots$ & 0.35       \\
\ion{He}{1}          &8156	& $\cdots$ & 0.09       \\
$[$\ion{Cr}{2}$]$    &8230      & $\cdots$ & 0.14       \\
\enddata
\tablecomments{Fluxes of Lines Indicated in Figure~\ref{6jdspeccomp}.}
\end{deluxetable}

\newpage

\clearpage

\clearpage

\clearpage

\clearpage

\clearpage

\clearpage

\clearpage

\clearpage

\clearpage

\clearpage

\clearpage

\clearpage

\clearpage

\clearpage

\clearpage

\clearpage

\clearpage

\clearpage
\begin{figure}[t]
\plottwo{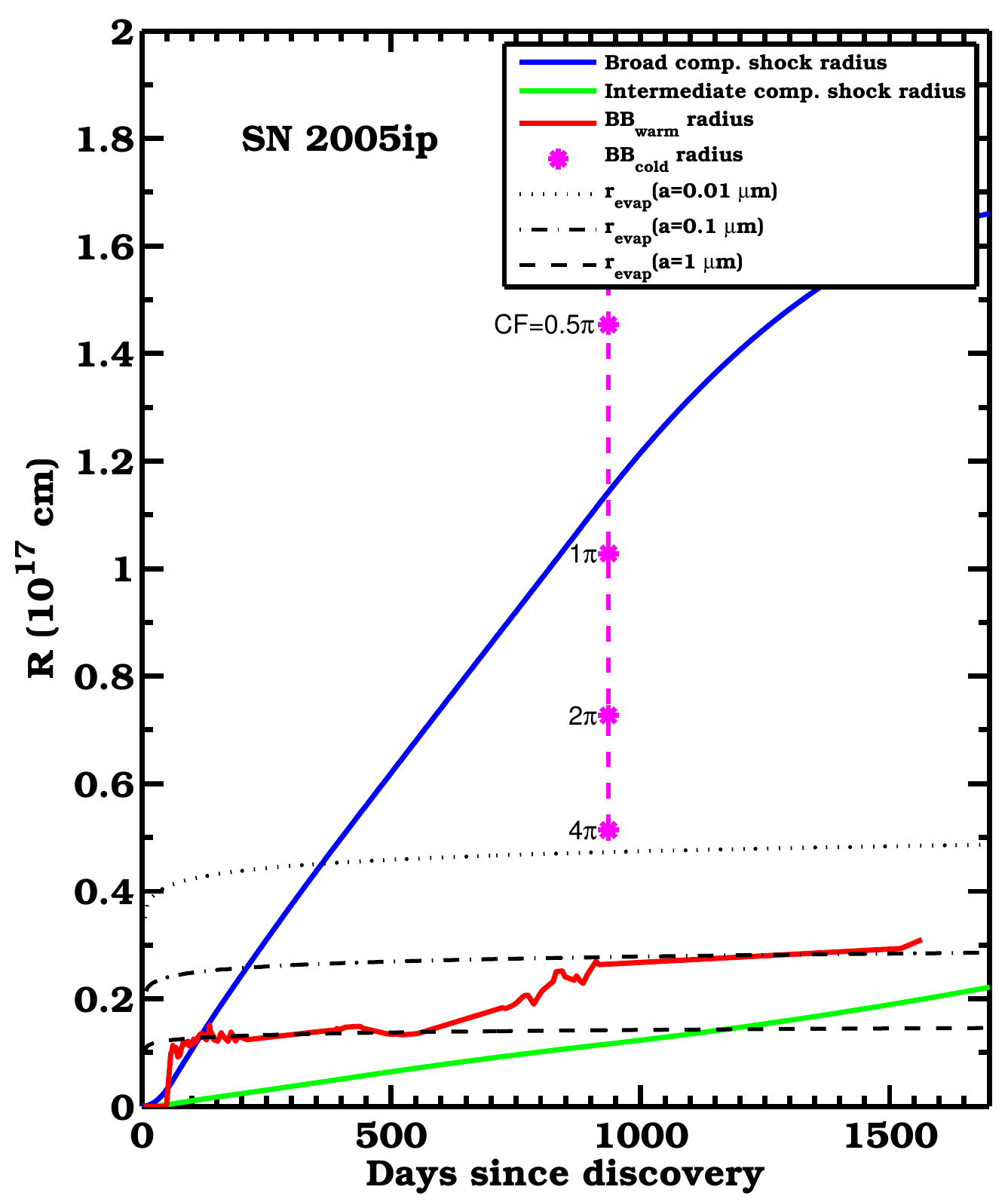}{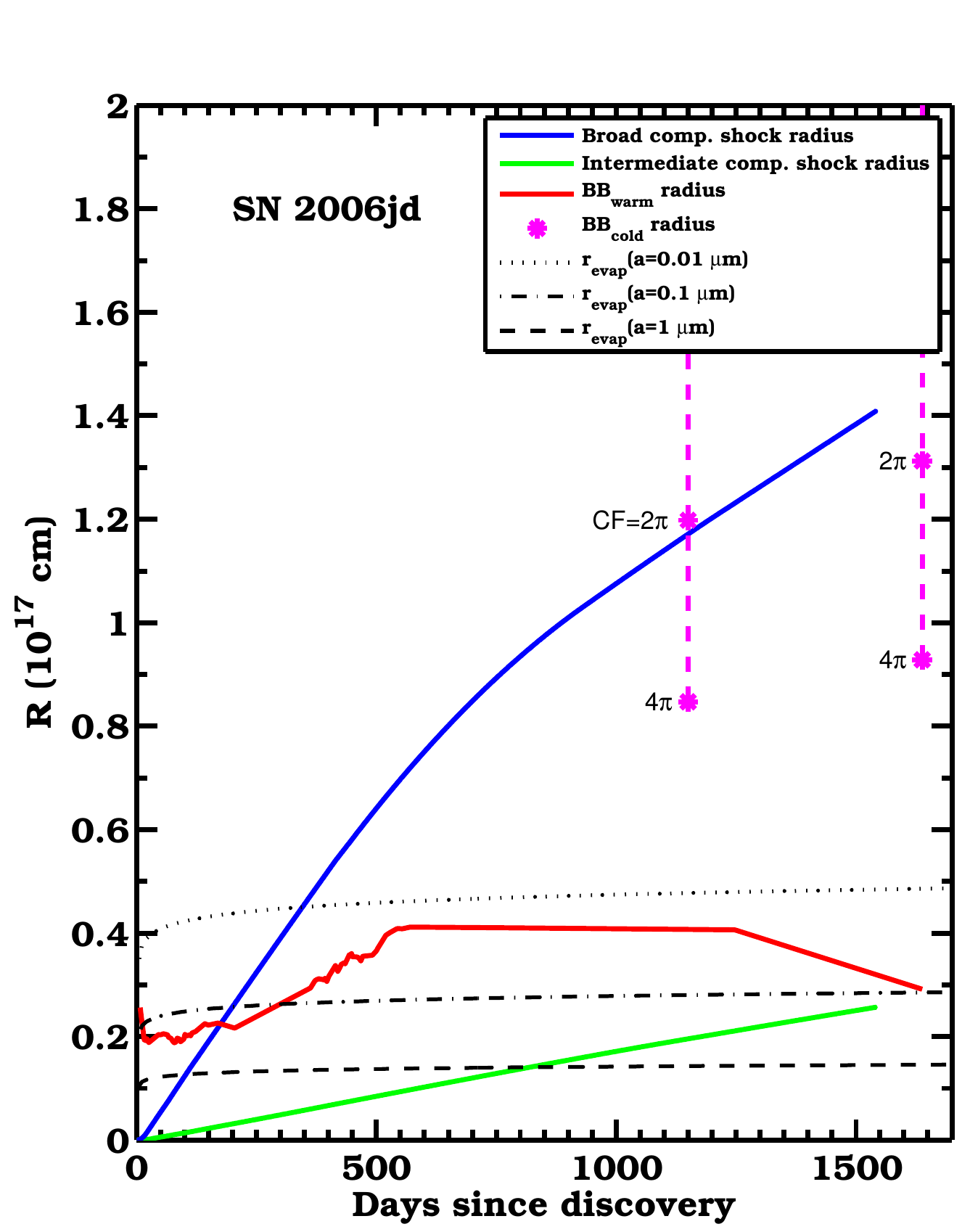}
\figcaption[]{ $BB$ radii of the warm (red line) and cold (vertical, dot-dash line) dust components plotted as a function of days. The continuum of cold dust $BB$ radii are indicated for various covering factors (CF). Also plotted are the 
locations of the fasting moving ejecta (blue line) and post-shocked gas (green line)
estimated from the BVZI of the broad- and intermediate-width H$\alpha$ emission profile. Horizontal dashed, dot-dashed, and doted lines are graphite dust evaporation radii for various grain sizes. \label{dustposition}} 
\end{figure}


\begin{thebibliography}{}


\bibitem[Agnoletto et al.(2009)]{agnoletto09}
Agnoletto, I., Benetti, S., Cappellaro, E., et al.\ 2009, \apj, 691, 1348

\bibitem[Andrews et al.(2011)]{andrews11}
Andrews, J.~E., Clayton, G.~C., Wesson, R., et al.\ 2011, \aj, 142, 45 

\bibitem[Asplund et al.(2009)]{asplund09}
Asplund, M., Grevesse, N., Sauval, A.~J., \& Scoot, P.\ 2009, ARA\&A, 47, 481

\bibitem[Benetti et al.(2006)]{benetti06}
Benetti, S., Cappellaro, E., Turatto, M., et al.\ 2006, \apjl, 653, L129 
 
\bibitem[Blondin, Lundqvist \& Chevalier(1996)]{blondin96}
Blondin, J.~M., Lundqvist, P., \& Chevalier, R.\ 1996, \apj, 472, 257

\bibitem[Blondin et al.(2006)]{blondin06}
Blondin, S., Modjaz, M., Kirshner, R., et al.\ 2006, CBET , 679, 1 

\bibitem[Boles et al.(2005)]{boles05}
Boles,T., Nakano, S., \& Itagaki, K.\ 2005, IAU Circ., 275, 1

\bibitem[Branch et al.(1981)]{branch81}
Branch, D., Falk, S.~W., Uomoto, A.~K., et al.\ 1981, \apj, 244, 780 

\bibitem[Breeveld  et al.(2011)]{breeveld11}
Breeveld, A. A., Landsman, W., Holland, S. T., et al. 2011, in AIP Conf. Proc. 1358,  
Gamma-Ray Bursts 2010, ed. J. E. McEnergy, J. L. Racusin, \& N. Gehrels
(Melville, NY: AIP), 373; arXiv:1102.4717 

\bibitem[Brown et al.(2009)]{brown09}
Brown, P.~J., Holland, S.~T., Immler, S., et al.\ 2009, \aj, 137, 4517 

\bibitem[Cardelli, Clayton, \& Mathis(1989)]{cardelli89}
Cardelli J. A., Clayton G. C., \& Mathis J. S.\ 1989, \apj, 345, 245

\bibitem[Chandra \& Soderberg(2007)]{chandra07}
Chandra, P., \& Soderberg, A.\ 2007, ATel, 1290, 1

\bibitem[Chandra et al.(2012)]{chandra12}
Chandra, P., et al. 2012, submitted to ApJ

\bibitem[Chatzopoulos et al.(2011)]{chatzopoulos11}
Chatzopoulos, E., Wheeler, J.~C., Vinko, J., et al.\ 2011, \apj, 729, 143 

\bibitem[Chevalier \& Fransson(1994)]{chevalier94}
Chevalier, R.~A., \& Fransson, C.\ 1994, \apj, 420, 268

\bibitem[Chevalier \& Irwin(2011)]{chevalier11}
Chevalier, R.~A., \& Irwin, C.~M.,\ 2011, \apj, 729, L6

\bibitem[Chevalier \& Irwin(2012)]{chevalier12}
Chevalier, R.~A., \& Irwin, C.~M.,\ 2012, \apj, 747, L17

\bibitem[Chugai \& Danziger(1994)]{chugai94}
Chugai, N.~N., \& Danziger, I. J.\ 1994, \mnras, 268, 173

\bibitem[Chugai et al.(2004)]{chugai04}
Chugai, N.~N., Blinnikov, S., Cumming, R.~J., et al.\ 2004, \mnras, 352, 1213

\bibitem[Contreras et al.(2010)]{contreras10}
Contreras, C., Hamuy, M., Phillips, M.~M., et al.\ 2010, \aj, 139, 519 

\bibitem[Deng et al.(2004)]{deng04}
Deng, J., Kawabata, K.~S., Ohyama, Y., et al.\ 2004, \apjl, 605, L37 

\bibitem[Dessart et al.(2009)]{dessart09}
Dessart, L., Hillier, J.~D., Suvi, G., et al.\ 2009, \mnras, 394, 21

\bibitem[Dopita et al.(1984)]{dopita84}
Dopita, M. A., Cohen, M., Schwartz, R. D., \& Evans, R.\ 1984, \apj, 287, 69

\bibitem[Dwek(1985)]{dwek85}
Dwek, E.\ 1985, \apj, 297, 719

\bibitem[Foley et al.(2007)]{foley07}
Foley, R.~J., Smith, N., Ganeshalingam, M., et al.\ 2007, \apj, 657, 105

\bibitem[Fox et al.(2009)]{fox09}
Fox, O. D., Skrutskie, M.~F., Chevalier, R.~A., et al.\ 2009, \apj, 691, 650 

\bibitem[Fox et al.(2010)]{fox10}
Fox, O.~D., Chevalier, R.~A., Dwek, E., et al.\ 2010, \apj, 725, 1768 

\bibitem[Fox et al.(2011)]{fox11}
Fox, O.~D., Chevalier, R.~A., Skrutskie, M.~F., et al.\ 2011, \apj, 741, 7 

\bibitem[Fransson et al.(2002)]{fransson02}
Fransson, C., Chevalier, R.~A., Filippenko, A.~V., et al.\ 2002, \apj, 572, 350 

\bibitem[Gall, Hjorth, \& Andersen(2011)]{gall11}
Gall, C., Hjorth, J., \& Andersen, A.~C.\ 2011, \aapr, 19, 43 

\bibitem[Gerardy et al.(2002)]{gerardy02}
Gerardy, C.~L., Fesen, R.~A., Nomoto, K., et al.\ 2002, \apj, 575, 1007 

\bibitem[Germany et al.(2000)]{germany00}
Germany, L.~M., Reiss, D.~J., Sadler, E.~M., Schmidt, B.~P., \& Stubbs, C.~W.\ 2000, \apj, 533, 320 

\bibitem[Gr\"oningsson et al.(2006)]{groningsson06}
Gr\"oningsson, P., Fransson, C., Lundqvist, P., et al.\ 2006, \aap, 456, 581

\bibitem[Hamuy et al.(2003)]{hamuy03} 
Hamuy, M., Phillips, M.~M., Suntzeff, N.~B., et al.\ 2003, \nat, 424, 651

\bibitem[Hamuy et al.(2006)]{hamuy06} 
Hamuy, M., Folatelli, G., Morrell, N.~I., et al.\ 2006, \pasp, 118, 2 

\bibitem[Heger et al.(1997)]{heger97}
Heger, A., Jeannin, L., Langer, N., \& Baraffe, I.\ 1997, \aap, 327, 224 

\bibitem[Henry \& Branch(1987)]{henry87}
Henry, R.~B.~C., \& Branch, D.\ 1987, \pasp, 99, 112

\bibitem[Hoffman et al.(2008)]{hoffman08}
Hoffman, J.~L., Leonard, D.~C., Chornock, R., et al.\ 2008, \apj, 688, 1186 

\bibitem[Hutsemek\'ers et al.(1994)]{hutsemekers94}
Hutsemekers, D., van Drom, E., Gosset, E., \& Melnick, J.\ 1994, \aap, 290, 906 

\bibitem[Immler \& Pooley(2007)]{immlerpooley07}
Immler, S., \& Pooley, D.\ 2007, ATel, 1004, 1

\bibitem[Immler et al.(2007)]{immler07}
Immler, S., Brown, P.~J., Filippenko, A.~V., \& Pooley, D.\ 2007, ATel, 1290, 1 

\bibitem[Kankare et al.(2012)]{kankare12}
Kankare, E., et al. 2012, \mnras, submitted

\bibitem[Kewley \& Dopita(2002)]{kewley02}
Kewley, L. J., \& Dopita, M. A.\ 2002, \apjs, 142, 35

\bibitem[Kiewe et al.(2012)]{kiewe12}
Kiewe, M., Gal-Yam, A., Arcavi, I., et al.\ 2012, \apj, 744, 10 

\bibitem[Landolt(1992)]{landolt92} 
Landolt,~A.~U.\ 1992, \aj, 104, 340

\bibitem[Meikle et al.(2007)]{meikle07}
Meikle, W.~P.~S., Mattila, S., Pastorello, A., et al.\ 2007, \apj, 665, 608 

\bibitem[Miller et al.(2010)]{miller10}
Miller, A.~A., Silverman, J.~M., Butler, N.~R., et al.\ 2010, \mnras, 404, 305 

\bibitem[Moriya et al.(2012)]{moriya12}
Moriya, T. J., Blinnikov, S. I., Tominaga, N., et al. 2012, submitted, arXiv:1204.6109

\bibitem[Nota et al.(1995)]{nota95}
Nota, A., Livio, M., Clampin, M., \& Schulte-Ladbeck, R.\ 1995, \apj, 448, 788 

\bibitem[O'hara et al.(2003)]{ohara03}
O'Hara, T.~B., Meixner, M., Speck, A.~K., Ueta, T., \& Bobrowsky, M.\ 2003, \apj, 598, 1255 

\bibitem[Ofek et al.(2007)]{ofek07}
Ofek, E.~O., Cameron, P.~B., Kasliwal, M.~M., et al.\ 2007, \apjl, 659, L13 

\bibitem[Pastorello et al.(2002)]{pastorello02}
Pastorello, A., Turatto, M., Benetti, S., et al.\ 2002, \mnras, 333, 27 

\bibitem[Pastorello et al. (2009)]{pastorello09}
Pastorello, A., Valenti, S., Zampieri, L., et al.\ 2009, \mnras, 394, 2266 


\bibitem[Persson et al.(1998)]{persson98}
Persson, S.~E., Murphy, D.~C., Krzeminski, W., Roth, M., \& Rieke, M.~J.\ 1998, \aj, 116, 2475 

\bibitem[Pettini \& Pagel(2004)]{pettini04}
Pettini, M., \& Pagel, B.~E.~J.\ 2004, \mnras, 348, 59

\bibitem[Pozzo et al.(2004)]{pozzo04}
Pozzo, M., Meikle, W.~P.~S., Fassia, A., et al.\ 2004, \mnras, 352, 457

\bibitem[Prasad \& Li(2006)]{prasad06}
Prasad, R.~R., \& Li, W.\ 2006, CBET, 673, 1

\bibitem[Quimby et al.(2007)]{quimby07}
Quimby, R.~M., Aldering, G., Wheeler, J.~C., et al.\ 2007, \apj, 668, L99s

\bibitem[Roming et al.(2005)]{roming05}
Roming, P.~W.~A., Kennedy, T.~E., Mason, K.~O., et al.\ 2005, \ssr, 120, 95

\bibitem[Schlegel(1990)]{schlegel90}
Schlegel, E.~M.\ 1990, \mnras, 244, 269 

\bibitem[Schlegel, Finkbeiner \& Davis(1998)]{schlegel98} 
Schlegel,~D.~J., Finkbeiner,~D.~P., \& Davis,~M.\ 1998, \apj, 500, 525 

\bibitem[Smith et al.(2002)]{smith02}  
Smith, J.~A., Tucker, D.~L., Kent, S., et al.\ 2002, \aj, 123, 2121 

\bibitem[Smith(2006)]{smith06}
Smith, N.\ 2006, \apj, 644, 1151

\bibitem[Smith \& McCray(2007)]{smithmccray07}
Smith, N., \& McCray, R. 2007, \apj, 671, L17

\bibitem[Smith et al.(2007)]{smith07}
Smith, N., Li, W., Foley, R.~J., et al.\ 2007, \apj, 666, 1116 

\bibitem[Smith et al.(2008)]{smith08}
Smith, N., Chornock, R., Li, W., et al.\ 2008, \apj, 686, 467 

\bibitem[Smith et al.(2009)]{smith09}
Smith, N., Silverman, J.~M., Chornock, R., et al.\ 2009, \apj, 695, 1334 

\bibitem[Smith et al.(2009)]{smith09b}
Smith, N., Hinkle, K. H., \& Ryde, N. 2009b, \apj, 137, 3558

\bibitem[Smith et al.(2012)]{smith12}
Smith, N., Silverman, J.~M., Filippenko, A.~V., et al.\ 2012, \aj, 143, 17 

\bibitem[Sollerman, Cumming, \& Lundqvist(1998)]{sollerman98}
Sollerman, J., Cumming, R. J., \& Lundqvist, P.\ 1998, \apj, 493, 933

\bibitem[Turatto et al.(1993)]{turatto93} 
Turatto, M., Cappellaro, E., Danziger, I.~J., et al.\ 1993, \mnras, 262, 128 

\bibitem[Turatto et al.(2000)]{turatto00}
Turatto, M., Suzuki, T., Mazzali, P.~A., et al.\ 2000, \apjl, 534, L57 

\bibitem[Turner \& Pearch(1992)]{turner92}
Turner, K., \& Pearce, G.\ 1992, \apss, 190, 1 

\bibitem[Woosley, Blinnikov, \& Heger(2007)]{woosley07}
Woosley, S. E., Blinnikov, S., \& Heger, A.\ 2007, Nature, 450, 390

\bibitem[Yoon \& Cantiello(2010)]{yooncantiello10}
Yoon, S.-C., \& Cantiello, M.\ 2007, \apj, 717, L62 

\end{thebibliography}
\end{document}